\documentclass{aastex63}
\usepackage{amsmath}
\usepackage{multirow}
\usepackage{subfigure}
\usepackage{booktabs}
\begin{document}

\title{GRB 221009A: An ordinary nearby GRB with extraordinary observational properties}
\correspondingauthor{He Gao}
\email{gaohe@bnu.edu.cn}

\author{Lin Lan}
\affil{Institute for Frontier in Astronomy and Astrophysics, Beijing Normal University, Beijing 102206, China;}
\affil{Department of Astronomy, Beijing Normal University, Beijing 100875, China;}

\author{He Gao}
\affil{Institute for Frontier in Astronomy and Astrophysics, Beijing Normal University, Beijing 102206, China;}
\affil{Department of Astronomy, Beijing Normal University, Beijing 100875, China;}

\author{An Li}
\affil{Institute for Frontier in Astronomy and Astrophysics, Beijing Normal University, Beijing 102206, China;}
\affil{Department of Astronomy, Beijing Normal University, Beijing 100875, China;}

\author{Shuo Xiao}
\affil{Guizhou Provincial Key Laboratory of Radio Astronomy and Data Processing, Guizhou Normal University, Guiyang 550001, People’s Republic of China;}
\affil{School of Physics and Electronic Science, Guizhou Normal University, Guiyang 550001, People’s Republic of China;}

\author{Shunke Ai}
\affil{School of Physics and Technology, Wuhan University, Wuhan 430072, China;}

\author{Zong-Kai Peng}
\affil{Institute for Frontier in Astronomy and Astrophysics, Beijing Normal University, Beijing 102206, China;}
\affil{Department of Astronomy, Beijing Normal University, Beijing 100875, China;}

\author{Long Li}
\affil{School of Astronomy and Space Science, University of Science and Technology of China, Hefei 230026, China;}

\author{Chen-Yu Wang}
\affil{Department of Astronomy, Tsinghua University, Beijing, China;}

\author{Nan Xu}
\affil{Institute for Frontier in Astronomy and Astrophysics, Beijing Normal University, Beijing 102206, China;}
\affil{Department of Astronomy, Beijing Normal University, Beijing 100875, China;}

\author{Shijie Lin}
\affil{Institute for Frontier in Astronomy and Astrophysics, Beijing Normal University, Beijing 102206, China;}
\affil{Department of Astronomy, Beijing Normal University, Beijing 100875, China;}

\author{Wei-Hua Lei}
\affil{Department of Astronomy , School of Physics, Huazhong University of Science and Technology, Wuhan, Hubei 430074, China;}

\author{Bing Zhang}
\affil{Nevada Center for Astrophysics, University of Nevada, Las Vegas, NV 89154, USA;}
\affil{Department of Physics and Astronomy, University of Nevada, Las Vegas, NV 89154, USA;}

\author{Yan-Qiu Zhang}
\affil{Key Laboratory of Particle Astrophysics, Institute of High Energy Physics, Chinese Academy of Sciences, Beijing 100049, China;}
\affil{University of Chinese Academy of Sciences, Chinese Academy of Sciences, Beijing 100049, China;}

\author{Chao Zheng}
\affil{Key Laboratory of Particle Astrophysics, Institute of High Energy Physics, Chinese Academy of Sciences, Beijing 100049, China;}
\affil{University of Chinese Academy of Sciences, Chinese Academy of Sciences, Beijing 100049, China;}

\author{Jia-Cong Liu}
\affil{Key Laboratory of Particle Astrophysics, Institute of High Energy Physics, Chinese Academy of Sciences, Beijing 100049, China;}
\affil{University of Chinese Academy of Sciences, Chinese Academy of Sciences, Beijing 100049, China;}

\author{Wang-Chen Xue}
\affil{Key Laboratory of Particle Astrophysics, Institute of High Energy Physics, Chinese Academy of Sciences, Beijing 100049, China;}
\affil{University of Chinese Academy of Sciences, Chinese Academy of Sciences, Beijing 100049, China;}

\author{Chen-Wei Wang}
\affil{Key Laboratory of Particle Astrophysics, Institute of High Energy Physics, Chinese Academy of Sciences, Beijing 100049, China;}
\affil{University of Chinese Academy of Sciences, Chinese Academy of Sciences, Beijing 100049, China;}

\author{Wen-Jun Tan}
\affil{Key Laboratory of Particle Astrophysics, Institute of High Energy Physics, Chinese Academy of Sciences, Beijing 100049, China;}
\affil{University of Chinese Academy of Sciences, Chinese Academy of Sciences, Beijing 100049, China;}

\author{Shao-Lin Xiong}
\affil{Key Laboratory of Particle Astrophysics, Institute of High Energy Physics, Chinese Academy of Sciences, Beijing 100049, China;}

\begin{abstract}
The gamma-ray burst GRB 221009A, known as the ``brightest-of-all-time" (BOAT), is the closest energetic burst detected so far, with an energy of $E_{\gamma,\rm iso} \sim 10^{55}$ ergs. This study aims to assess its compatibility with known GRB energy and luminosity distributions. Our analysis indicates that the energy/luminosity function of GRBs is consistent across various redshift intervals, and that the inclusion of GRB 221009A does not significantly impact the function at low redshifts. Additionally, our evaluation of the best-fitting result of the entire GRB sample suggests that the expected number of GRBs with energy greater than $10^{55}$ ergs at a low redshift is 0.2, so that the emergence of GRB 221009A is consistent with expected energy/luminosity functions within $\sim 2\sigma$ Poisson fluctuation error, still adhering to the principles of small number statistics. Furthermore, we find that GRB 221009A and other energetic bursts, defined as $E_{\gamma,\rm iso} \gtrsim10^{54} {\rm ergs}$, exhibit no significant differences in terms of distributions of $T_{90}$, minimum timescale, Amati relation, $E_{\rm \gamma,iso}$-$E_{\rm X,iso}$ relation, $L_{\gamma,\rm iso}-\Gamma_0$ relation, $E_{\gamma,\rm iso}-\Gamma_0$ relation, $L_{\gamma,\rm iso}-E_{\rm p,i}-\Gamma_0$ relation, and host galaxy properties, compared to normal long GRBs. This suggests that energetic GRBs (including GRB 221009A) and other long GRBs likely have similar progenitor systems and undergo similar energy dissipation and radiation processes. The generation of energetic GRBs may be due to more extreme central engine properties or, more likely, a rarer viewing configuration of a quasi-universal structured jet. 
\end{abstract}


\section{Introduction}

As one of the most violent explosions in the universe, gamma-ray bursts (GRBs) are detected in a wide redshift range (from $z=0.0085$ to $z=9.4$) and a wide energy distribution ($E_{\gamma,\rm iso}$ ranging from $\sim10^{46}$ ergs to $\gtrsim 10^{54}$ ergs) \cite[][for a review]{zhang2018book}. The distribution of $E_{\gamma,\rm iso}$ generally follows a simple power law distribution with a cutoff above $(1-3)\times10^{54}$ ergs \citep{Atteia17}. The cutoff feature, which should not be due to a selection effect because of their high brightness, may be related to some intrinsic limit of generating apparently energetic GRBs. In this paper, we define ``energetic GRBs'' as 
GRBs with the isotropic-equivalent energy $E_{\gamma,\rm iso}\gtrsim10^{54} {\rm ergs}$.

Most recently, the ``brightest-of-all-time'' (BOAT) gamma-ray burst, GRB 221009A, was detected by many space-borne and ground-based telescopes in all wavelength. 
The burst was located at redshift $z=0.151$, and had an isotropic radiation energy of $\sim10^{55}~\rm erg$, making it the most energetic GRB among energetic GRBs \citep{GECAM}.It first triggered the Fermi/Gamma-ray Burst Monitor (GBM) at 13:16:59 UT on 2022 October 9, with a fluence of $(2.12\pm0.05)\times10^{-5}{\rm erg~cm^{-2}}$ in 10–1000 keV within a duration of $T_{90}=327$ s \citep{GCN32642}. Around the similar time, it triggered Insight-HXMT \citep{HXMT} and had been detected by HEBS \citep{HEBS}. Later, it was registered by the Swift Burst Alert Telescope (BAT) at 14:10:17 UT \citep{swiftGCN}. Multiple ground- and space-based follow-up observations was performed, from radio to very high energy $\gamma$-ray \citep{HXMT,GCN32638,GCN32656,GCN32660,GCN32677,GCN32694,GCN32658,GCN32676,GCN32736,GCN32750,GCN32757}. 
Most interestingly, the Large High Altitude Air Shower Observatory (LHAASO) reported more than 5,000  very high energy photons within the first $\sim$2,000 s after the burst trigger with energies above 500 GeV all the way to 18 TeV, making them the most energetic photons ever observed from a GRB \citep{GCN32677}. Various physical models and radiation mechanisms have been proposed to explain the observed 18 TeV photon
\citep{Baktash22,Brdar22,Carenza22,Finke22,Galanti22,Gonzalez22,Li22,Ren22,Sahu22,Smirnov22,Troitsky22,Xia22,Zhang22,Zhao22,Zheng22}. 

With a geometric extrapolation of the total fluence and peak flux distributions, \cite{Burns23} argue that GRB 221009A appears to be a once-in-10,000-year event. Given that the most prominent characteristic of GRB 221009A is its exceedingly high total isotropic energy, our objective is to ascertain the predictability and low probability of its occurrence through a comprehensive examination of the energy/luminosity functions of GRBs. In particular, we intend to investigate the following questions:

\begin{itemize}
    \item Is there an obvious difference between the energy/luminosity distributions of high- and low-redshift energetic burst samples? 
    \item Due to the addition of GRB 221009A, does the energy distribution for the low redshift sample still follow a cutoff power law function? 
    \item Normalized to the existing sample size, what is the expected number of GRBs with energy greater than $10^{55}$ ergs at low redshifts? 
    \item Are GRB 221009A and other energetic GRBs systematically different from other GRBs in terms of statistics of various properties, including the prompt emission, afterglow, and host galaxy properties?
\end{itemize}

\section{GRB sample selection}

For this work, we extensively search for the sample of GRBs with measured redshift (both spectroscopic and photometric) peak energy $E_{\rm p}$, as well as isotropic-equivalent energy $E_{\gamma,\rm iso}$ from published papers or the Gamma-ray Coordinates Network Circulars\footnote{\url{https://gcn.gsfc.nasa.gov/gcn3_archive.html}} if no published paper is available. We eventually find 355 GRBs in total registered from 1997 February up to 2022 November, covering the redshift range from 0.0098 (GRB 170817A) to 9.4 (GRB 090429B). For each burst in our sample, we collect their temporal and spectral properties from previous statistical investigations \citep{Amati08,Zhang09,Qin13,Li16,Zhang18,Zou18,Minaev20,Jia22}. In Table \ref{table-1}, we list their prompt emission duration $T_{\rm 90}$, spectral peak energy in the rest frame $E_{\rm p,i} = E_{\rm p}(1+z)$ and isotropic equivalent energy $E_{\rm iso}$. For GRB 221009A, the value of $E_{\rm iso}$ and $E_{\rm p}$ are calculated based on the observational data from HXMT and HEBS \citep{GECAM}, which made an unprecedentedly accurate measurement of the prompt emission during the first$\sim$1800 s, including its precursor emission, main emission, flaring emission and early afterglow. A record-breaking isotropic equivalent energy $E_{\rm iso}=(1.5\pm0.2)\times10^{55}$ erg was measured based on the HEBS unsaturated data. The peak energy $E_{\rm p}=1247.4\pm91.2$ keV for the time-integrated spectral of full burst in prompt emission.

\begin{figure*}
\centering
\includegraphics    [angle=0,scale=0.7]     {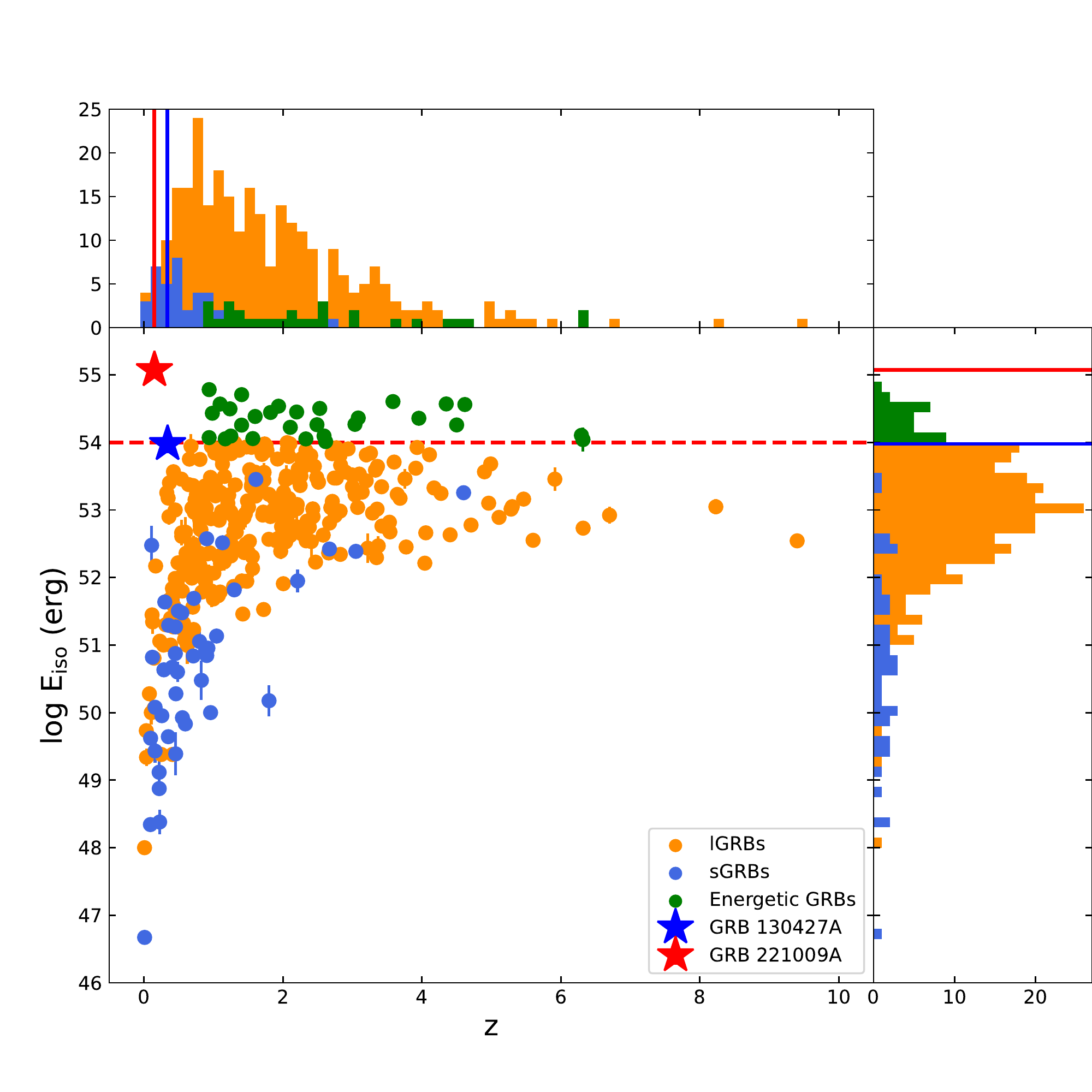}
\caption{The isotropic gamma-ray energy $E_{\gamma,\rm iso}$ for our collected GRBs as a function of the redshift $z$. The blue circles, orange circles, blue stars, red stars, and green circles represent the sGRBs, lGRBs, GRB 130427A, GRB 221009A, and other energetic GRBs, respectively. The red horizontal dotted line represent $E_{\gamma,\rm iso}=10^{54} {\rm ergs}$. The histograms on top and right represent the $E_{\gamma,\rm iso}$ and $z$ distribution for all GRBs, and the blue and red vertical lines represent GRB 130427A and GRB 221009A location, respectively.}  
\label{fig:z-Eiso}
\end{figure*}

The bursts in the sample are mainly observed by the Konus-\emph{Wind}, \emph{Swift}, and \emph{Fermi} satellites, including 36 sGRBs and 319 lGRBs, with 12 bursts being characterized by an extended emission (EE) component. Among the sample, 31 are energetic GRBs\footnote{We take GRB 130427A as an energetic GRB, although its isotropic energy is $9.51\times10^{53}~{\rm ergs}$, since it has similar characteristics to GRB 221009A in temporal and spectral properties from the prompt emission to afterglow.} with $E_{\gamma,\rm iso}\gtrsim10^{54}~{\rm ergs}$. Figure \ref{fig:z-Eiso} shows the distribution of redshift and $E_{\rm iso}$ for our sample. 

\setlength{\tabcolsep}{3mm}
\begin{deluxetable}{cccccccccc}
\tabletypesize{\scriptsize}
\tablecolumns{8}
\tablewidth{0pc}
\tablecaption{The sample of gamma-ray bursts. \label{table-1}}
\tablenum{1}
\tablehead{
\colhead{GRB$^a$} & \colhead{$z^b$} & \colhead{$T_{\rm 90}^c$} & \colhead{$E_{\rm iso}^d$} & \colhead{$E_{\rm p,i}^e$} & \colhead{Type$^f$} & \colhead{Detector$^g$} & \colhead{Refs$^h$}\\
\colhead{}        &  \colhead{}     & \colhead{(s)}             &  \colhead{($10^{52}$ erg)} & \colhead{(keV)}          & \colhead{}         &     \colhead{}         & \colhead{}}

\startdata
970228  & 0.695    & 56.0   & $1.65\pm0.12$       & $195\pm64$     & L   & SAX          & (1), (2), (4)   & \\
970508  & 0.835    & 14.0   & $0.61\pm0.13$       & $145\pm43$     & L   & SAX          & (1), (2), (5)   & \\
970828  & 0.9578   & 146.59 & $30.38\pm3.57$      & $586\pm117$    & L   & GRO          & (1), (2), (6)   & \\
971214  & 3.42     & 6.0    & $22.06\pm2.76$      & $685\pm133$    & L   & SAX          & (1), (2), (7)   & \\
980326  & 1.0      & 312.0  & $0.482\pm0.09$       & $35.5\pm18$   & L   & SAX          & (2), (3)        & \\
980425  & 0.0085   & 22.0   & $0.0001\pm0.00002$  & $136\pm12.7$   & L   & SAX/GRO      & (2), (3), (8)   & \\
980613  & 1.0969   & 42.0   & $0.61\pm0.09$       & $194\pm89$     & L   & SAX          & (1), (2), (9)   & \\
980703  & 0.966    & 102.43 & $7.42\pm0.71$       & $503\pm64$     & L   & GRO          & (1), (3), (10)  & \\
\enddata
\end{deluxetable}

\setlength{\tabcolsep}{3mm}
\begin{deluxetable}{cccccccccc}
\tabletypesize{\scriptsize}
\tablecolumns{8}
\tablewidth{0pc}
\label{table1}
\tablecaption{--- continued \label{table-1}}
\tablenum{1}
\tablehead{
\colhead{GRB$^a$} & \colhead{$z^b$} & \colhead{$T_{\rm 90}^c$} & \colhead{$E_{\rm iso}^d$} & \colhead{$E_{\rm p,i}^e$} & \colhead{Type$^f$} & \colhead{Detector$^g$} & \colhead{Refs$^h$}\\
\colhead{}        &  \colhead{}     & \colhead{(s)}             &  \colhead{($10^{52}$ erg)} & \colhead{(keV)}          & \colhead{}         &     \colhead{}         & \colhead{}}

\startdata
060121  & 4.6      & 1.57   & $18\pm1.20$         & $767\pm84$     & L   & KW           & (3), (64)       & \\
060124  & 2.297    & 658.2  & $43.85\pm6.45$      & $784\pm285$    & L   & SWI/KW       & (1), (2), (59)  & \\
060206  & 4.0559   & 7.55   & $4.59\pm0.98$       & $394\pm46$     & L   & SWI          & (1), (2), (59)  & \\
060210  & 3.9122   & 242.18 & $41.53\pm5.7$       & $575\pm186$    & L   & SWI          & (2), (3), (59)  & \\
060218  & 0.03351  & 2100   & $0.0054\pm0.0003$   & $4.9\pm0.3$    & L   & SWI          & (1), (2), (65)  & \\
060223A & 4.406    & 11.32  & $4.29\pm0.66$       & $339\pm63$     & L   & SWI          & (2), (3), (66)  & \\
060306  & 1.5597   & 60.94  & $2.03\pm0.7$        & $171\pm79$     & L   & SWI          & (2), (3), (48)  & \\
060418  & 1.4901   & 109.17 & $13.63\pm2.96$      & $572\pm143$    & L   & SWI/KW       & (1), (2), (67)  & \\
\enddata
\end{deluxetable}

\setlength{\tabcolsep}{3mm}
\begin{deluxetable}{cccccccccc}
\tabletypesize{\scriptsize}
\tablecolumns{8}
\tablewidth{0pc}
\label{table1}
\tablecaption{--- continued \label{table-1}}
\tablenum{1}
\tablehead{
\colhead{GRB$^a$} & \colhead{$z^b$} & \colhead{$T_{\rm 90}^c$} & \colhead{$E_{\rm iso}^d$} & \colhead{$E_{\rm p,i}^e$} & \colhead{Type$^f$} & \colhead{Detector$^g$} & \colhead{Refs$^h$}\\
\colhead{}        &  \colhead{}     & \colhead{(s)}             &  \colhead{($10^{52}$ erg)} & \colhead{(keV)}          & \colhead{}         &     \colhead{}         & \colhead{}}

\startdata
080913  & 6.7      & 7.46   & $8.36\pm2.44$       & $710\pm350$    & L   & SWI/KW       & (1), (2), (59)  & \\
080916A & 0.6887   & 61.35  & $0.98\pm0.09$       & $184\pm18$     & L   & SWI/FER      & (1), (2), (59)  & \\
080916C & 4.35     & 62.98  & $371.24\pm78.06$    & $2646\pm566$   & L   & FER/KW       & (1), (2), (94)  & \\
081007  & 0.5295   & 9.01   & $0.18\pm0.02$       & $61\pm15$      & L   & FER          & (1), (2), (95)  & \\
081024B & 3.05     & 0.648  & $2.44\pm0.22$       & $5690\pm4321$  & S   & FER          & (3), (96)       & \\
081028  & 3.038    & 284.42 & $16.75\pm1.96$      & $234\pm93$     & L   & SWI          & (1), (2), (97)  & \\
081118  & 2.58     & 49.28  & $4.25\pm0.89$       & $147\pm14$     & L   & SWI/FER      & (1), (2), (98)  & \\
081121  & 2.512    & 17.67  & $25.73\pm4.97$      & $871\pm123$    & L   & SWI/KW       & (1), (2), (99)  & \\
\enddata
\end{deluxetable}

\setlength{\tabcolsep}{3mm}
\begin{deluxetable}{cccccccccc}
\tabletypesize{\scriptsize}
\tablecolumns{8}
\tablewidth{0pc}
\label{table1}
\tablecaption{--- continued \label{table-1}}
\tablenum{1}
\tablehead{
\colhead{GRB$^a$} & \colhead{$z^b$} & \colhead{$T_{\rm 90}^c$} & \colhead{$E_{\rm iso}^d$} & \colhead{$E_{\rm p,i}^e$} & \colhead{Type$^f$} & \colhead{Detector$^g$} & \colhead{Refs$^h$}\\
\colhead{}        &  \colhead{}     & \colhead{(s)}             &  \colhead{($10^{52}$ erg)} & \colhead{(keV)}          & \colhead{}         &     \colhead{}         & \colhead{}}

\startdata
120119A & 1.7291   & 253.8  & $36.00\pm11.70$     & $417\pm55$     & L   & SWI/KW       & (1), (2), (48)    & \\
120326A & 1.798    & 69.6   & $3.68\pm0.17$       & $130\pm10$     & L   & SWI/FER      & (1), (2), (143)   & \\
120624B & 2.1974   & 300    & $282.0\pm1.20$      & $1791\pm134$   & L   & KW           & (1), (2), (48)    & \\
120711A & 1.405    & 40.033 & $180.41\pm18.04$    & $2340\pm230$   & L   & FER          & (1), (2), (144)   & \\
120712A & 4.1745   & 14.7   & $21.19\pm1.84$      & $641\pm130$    & L   & SWI          & (1), (2), (145)   & \\
120716A & 2.486    & 234    & $30.15\pm0.27$      & $397\pm40$     & L   & FER          & (1), (2), (146)   & \\
120724A & 1.48     & 72.8   & $0.88\pm0.12$       & $68\pm19$      & L   & SWI          & (1), (2), (147)   & \\
120729A & 0.80     & 71.5   & $2.3\pm1.5$         & $559\pm57$     & L   & FER          & (2), (3), (148)   & \\
\enddata
\end{deluxetable}

\setlength{\tabcolsep}{3mm}
\begin{deluxetable}{cccccccccc}
\tabletypesize{\scriptsize}
\tablecolumns{8}
\tablewidth{0pc}
\label{table1}
\tablecaption{--- continued \label{table-1}}
\tablenum{1}
\tablehead{
\colhead{GRB$^a$} & \colhead{$z^b$} & \colhead{$T_{\rm 90}^c$} & \colhead{$E_{\rm iso}^d$} & \colhead{$E_{\rm p,i}^e$} & \colhead{Type$^f$} & \colhead{Detector$^g$} & \colhead{Refs$^h$}\\
\colhead{}        &  \colhead{}     & \colhead{(s)}             &  \colhead{($10^{52}$ erg)} & \colhead{(keV)}          & \colhead{}         &     \colhead{}         & \colhead{}}

\startdata
150120A & 0.46     & 1.20   & $0.019\pm0.004$     & $190\pm73$     & S   & FER          & (3), (167), (199) & \\
150206A & 2.087    & 35.19  & $61.9\pm4.5$        & $704\pm71$     & L   & KW           & (3), (200)        & \\
150301B & 1.517    & 13.31  & $3.43\pm0.59$       & $461\pm29$     & L   & FER          & (1), (167), (201) & \\
150314A & 1.758    & 14.79  & $89.16\pm2.15$      & $957\pm8$      & L   & FER          & (1), (167), (202) & \\
150323A & 0.593    & 149.6  & $1.256\pm0.069$     & $151\pm14$     & L   & FER/KW       & (3), (167), (203) & \\
150403A & 2.06     & 22.27  & $99.28\pm2.42$      & $1312\pm21$    & L   & FER          & (1), (167), (204) & \\
150413A & 3.2      & 263.6  & $65.3\pm16.2$       & $397\pm232$    & L   & FER/KW       & (3), (167), (205) & \\
150423A & 0.22     & 1.39   & $0.00075\pm0.00013$ & $146\pm43$     & S   & SWI          & (3), (167), (206) & \\
\enddata
\end{deluxetable}

\setlength{\tabcolsep}{5.5mm}
\begin{deluxetable}{ccccccccccc}
\tabletypesize{\scriptsize}
\tablecolumns{8}
\label{table1}
\tabletypesize{\tiny}
\tablecaption{--- continued \label{table-1}}
\tablenum{1}
\tablehead{
\colhead{GRB$^a$} & \colhead{$z^b$} & \colhead{$T_{\rm 90}^c$} & \colhead{$E_{\rm iso}^d$} & \colhead{$E_{\rm p,i}^e$} & \colhead{Type$^f$} & \colhead{Detector$^g$} & \colhead{Refs$^h$}\\
\colhead{}        &  \colhead{}     & \colhead{(s)}             &  \colhead{($10^{52}$ erg)} & \colhead{(keV)}          & \colhead{}         &     \colhead{}         & \colhead{}}

\startdata
201015A & 0.426    & 9.78   & $0.386\pm0.03$      & $58\pm20$      & L   & SWI          & (271), (272)      & \\
201020B & 0.804    & 16.0   & $7.02\pm0.07$       & $247\pm6$      & L   & FER          & (273), (274)      & \\
201021C & 1.07     & 35.3   & $0.54\pm0.06$       & $271\pm58$     & L   & FER          & (275), (276)      & \\
201103B & 1.105    & 105.0  & $18.0\pm2.0$        & $848\pm137$    & L   & KW           & (277), (278)      & \\
201216C & 1.10     & 29.95  & $63.2\pm6.3$        & $739\pm27$     & L   & FER          & (271), (279)      & \\
201221D & 1.045    & 0.15   & $0.136\pm0.015$     & $231\pm16$     & S   & SWI          & (280), (281)      & \\
210204A & 0.876    & 197.0  & $22.6\pm1.7$        & $257\pm23$     & L   & FER          & (282), (283)      & \\
210210A & 0.715    & 6.60   & $0.17\pm0.02$       & $28\pm15$      & L   & SWI/KW       & (284), (285)      & \\
\enddata
\end{deluxetable}

\section{Energy distribution function}

\begin{figure*}
\centering
\includegraphics    [angle=0,scale=0.55]     {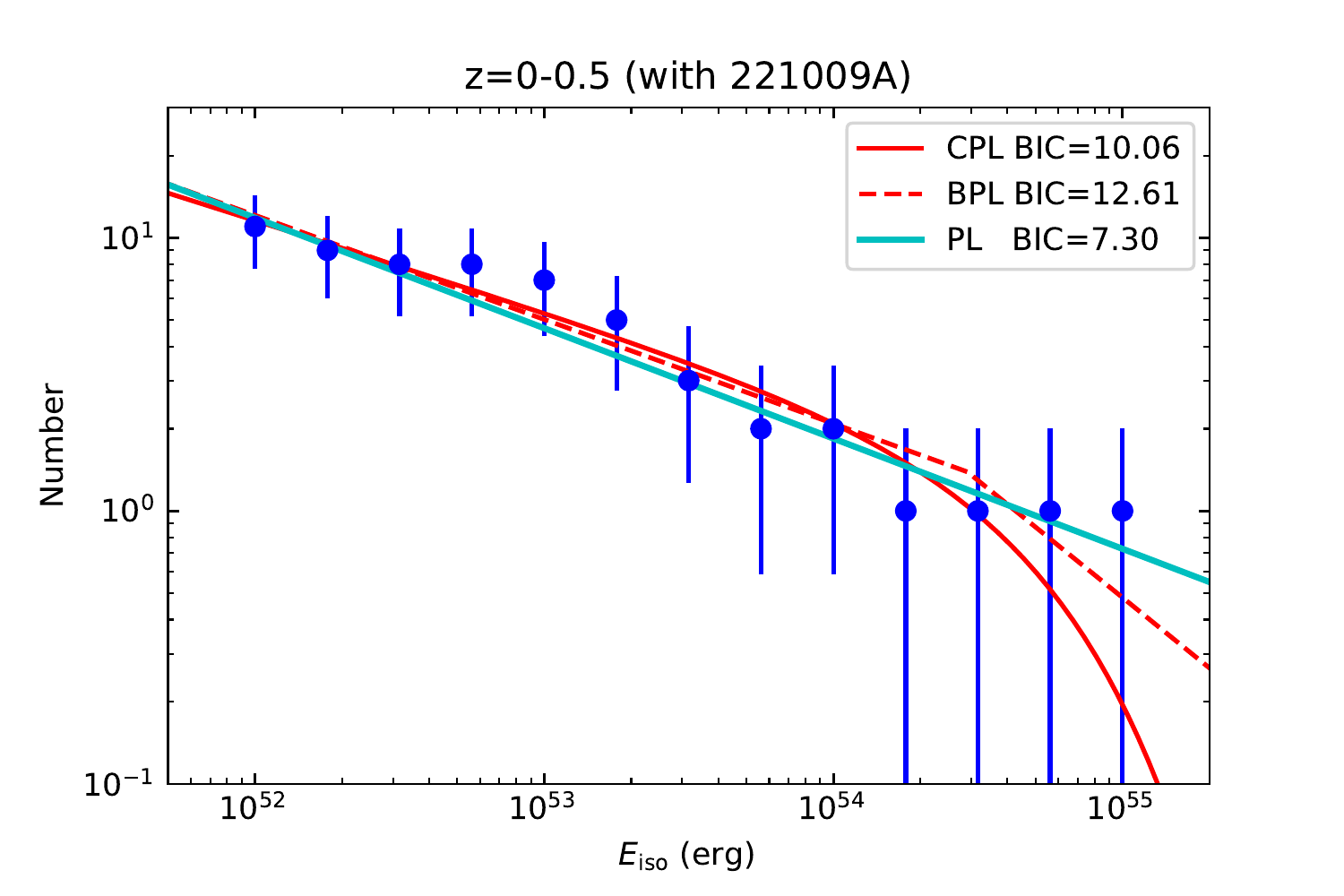}
\includegraphics    [angle=0,scale=0.55]     {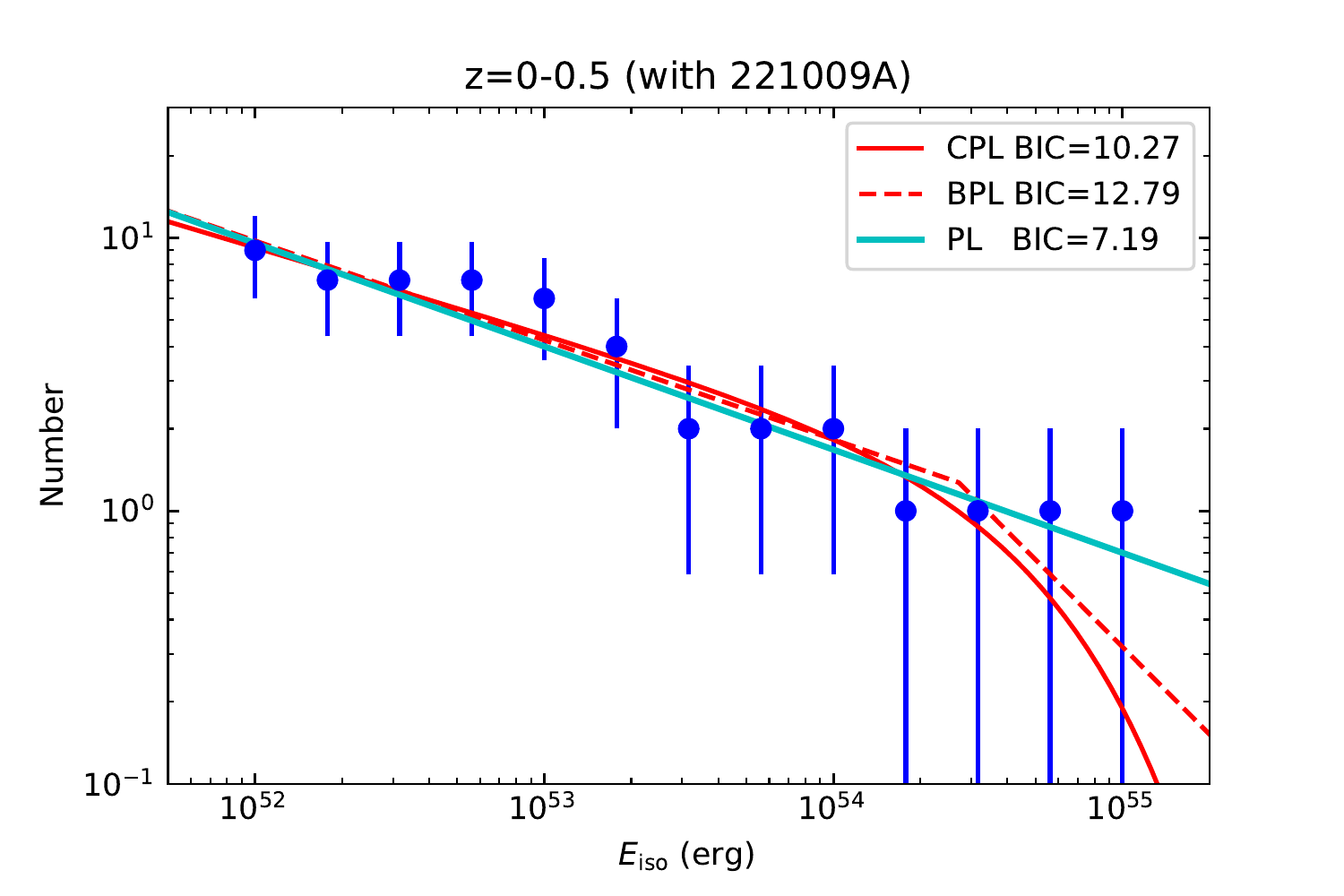}\\
\includegraphics    [angle=0,scale=0.55]     {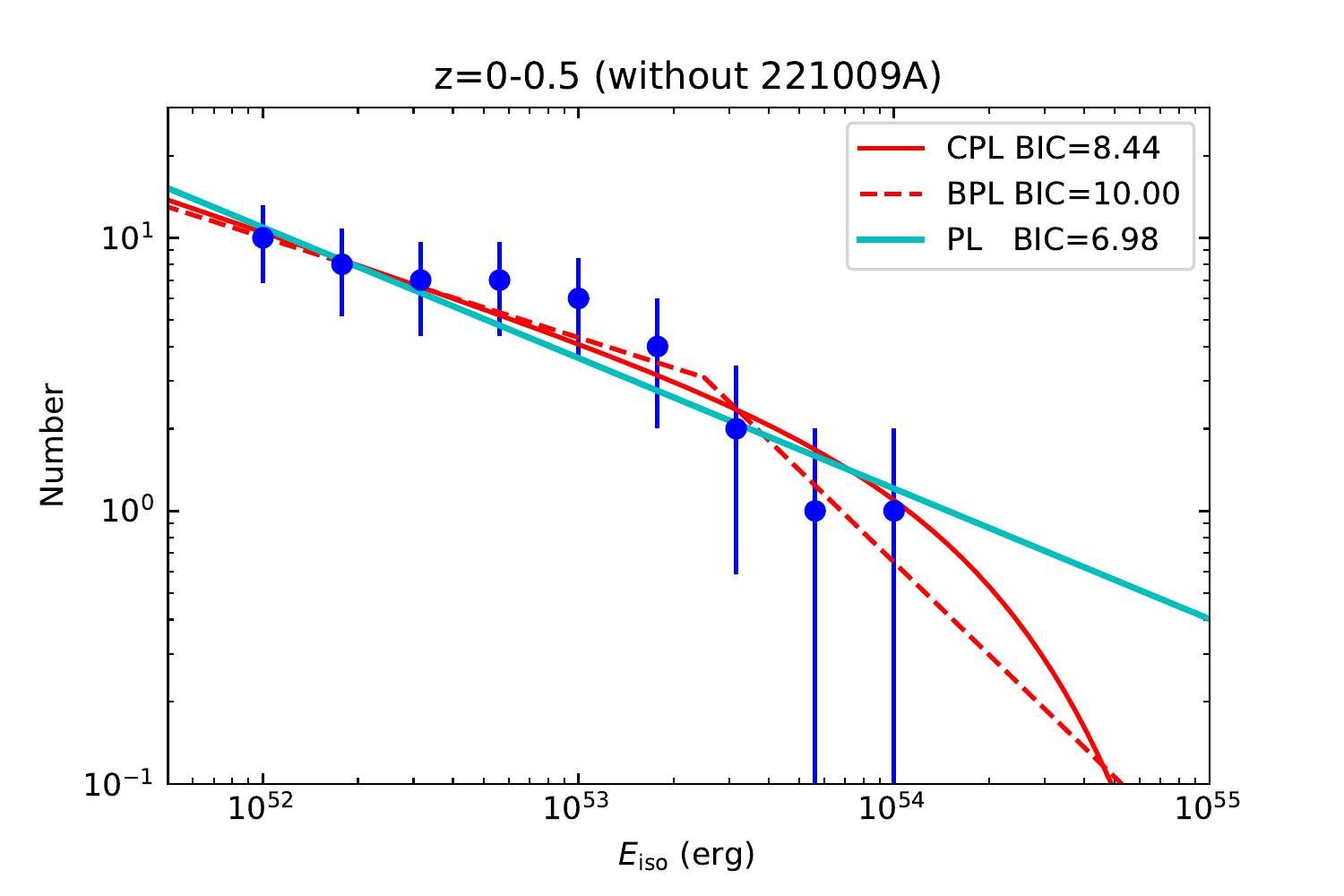}
\includegraphics    [angle=0,scale=0.55]     {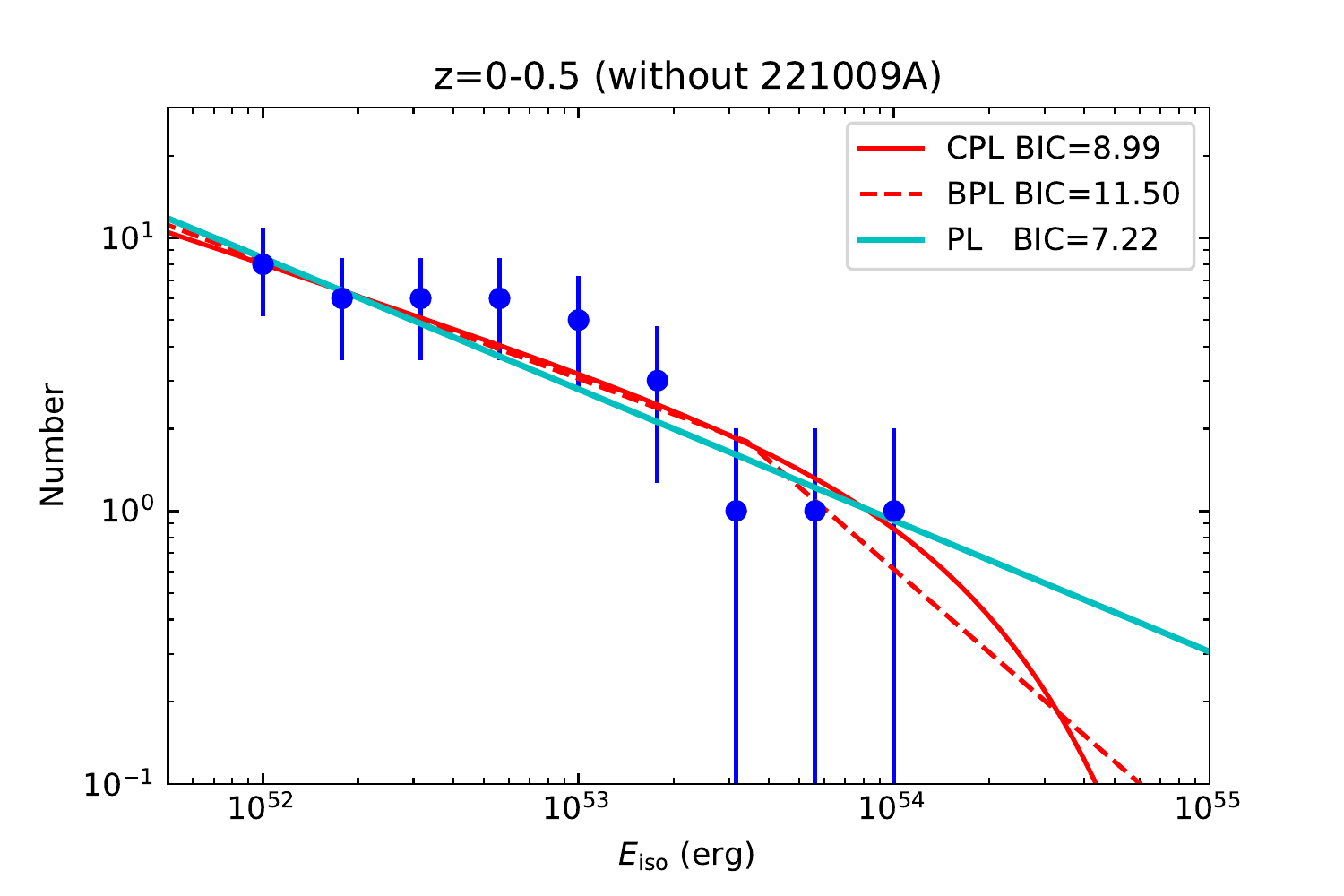}\\
\includegraphics    [angle=0,scale=0.55]     {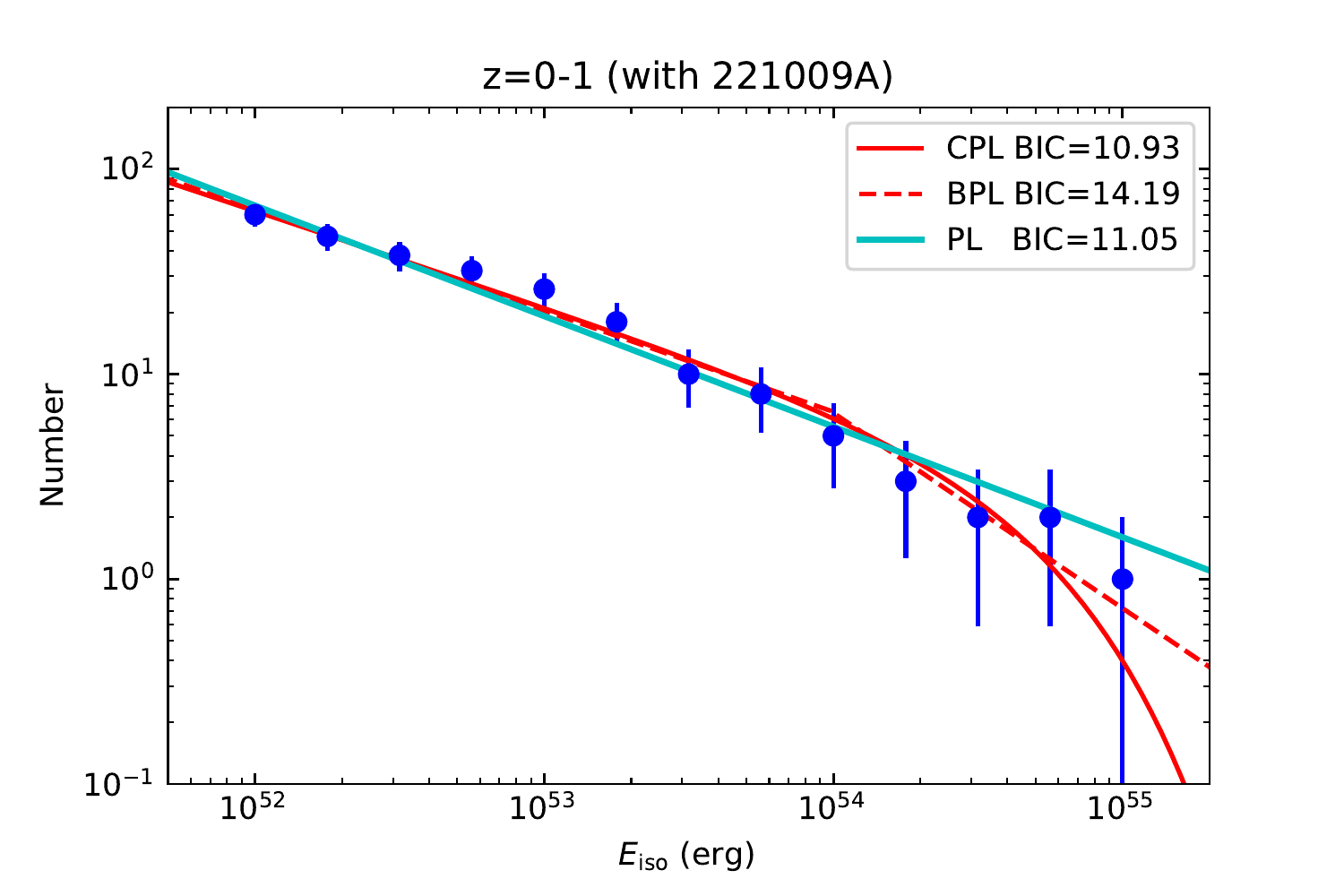}
\includegraphics    [angle=0,scale=0.55]     {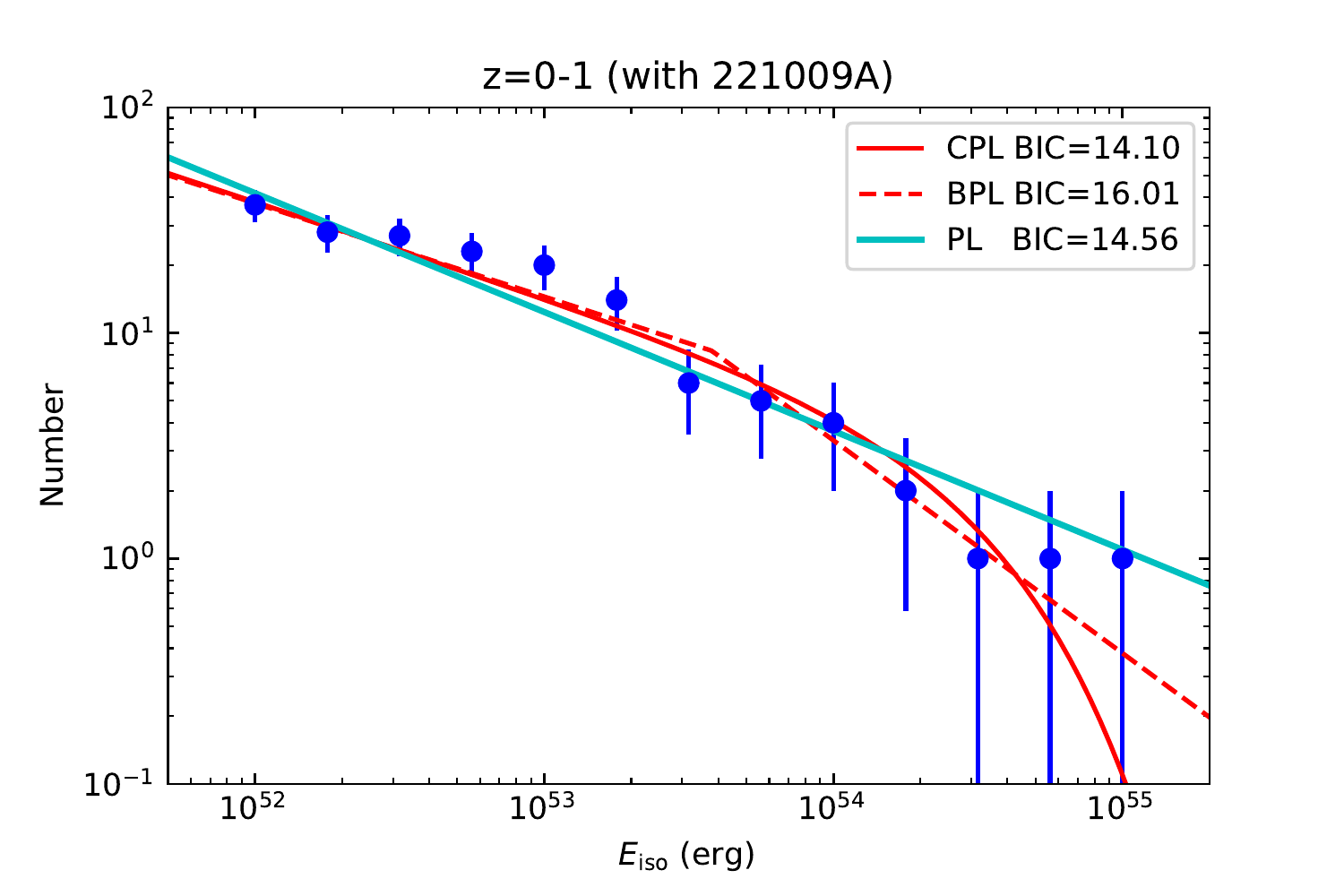}\\
\includegraphics    [angle=0,scale=0.55]     {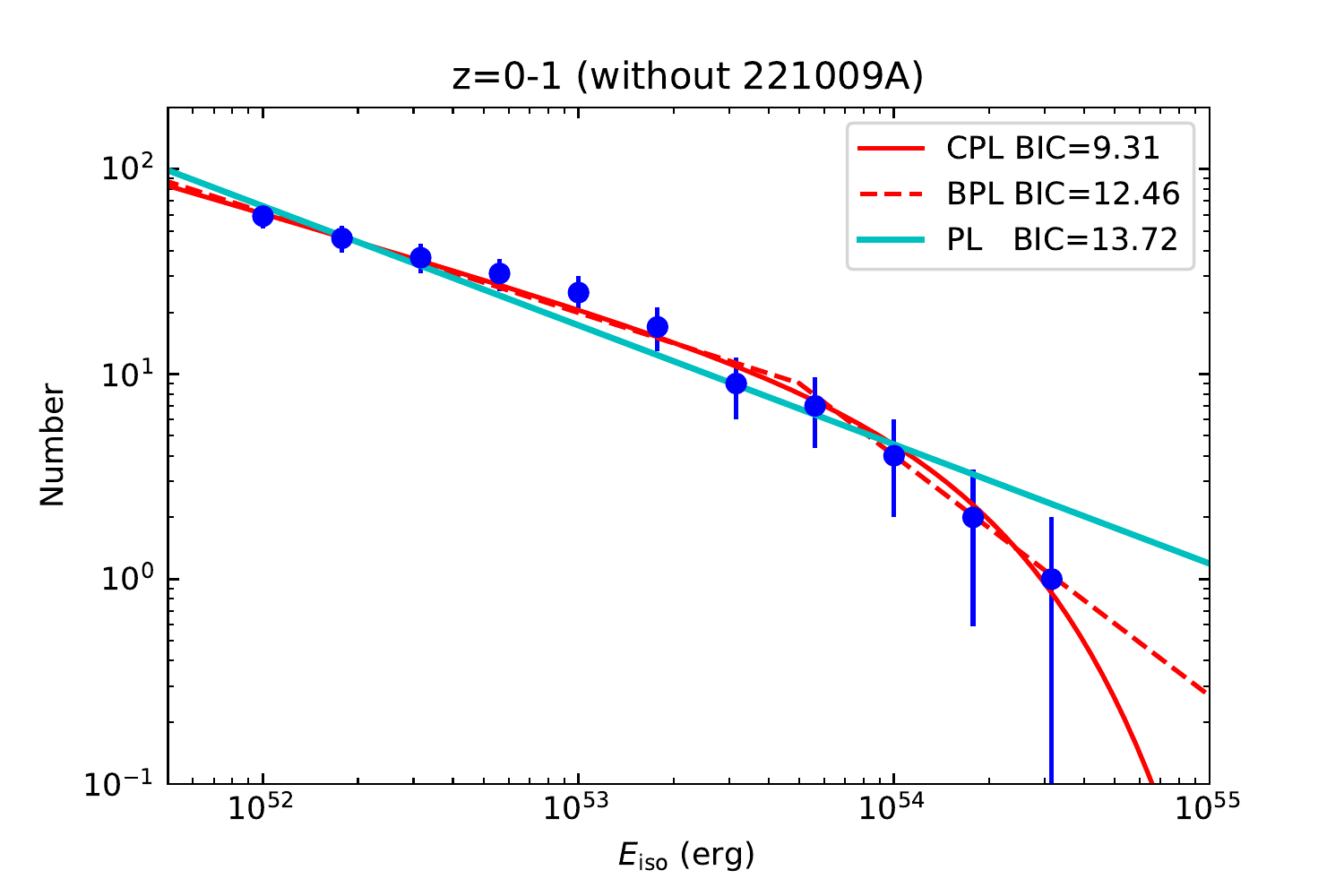}
\includegraphics    [angle=0,scale=0.55]     {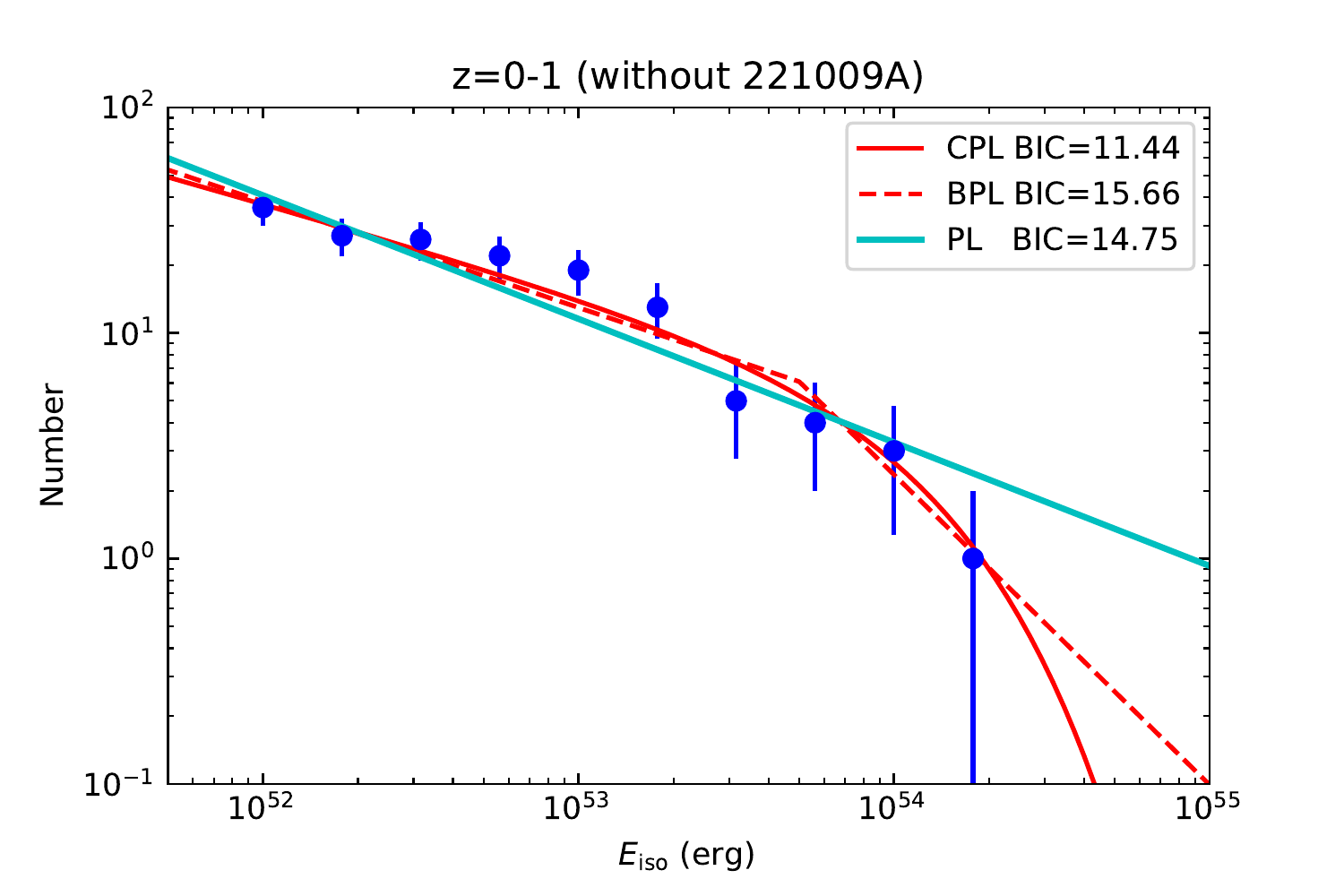}\\
\caption{Cumulative distribution of GRB isotropic energy for total samples (left panels) and pure Konus-\emph{Wind} samples (right panels) in the low redshift bins. The blue circles show the observed distribution, and the red solid lines, red dashed lines, and blue solid lines represent the best-fitting CPL model, BPL model, and PL model, respectively.}
\label{fig:low-z Eiso}
\end{figure*}

\begin{figure*}
\centering
\includegraphics    [angle=0,scale=0.48]     {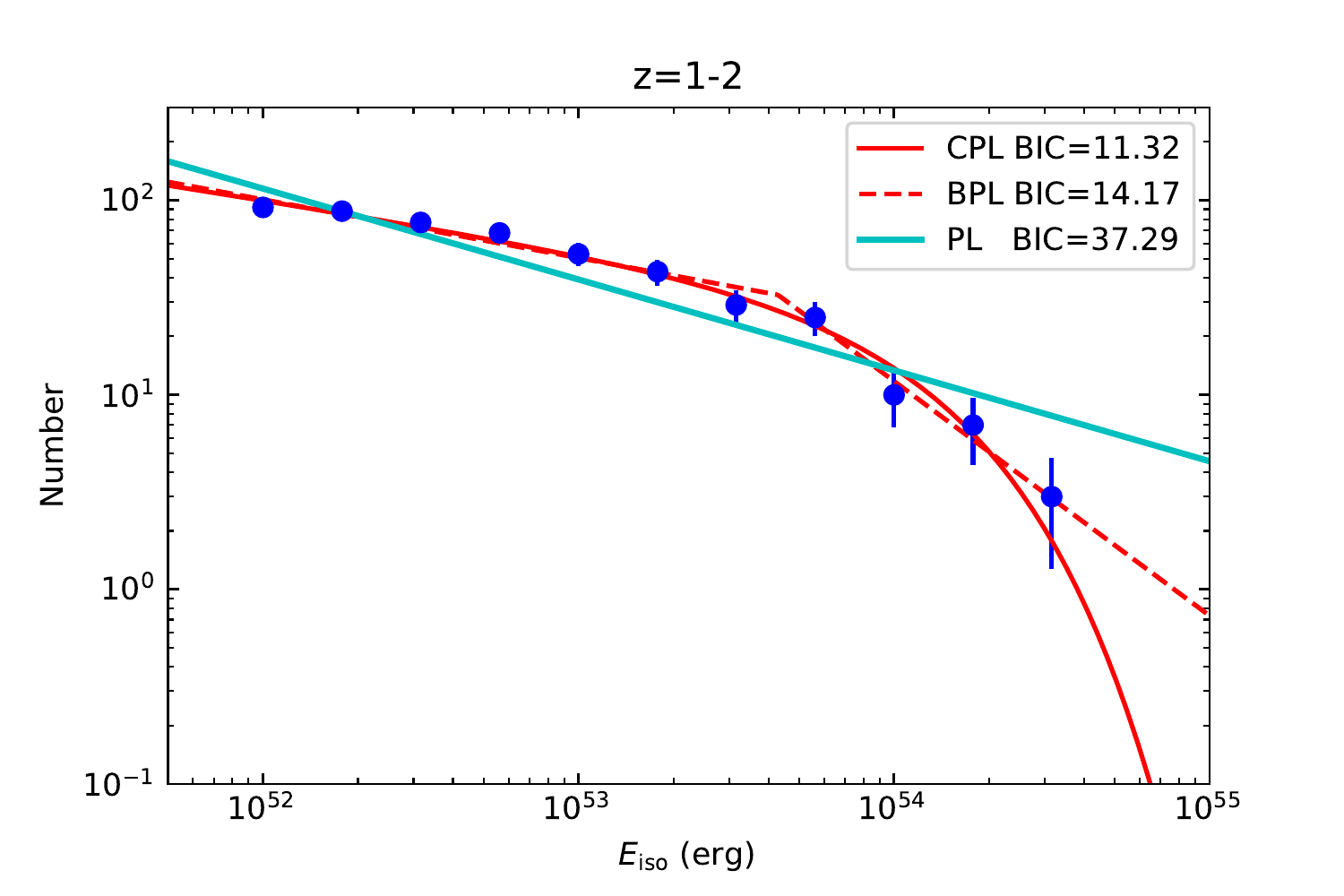}
\includegraphics    [angle=0,scale=0.48]     {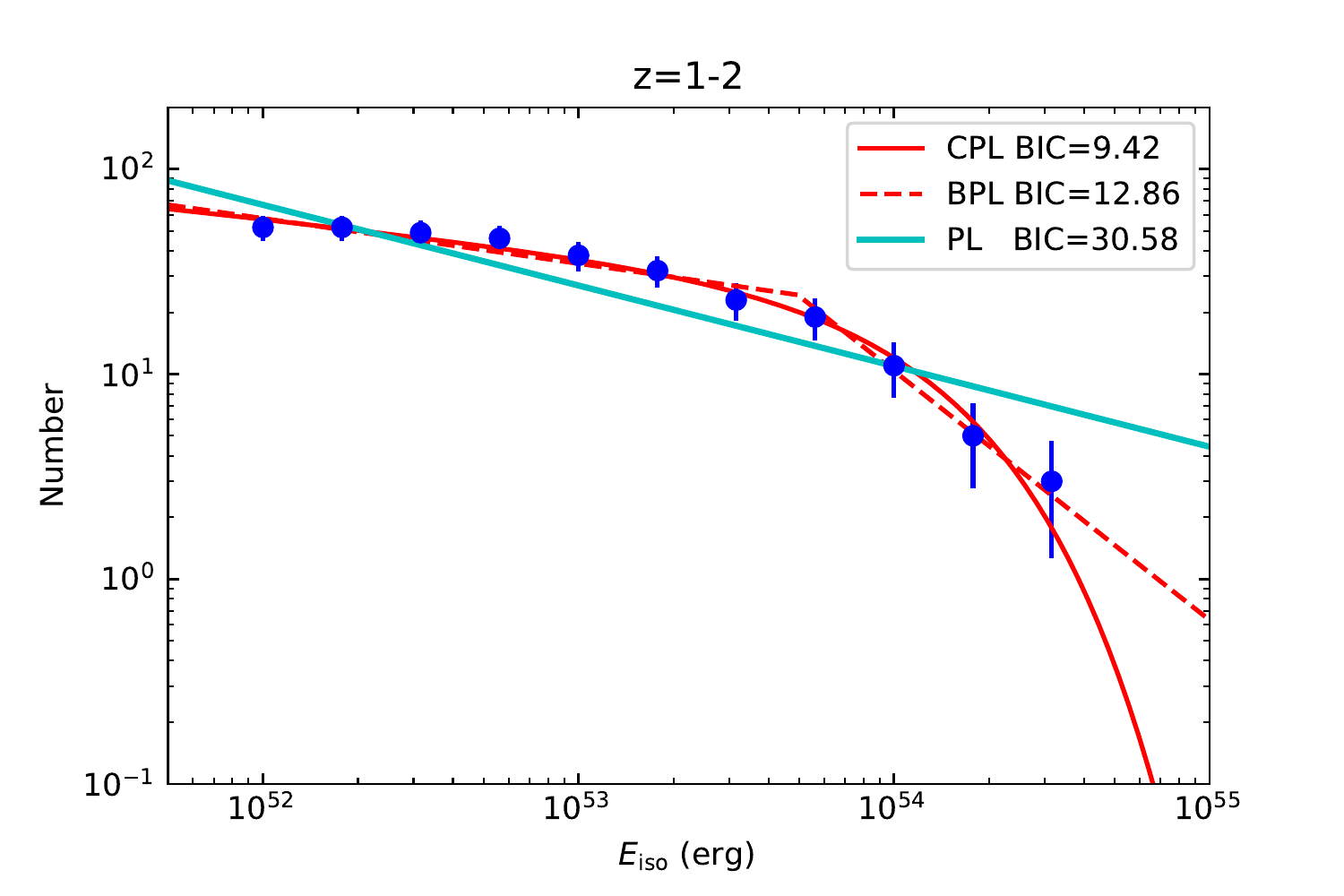}\\
\includegraphics    [angle=0,scale=0.48]     {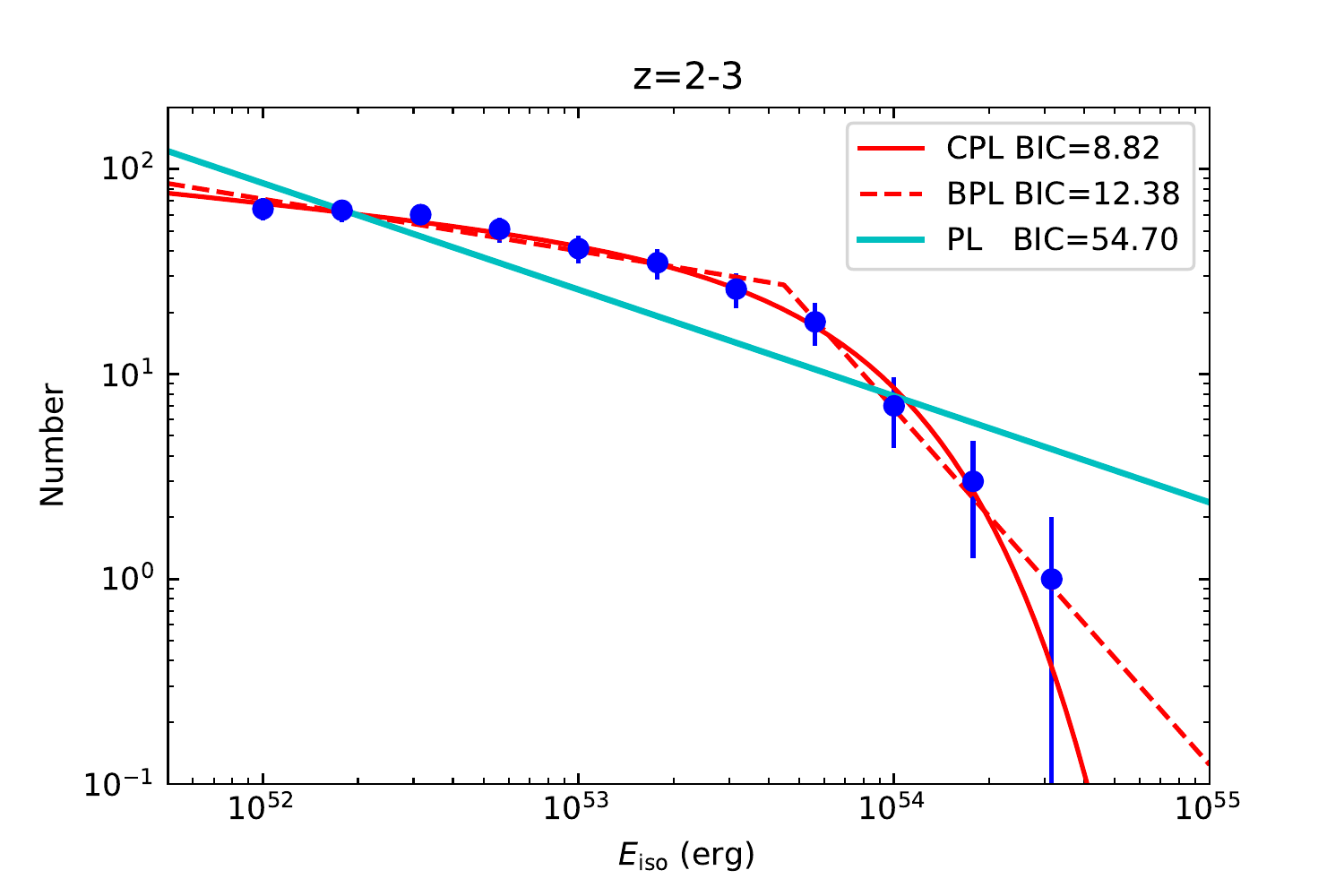}
\includegraphics    [angle=0,scale=0.48]     {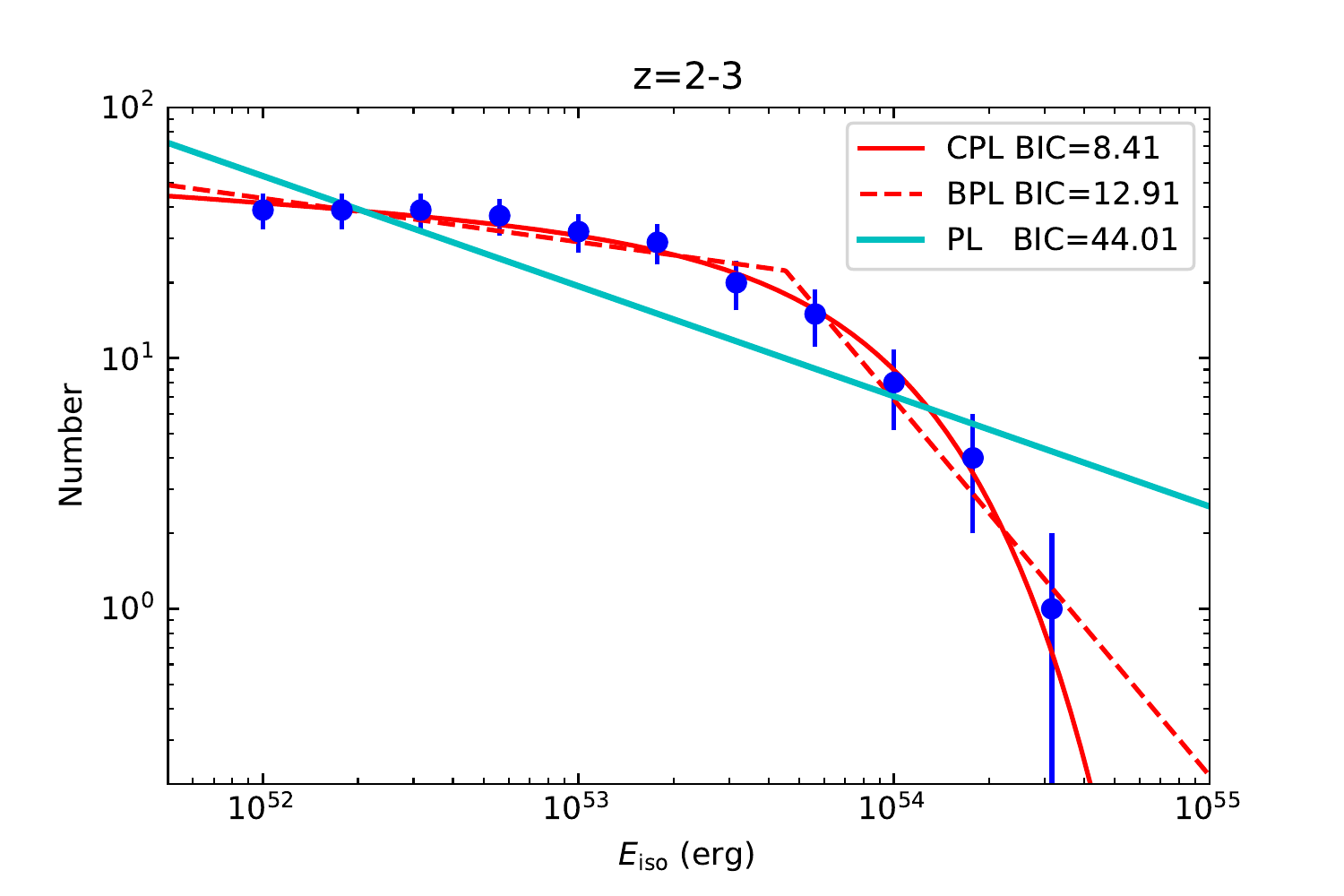}\\
\includegraphics    [angle=0,scale=0.48]     {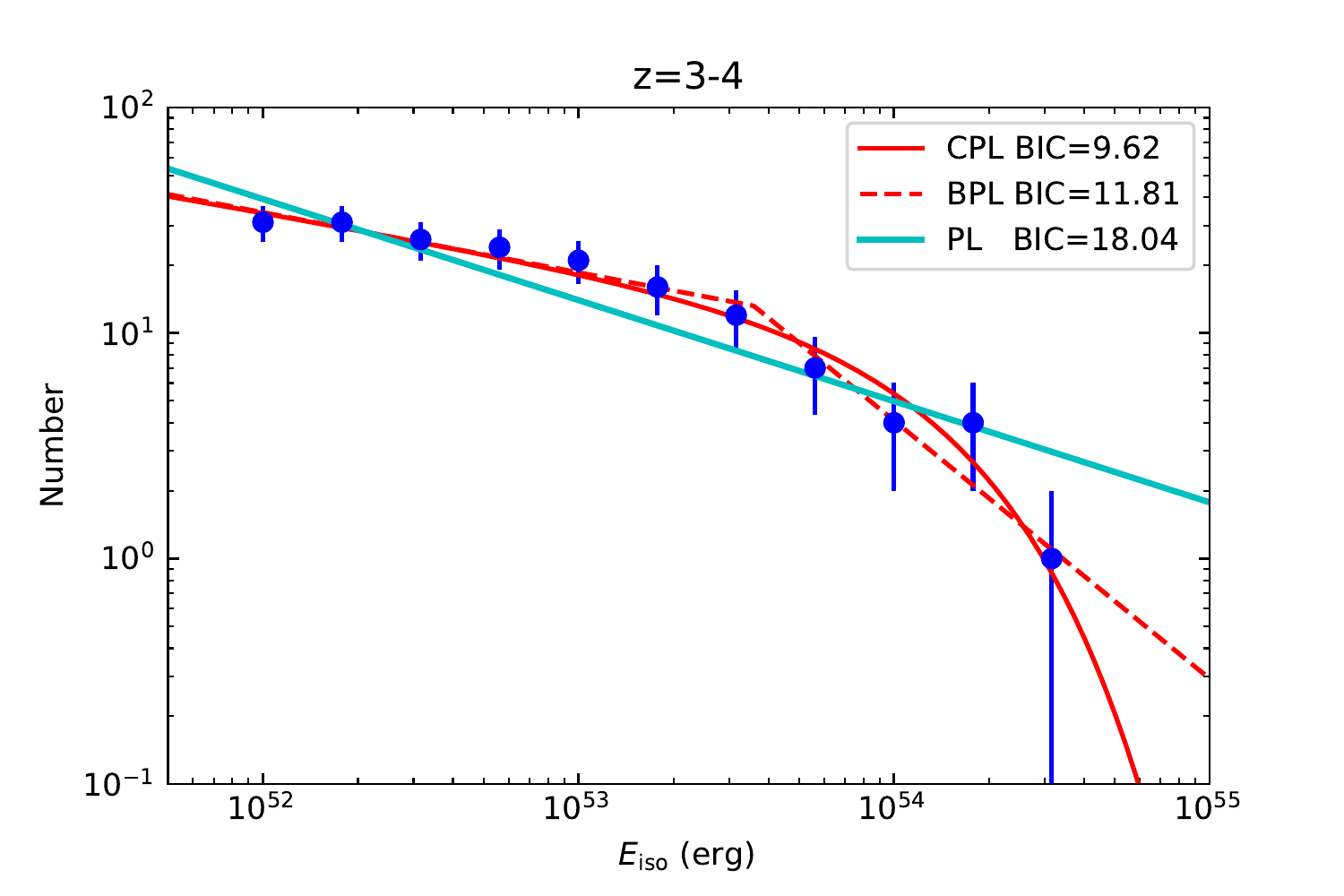}
\includegraphics    [angle=0,scale=0.48]     {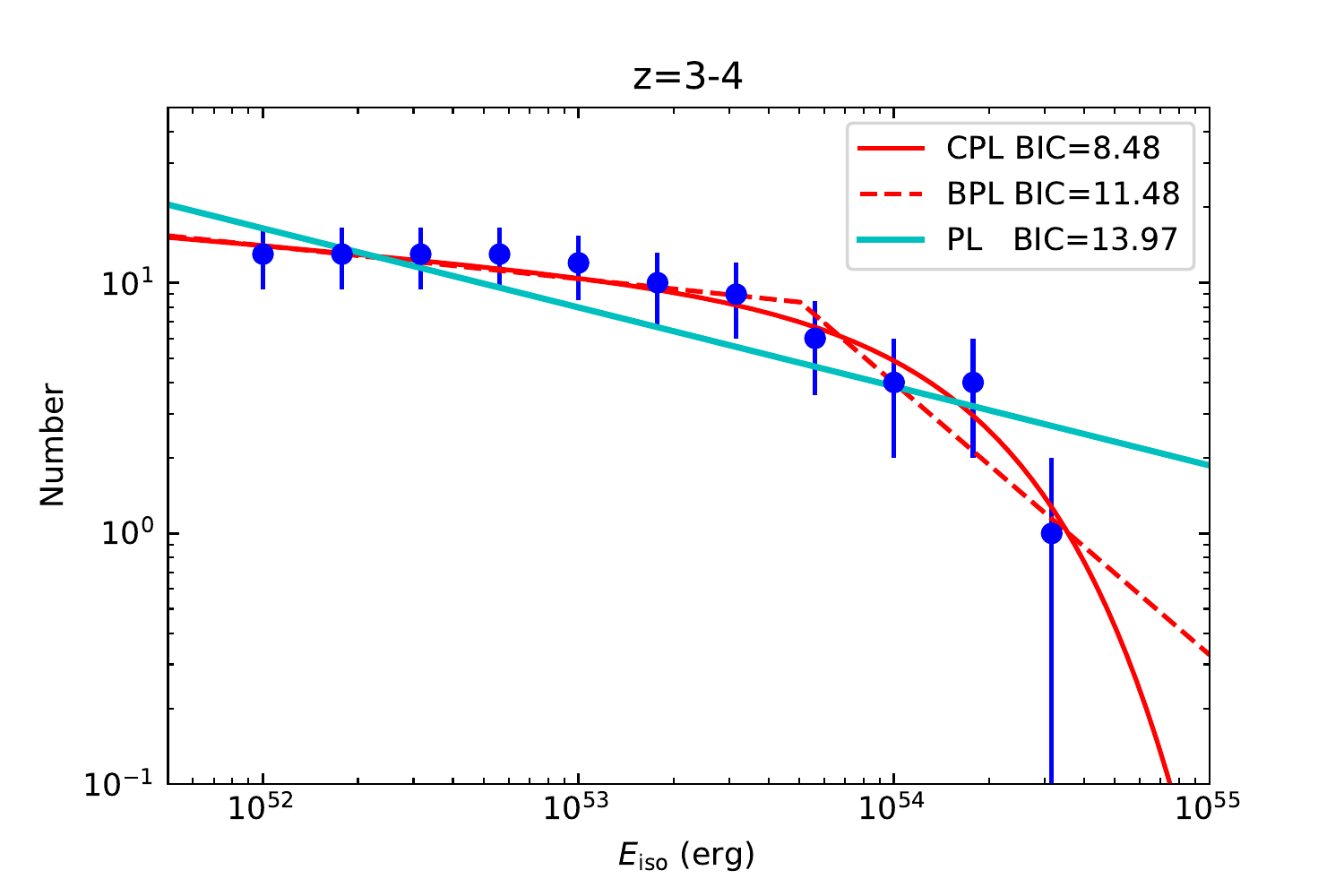}\\
\includegraphics    [angle=0,scale=0.48]     {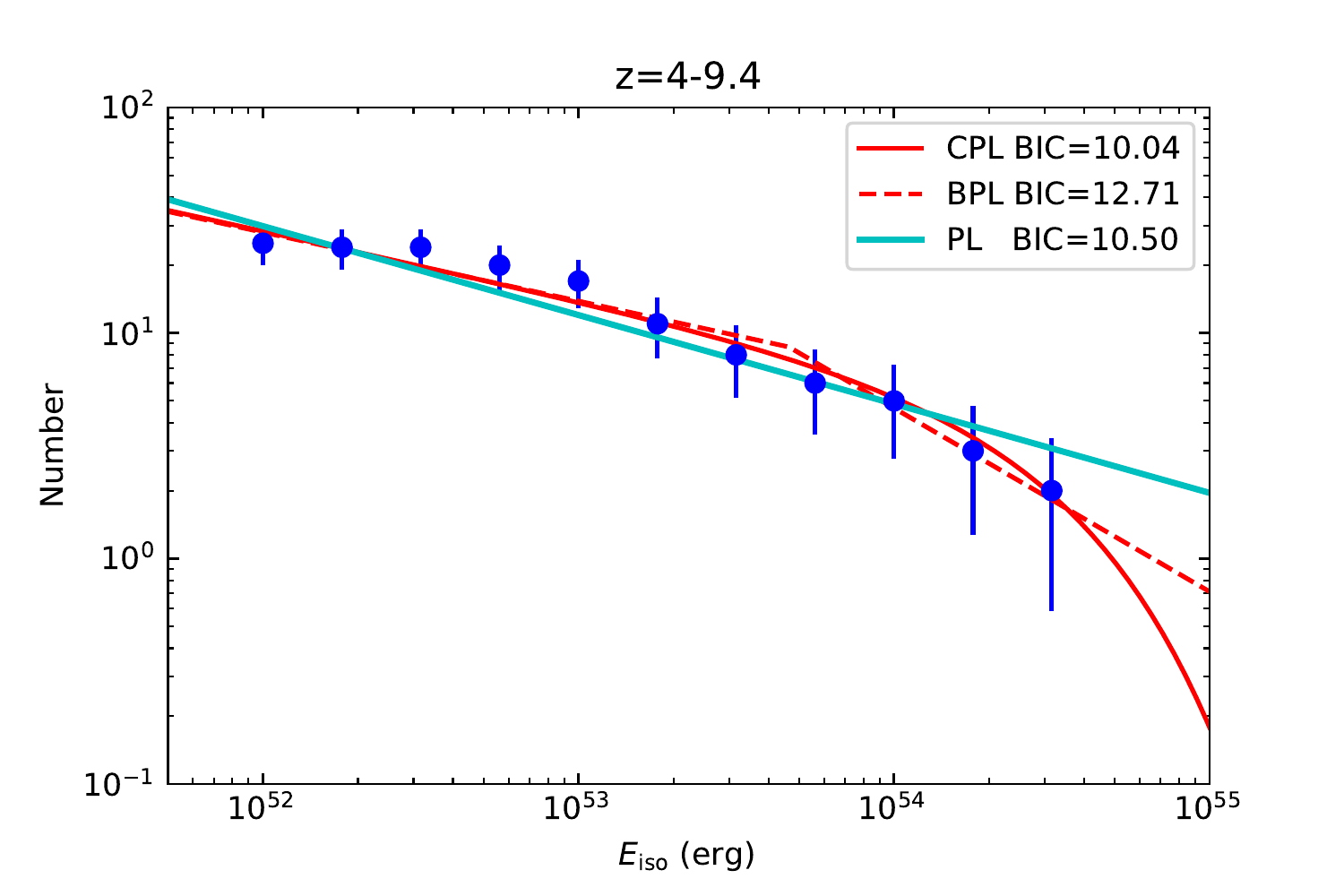}
\includegraphics    [angle=0,scale=0.48]     {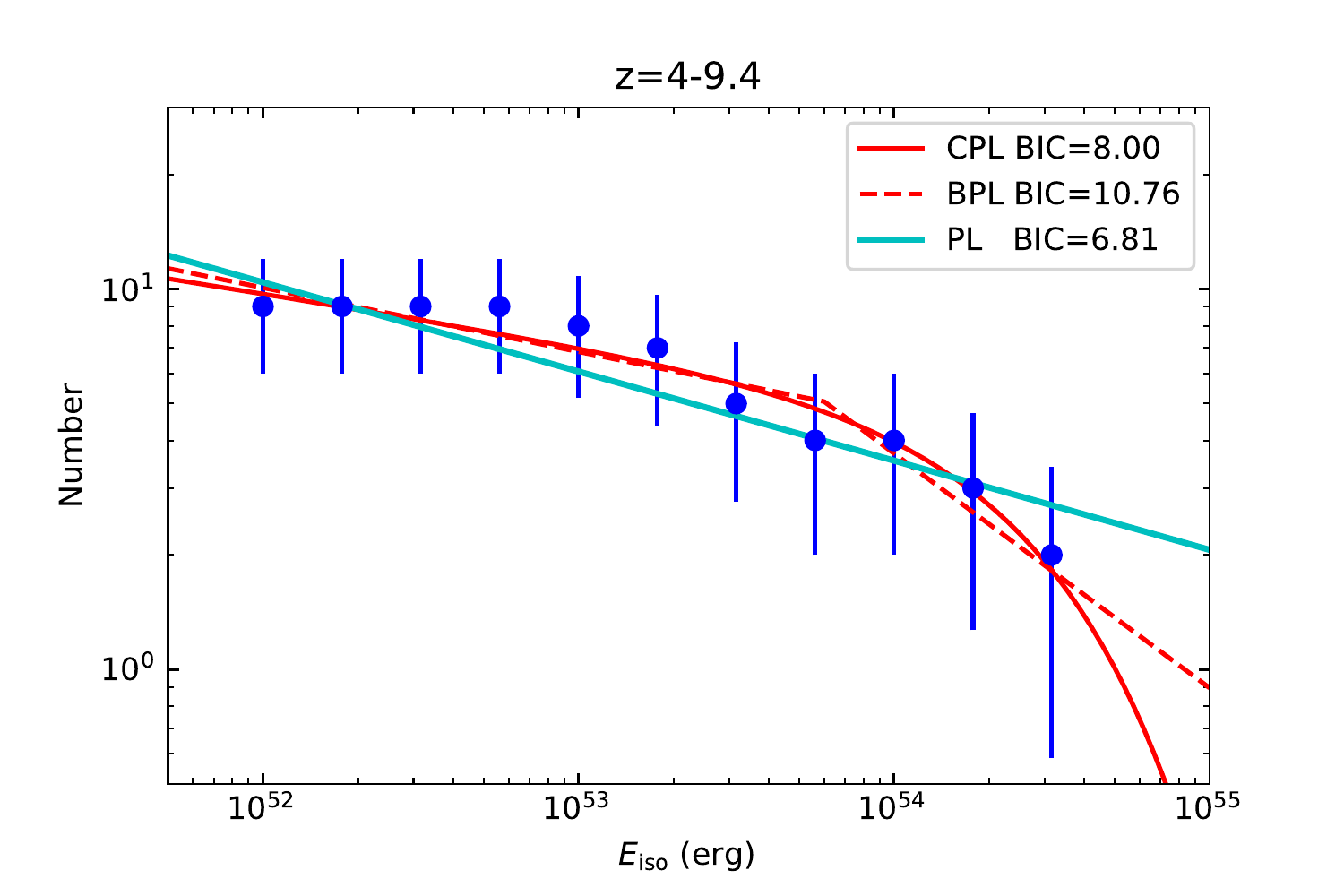}\\
\includegraphics    [angle=0,scale=0.48]     {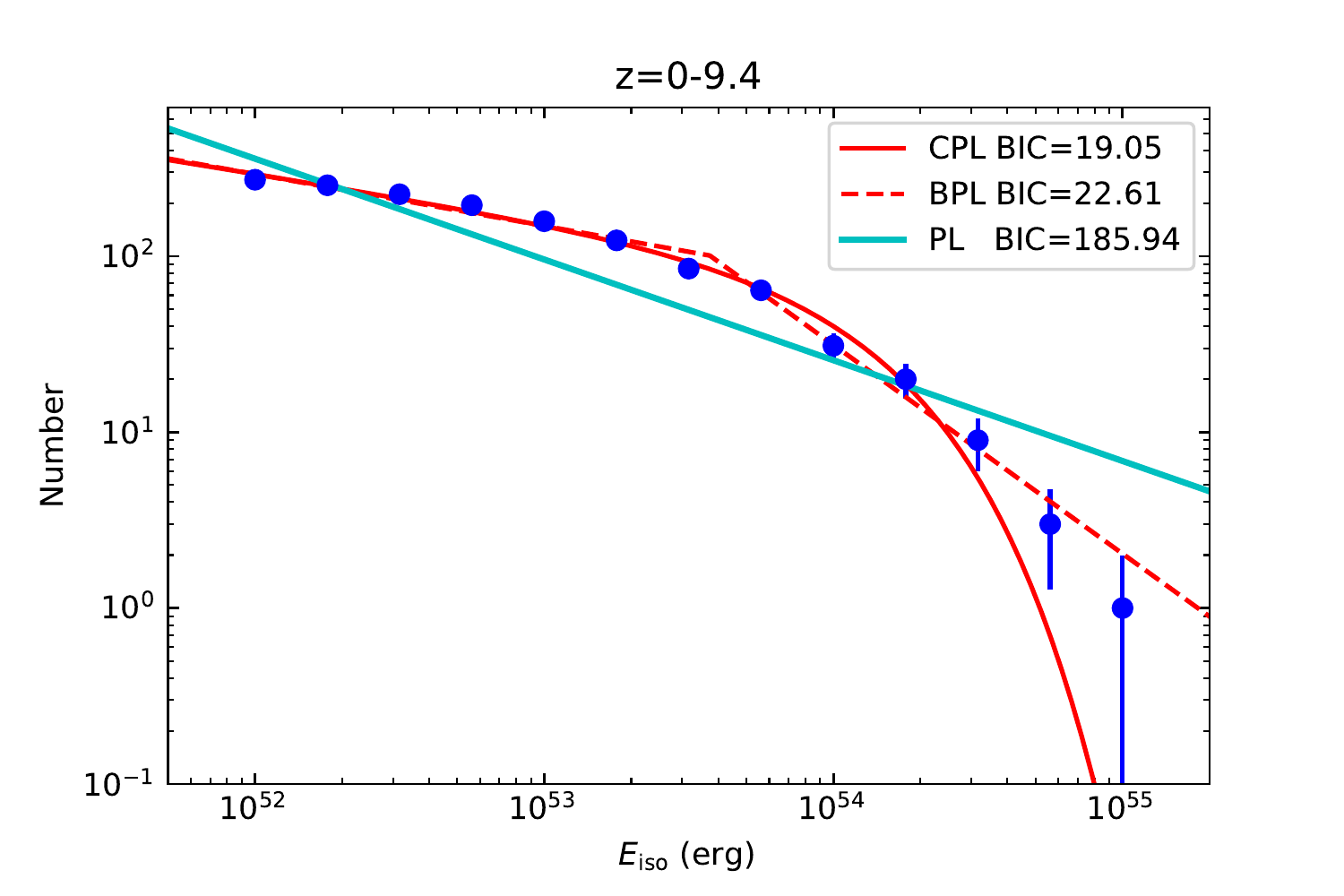}
\includegraphics    [angle=0,scale=0.48]     {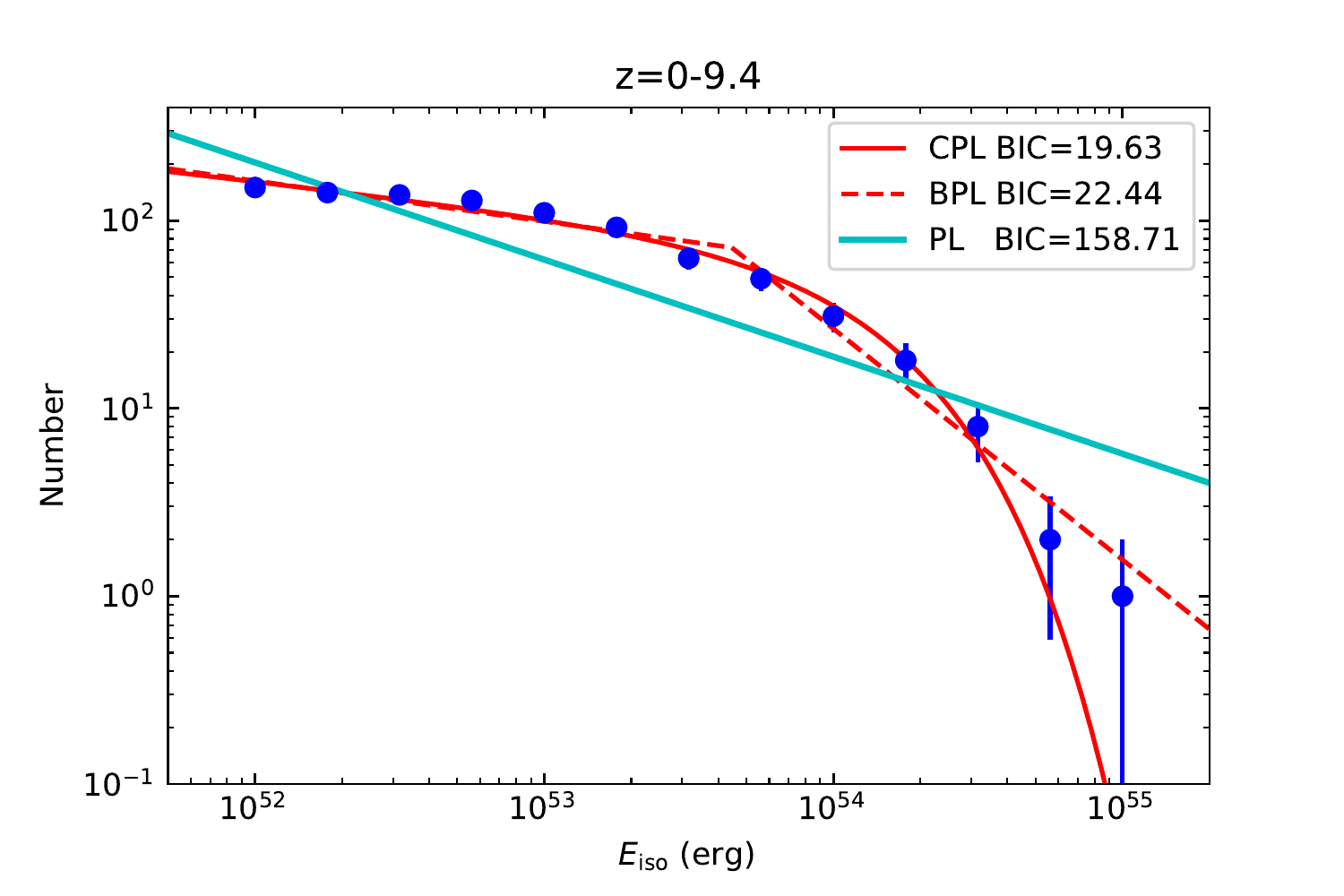}\\
\caption{Similar to Figure \ref{fig:low-z Eiso}, but for high redshift bins.}
\label{fig:high-z Eiso}
\end{figure*}

For the purpose of this work, we first divide the collected sample into several subsamples with different redshift bins: [0$\sim$0.5], [0$\sim$1], [1$\sim$2], [2$\sim$3], [3$\sim$4], [4$\sim$9.4], [0$\sim$9.4]. For each subsample, we fit
the distribution function of $E_{\rm iso}$ with three models of the energy function: a simple power-law function (PL), a broken power-law
function (BPL), and a power-law with a high energy cutoff feature (CPL), whose expressions are as follows
\begin{equation}
\phi (E_{\rm iso})=A_{0}E_{\rm iso}^{-\alpha},
\end{equation}
\begin{equation}
 \phi (E_{\rm iso})  = A_{0}\left\{\begin{array}{l}
{\left(\frac{{{E_{\rm iso}}}}{{{E_b}}}\right)^{-\alpha}};\,\,{E_{\rm iso}} \le {E_b}\\
{\left(\frac{{{E_{\rm iso}}}}{{{E_b}}}\right)^{-\beta}};\,\,{E_{\rm iso}} > {E_b}\;,
\end{array} \right.
\end{equation}
\begin{equation}
\phi (E_{\rm iso})=A_{0}\left(\frac{E_{\rm iso}}{E_c}\right)^{-\alpha}{\rm exp}\left(\frac{E_{\rm iso}}{E_c}\right),
\end{equation}
where $A_0$ is a normalization factor, $\alpha$ and $\beta$ are the power-law indices, and $E_b$ and $E_c$ are the break energy and cutoff energy for the BPL and CPL model, respectively. For different models, the Markov Chain Monte Carlo method through the \emph{emcee} package \citep{Foreman-Mackey13} is employed to obtain the best-fitting parameters and their uncertainties. All fitting results for different subsamples are presented in Figure \ref{fig:low-z Eiso}-\ref{fig:high-z Eiso} and Table \ref{table-2}. In order to justify which model is best fitted to the distribution of $E_{\rm iso}$ for a given subsample, we compared the goodness of the fits by invoking the Bayesian information criteria (BIC)\citep{Schwarz78}. BIC is a criterion to evaluate the best-fitted model among a finite set of models, and the model with the lowest BIC is preferred. The definition of BIC can be written as: $BIC=\rm -2ln L+k\cdot ln(n)$, where $k$ is the number of model parameters, $n$ is the number of data points, and $L$ is the maximum value of the likelihood function of the estimated model. (1) if $0<\Delta BIC<2$, the evidence against the model with higher BIC is not worth more than a bare mention; (2) if $2<\Delta BIC<6$, the evidence against the model with higher BIC is positive; (3) if $6<\Delta BIC<10$, the evidence against the model with higher BIC is strong; (4) if $10<\Delta BIC$, the evidence against the model with higher BIC is very strong. 

Before discussing the fitting results, we would like to illustrate two things in advance: firstly, due to the sensitivity of the $\gamma$-ray detectors, only sources with $E_{\rm iso}>10^{52}$ erg can be detected at high redshifts (see Figure 1). To fairly compare the fitting results, for all the samples at different redshift, we only adopt GRBs with $E_{\rm iso}>10^{52}$ erg. Secondly, GRBs in our collected sample are detected by different detectors, e.g., Konus-\emph{Wind}, \emph{Swift}, \emph{Fermi}, \emph{CGRO}/BATSE, and HETE-2, which may cause concern about systematic uncertainty when we put them together to study the energy distribution. Because of this, we perform all analyses twice, one for the total samples and one for pure Konus-\emph{Wind} samples (the majority of our collected sample are detected by Konus-\emph{Wind}). We find that the results of the two analysis are consistent, which proves that there is no clear systematic error caused by the difference in detection sensitivity for different instruments. 

We first focus on the low redshift subsample with $0<z<0.5$. We find that when GRB 221009A is not introduced, the best fitting model of the energy function in this redshift interval is PL, although the other two models are not significantly excluded from the BIC analysis results (see Figure \ref{fig:low-z Eiso}). The BIC difference between PL and CPL is less than 3, and the BIC difference between PL and BPL is larger than 2 but smaller than 6. With the addition of GRB 221009A (currently the highest energy source in this redshift bin), the best fit model of the energy function is still PL. Again, the BIC difference between PL and CPL is only around 3, and the BIC difference between PL and BPL is larger than 2 but smaller than 6.

In order to exclude the influence of the sample size, we also analyzed the subsample with $0<z<1$. In this case, we find that when GRB 221009A is not introduced, the best fitting model of energy distribution is CPL. The BIC difference between PL and CPL is around 3, and the BIC difference between PL and BPL is around 2. With the addition of GRB 221009A, the best fit model of the energy function is also CPL. The BIC difference between CPL and PL is less than 2, and the BIC difference between PL and BPL is only around 2.

In general, the energy distribution of low redshift GRBs does not show an obvious cutoff at the high energy end, which may be due to the small sample size, especially the small number of high energy sources. With the expansion of the sample, from $0<z<0.5$ to $0<z<1$, the CPL model does change from inferior to slightly better than the PL model. The addition of GRB 201009A has no significant influence on the above conclusions. 

For comparison, we have made similar analyses on other subsamples with high redshift. We find that for subsamples with sufficient sample size (e.g., $1<z<2$ and $2<z<3$), the best fitting model of the energy function is the CPL model, although the BIC difference between the BPL model and the CPL model is larger than 2 but smaller than 6. In these cases, the PL model is significantly excluded (the BIC difference between the PL model and the CPL model is much more than 10). For subsamples with higher redshift, the sample size decreases again, and the goodness of fit of the three models for energy distribution becomes indistinguishable again. For the subsample with $3<z<4$, the best fitting model is the CPL, but the BIC difference between PL/BPL and CPL is smaller than 6. For the subsample with $z>4$, the best fitting model is PL, but the BIC difference between CPL/BPL and PL is also smaller than 6.  

\begin{figure*}
\centering
\includegraphics    [angle=0,scale=0.8]     {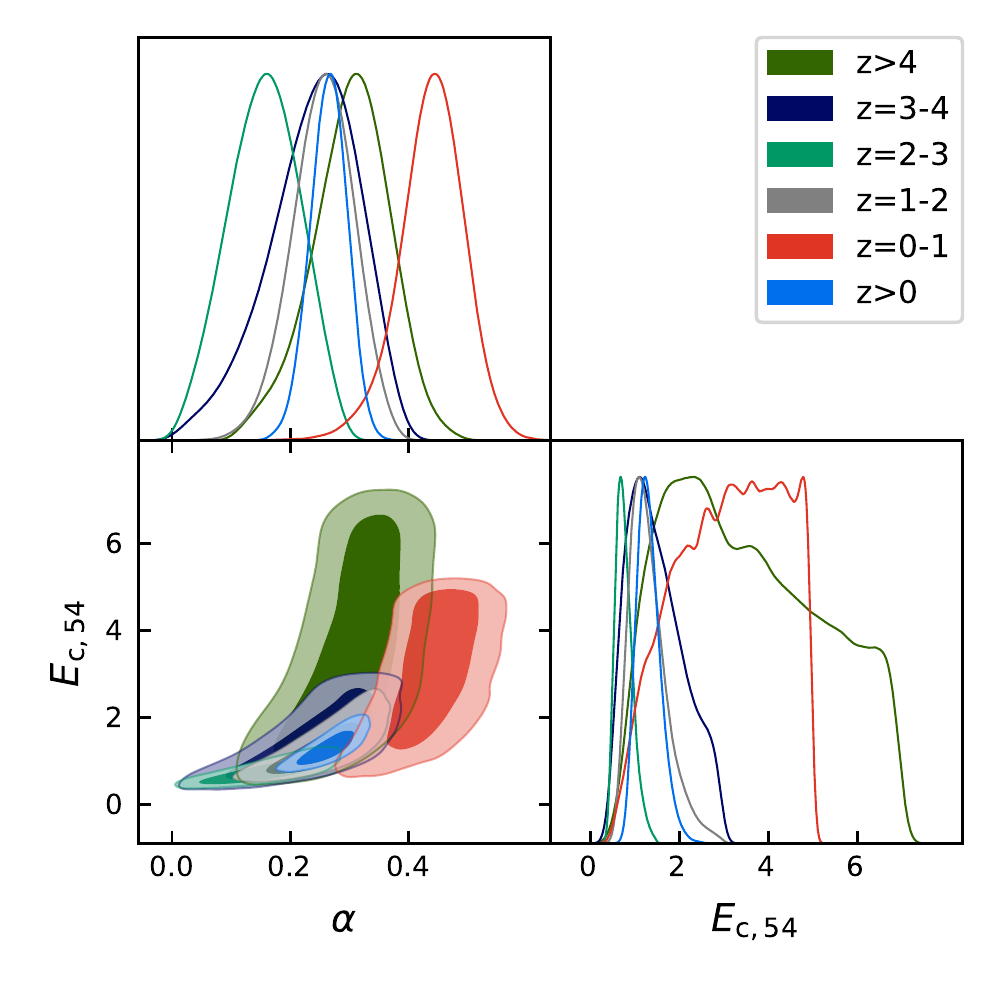}
\includegraphics    [angle=0,scale=0.8]     {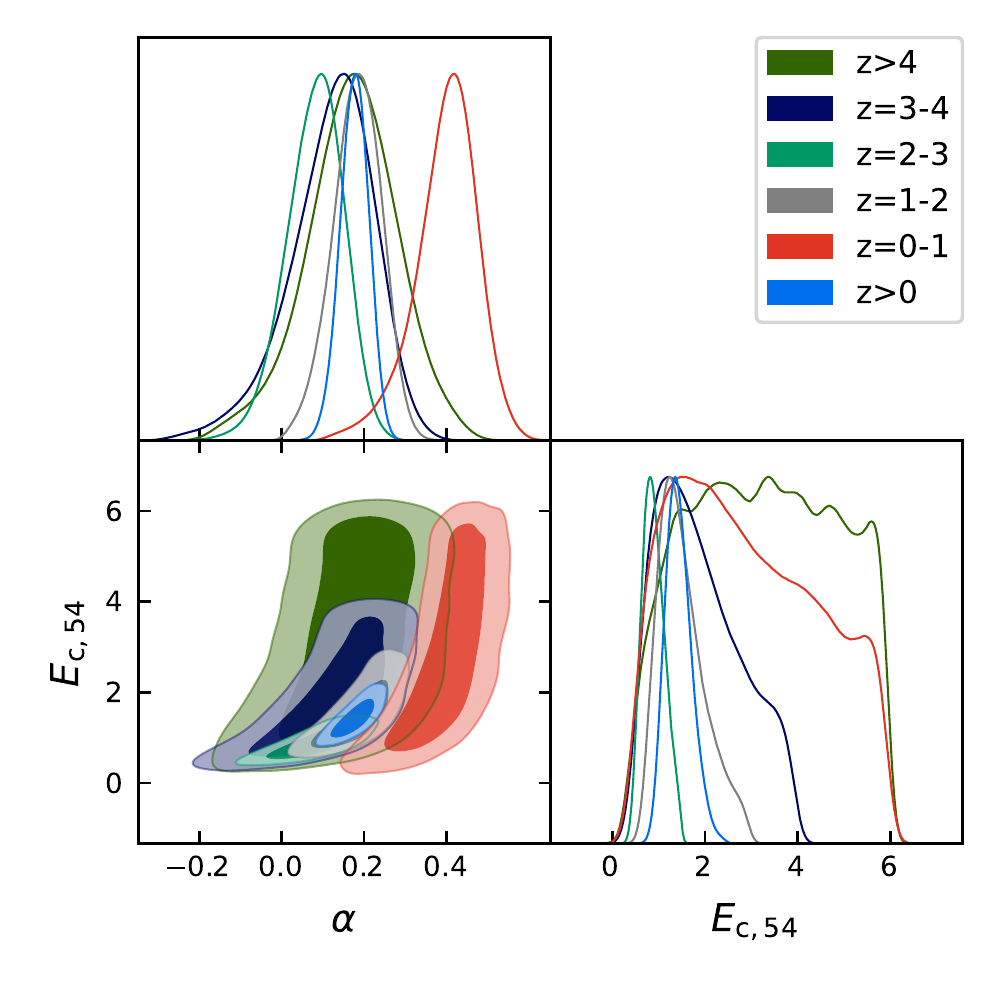}
\caption{The likehood distribution of $\alpha$ and $E_{\rm c}$ in the CPL model, from the total samples (left panel) and pure Konus-\emph{Wind} samples (right panel) in different redshift bins.}
\label{fig:CPL-corner}
\end{figure*}

\begin{figure*}
\centering
\includegraphics    [angle=0,scale=0.55]     {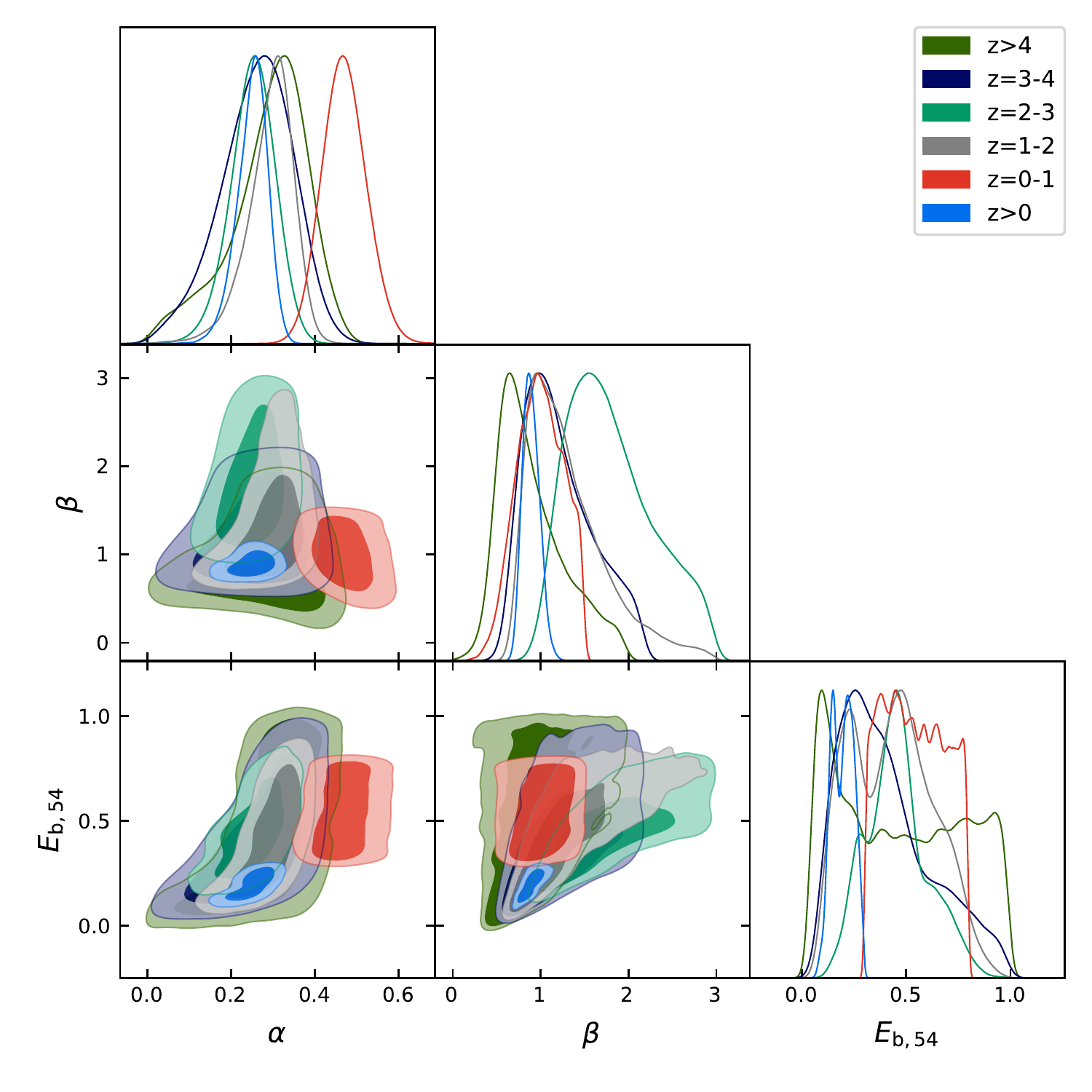}
\includegraphics    [angle=0,scale=0.55]     {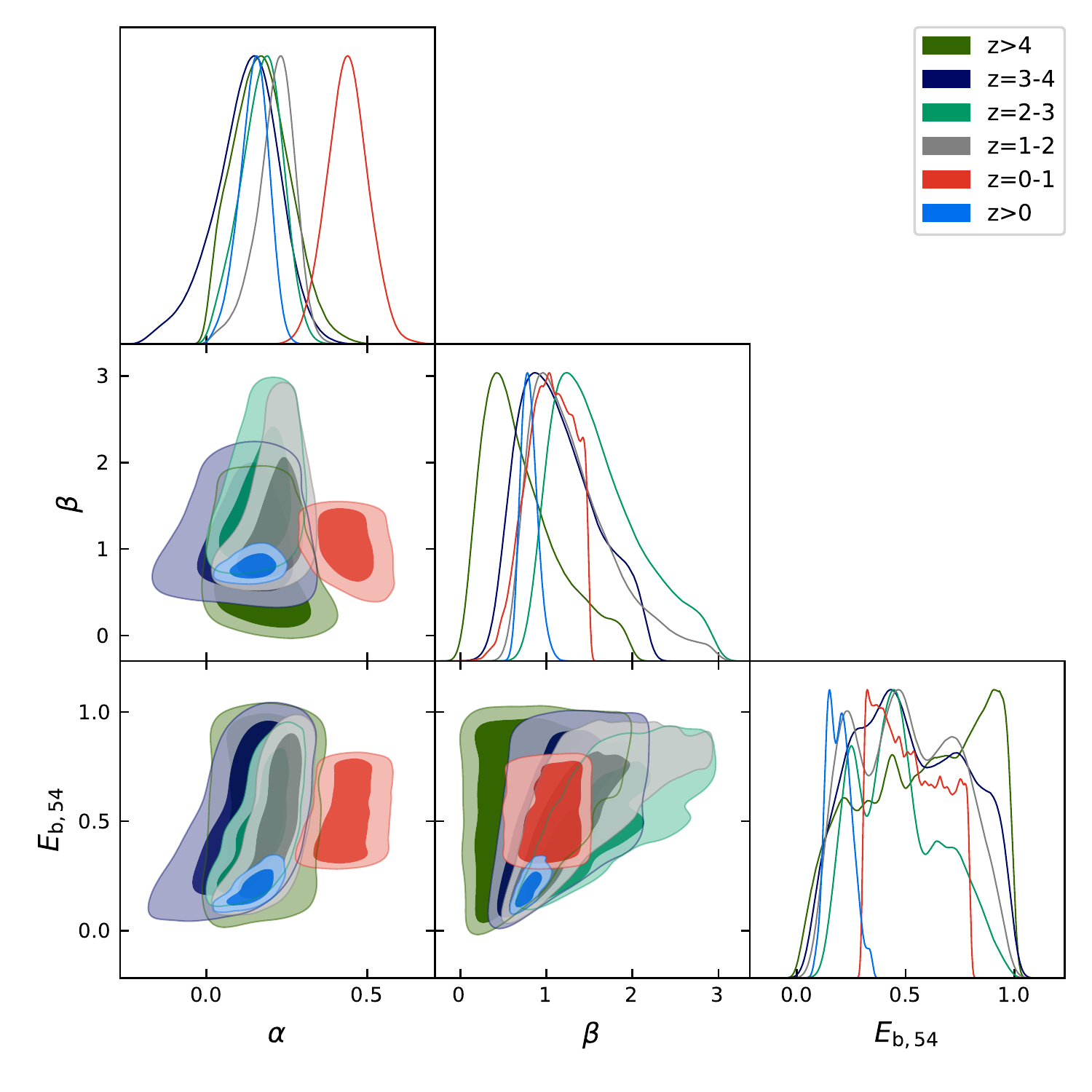}
\caption{The likehood distribution of $\alpha$, $\beta$ and $E_{\rm b}$ in the BPL model, from the total samples (left panel) and pure Konus-\emph{Wind} samples (right panel) in different redshift bins.}
\label{fig:BPL-corner}
\end{figure*}

In Figure \ref{fig:CPL-corner}-\ref{fig:BPL-corner}, we plot the best fit parameter contours for CPL and BPL models at different redshift for total samples and pure Konus-\emph{Wind} samples. We find that the best fitting parameters for different redshift samples are in good agreement with each other, supporting the hypothesis that the energy function of GRBs does not evolve with redshift. Under such a hypothesis, we compare the distribution of $E_{\rm iso}$ for the total sample ($0<z<9.4$) with PL/BPL/CPL models. We find that the best fitting model for the total sample is the CPL model, with $\alpha=0.18^{+0.03}_{-0.04}$ and $E_{c,54}=1.41^{+0.30}_{-0.26}$. The results are consistent with the previous results from \cite{Atteia17}. In this case, the PL model is significantly excluded, since the BIC difference between the PL model and the CPL model is larger than 100. However, it is worth to noting that the BIC difference between BPL and CPL is smaller than 3, indicating that there is no clear evidence to distinguish between these two models.

With our collected sample, we also studied the luminosity function of GRBs\footnote{To construct the luminosity funciton, we use the average luminosity ($E_{\rm iso}/T_{90}$) instead of peak luminosity. It is found that the results from two approaches have good agreement for the low-luminosity power-law index and the break luminosity, but a discrepancy for the high-luminosity power-law index \citep{Wanderman10}. Therefore, the utilization of the average luminosity will not significantly impact our discussion here.}. In Figure \ref{fig:Liso}, we plot the best fitting results of the luminosity function for PL, BPL, and CPL models in different redshift bins. Similar to the results of the energy function, the luminosity distribution of low redshift GRBs ($z<1$) does not show a clear cutoff feature, which may be also due to the small sample size, especially the small number of high luminosity sources. For samples with sufficient size (subsamples with $z>1$ and the total sample), the luminosity function also follows the CPL or BPL model. Again, the addition of GRB 201009A has no significant influence on the above conclusions.  All best fitting parameters are collected in Table 3. Our results are generally consistent with previous studies. For instance, our best fitting results of the BPL model to the entire sample are $\alpha=0.26^{+0.05}_{-0.07},~\beta = 0.75^{+0.11}_{-0.08}$ and  $L_{b,54}=0.01^{+0.01}_{-0.01}$. our low-luminosity power-law index and the break luminosity is well consistent with \cite{Wanderman10} ($\alpha=0.2^{+0.2}_{-0.1},~L_{b,54}=0.03^{+0.02}_{-0.02}$) and \cite{Sun15} ($\alpha=0.3, L_{b,54}=0.01$), while our high-luminosity power-law index is smaller than \cite{Wanderman10} ($\beta =1.4^{+0.3}_{-0.6}$) and \cite{Sun15} ($\beta =1.3$), which may be because we invoke more energetic GRBs in our sample.

Overall, based on the results of different subsamples, it is more likely that the energy/luminosity function of GRBs always follows the CPL or BPL model, namely, there is always a cutoff or break at the high energy end. This conclusion is consistent with results from previous studies both on luminosity function \citep{Liang07,Virgili09,Wanderman10,Sun15} and energy function \citep{Atteia17,Lan22}. 

\begin{figure*}
\centering
\includegraphics    [angle=0,scale=0.55]     {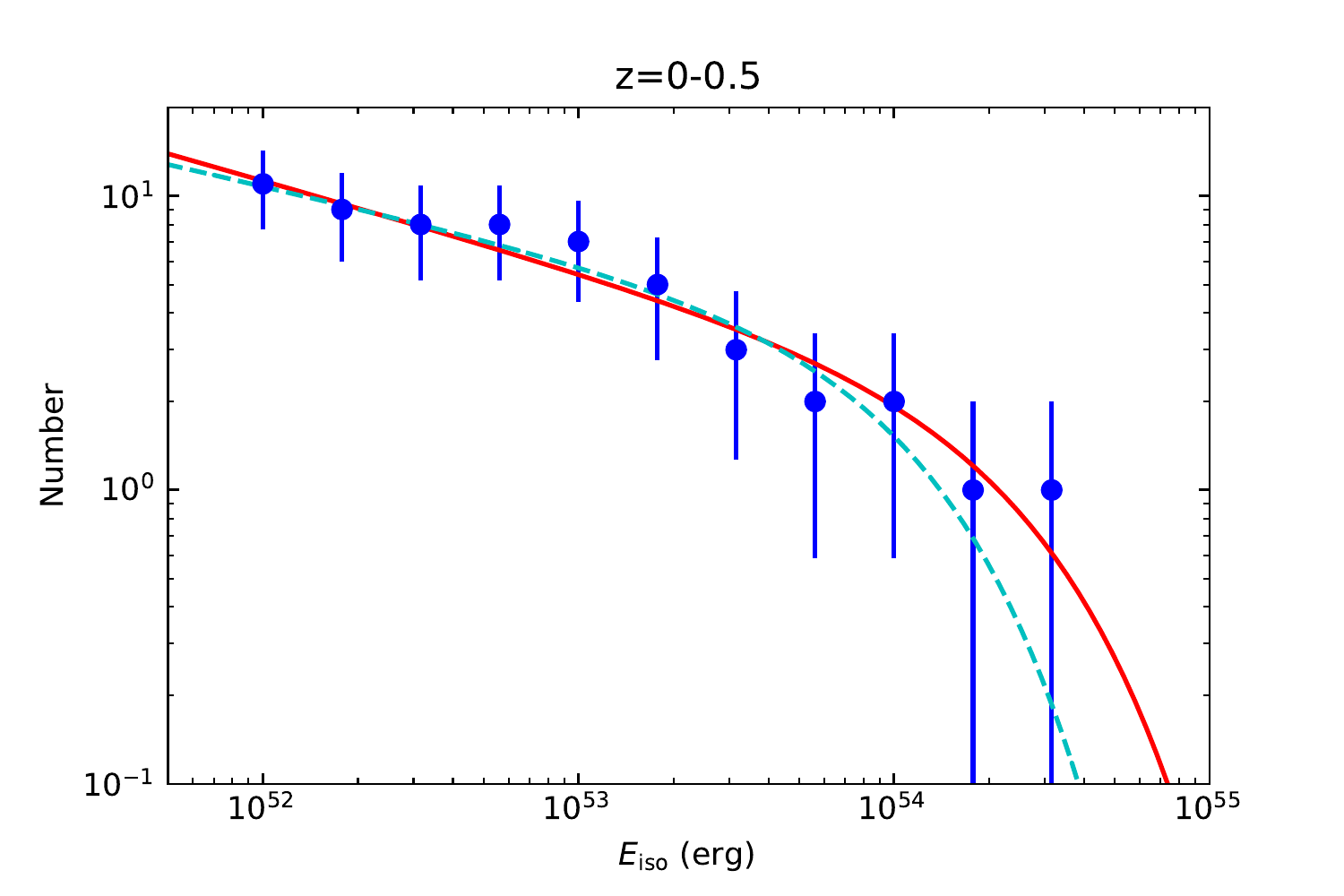}
\includegraphics    [angle=0,scale=0.55]     {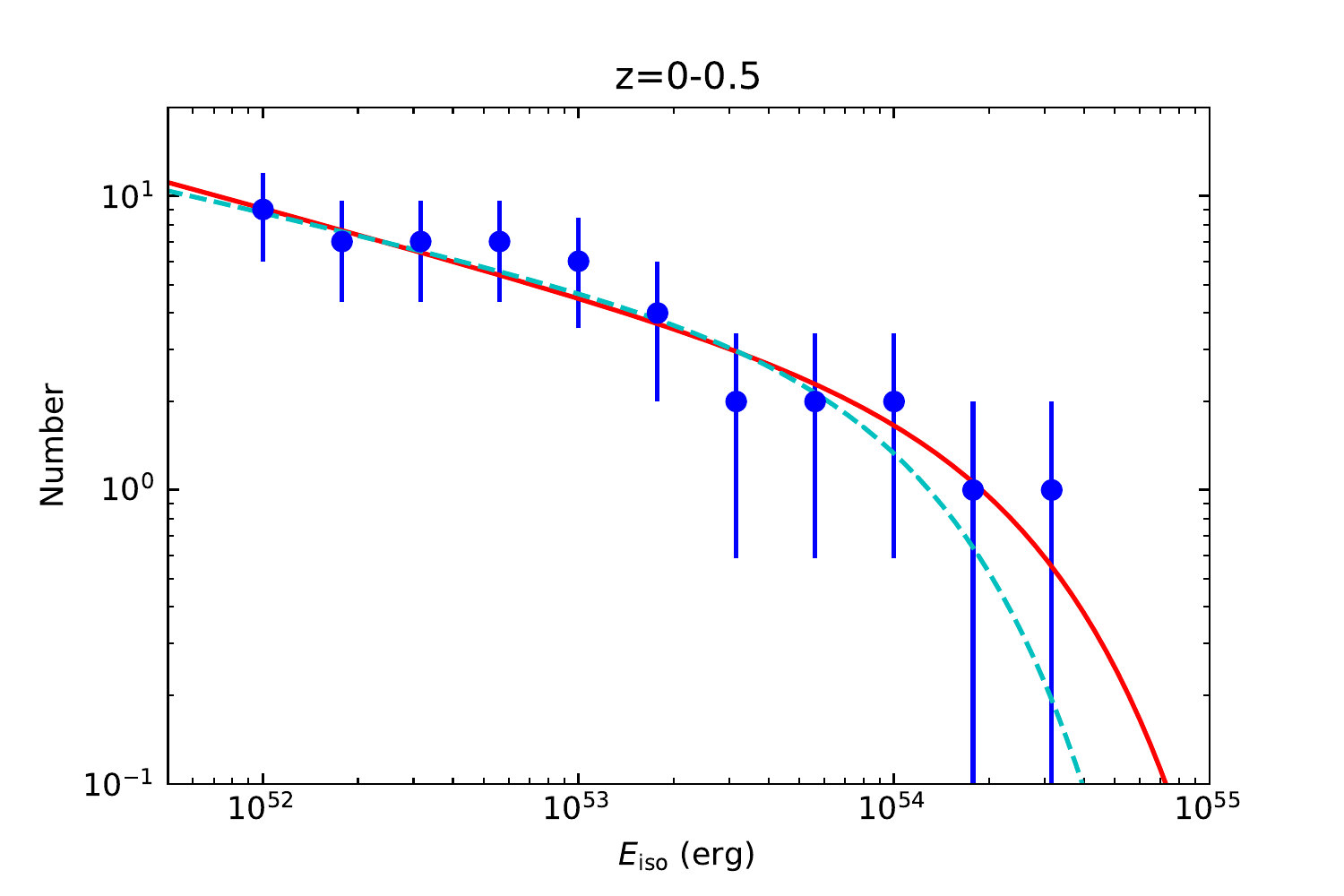}\\
\caption{Comparison between the normalized best-fitting CPL model of the total redshift bin ($0<z<9.4$) and the best-fitting CPL model of the low redshift bin ($0<z<0.5$) for total samples (left panel) and pure Konus-\emph{Wind} samples (right panel). The blue dashed lines and red solid lines represent the best-fitting CPL model for total redshift bin and low redshift bin.}
\label{fig:CPL-compare}
\end{figure*}

\begin{figure*}
\centering
\includegraphics    [angle=0,scale=0.55]     {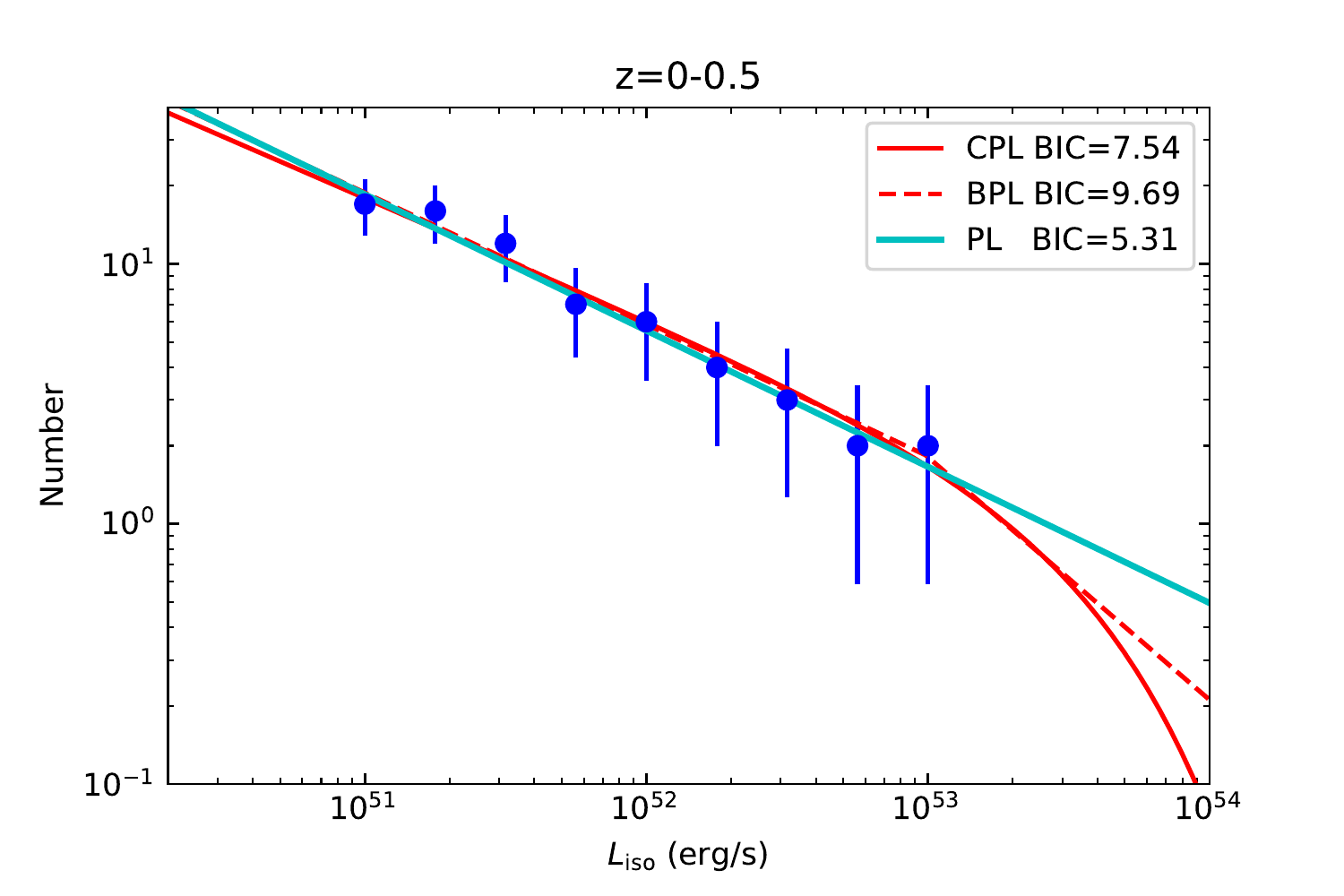}
\includegraphics    [angle=0,scale=0.55]     {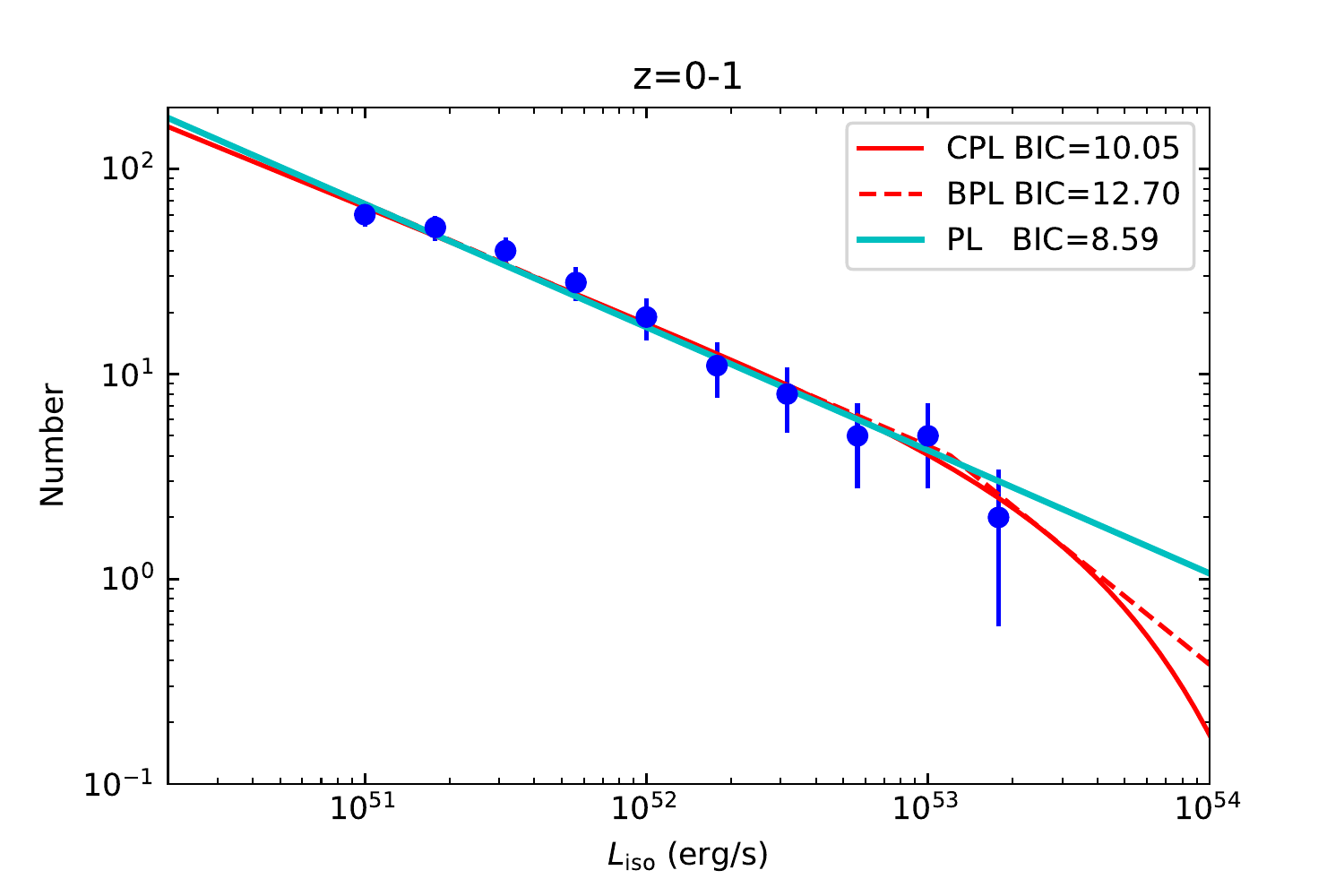}\\
\includegraphics    [angle=0,scale=0.55]     {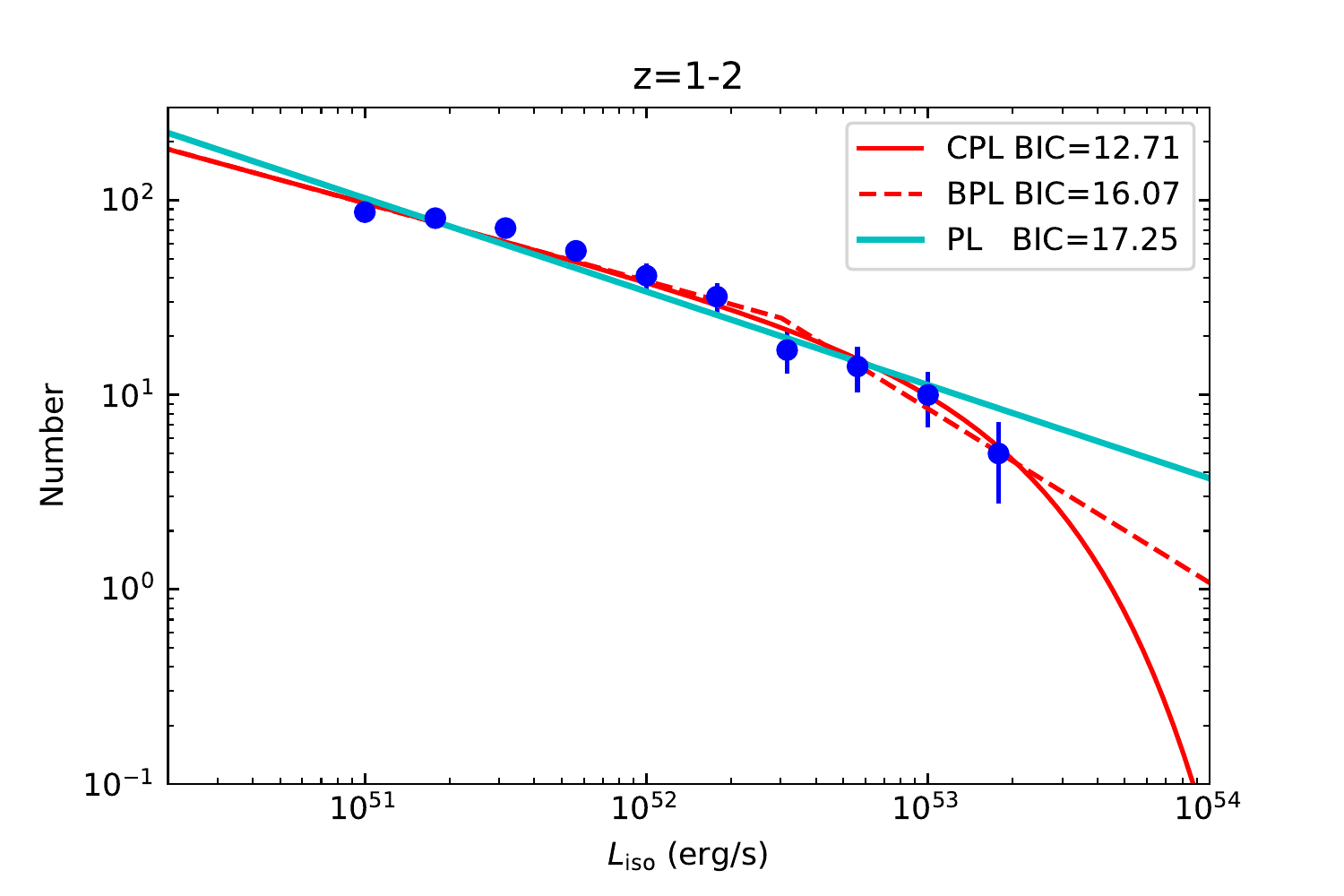}
\includegraphics    [angle=0,scale=0.55]     {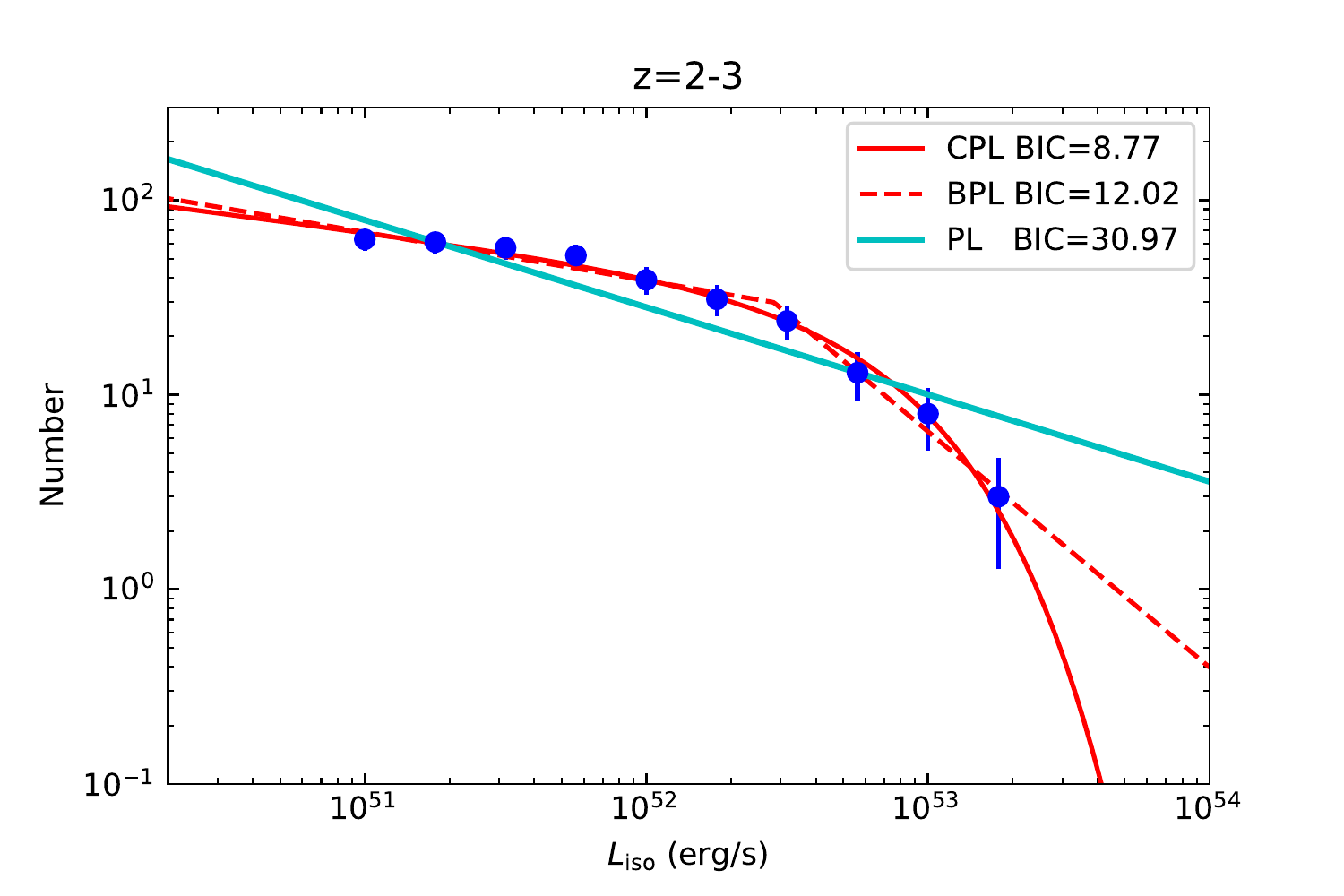}\\
\includegraphics    [angle=0,scale=0.55]     {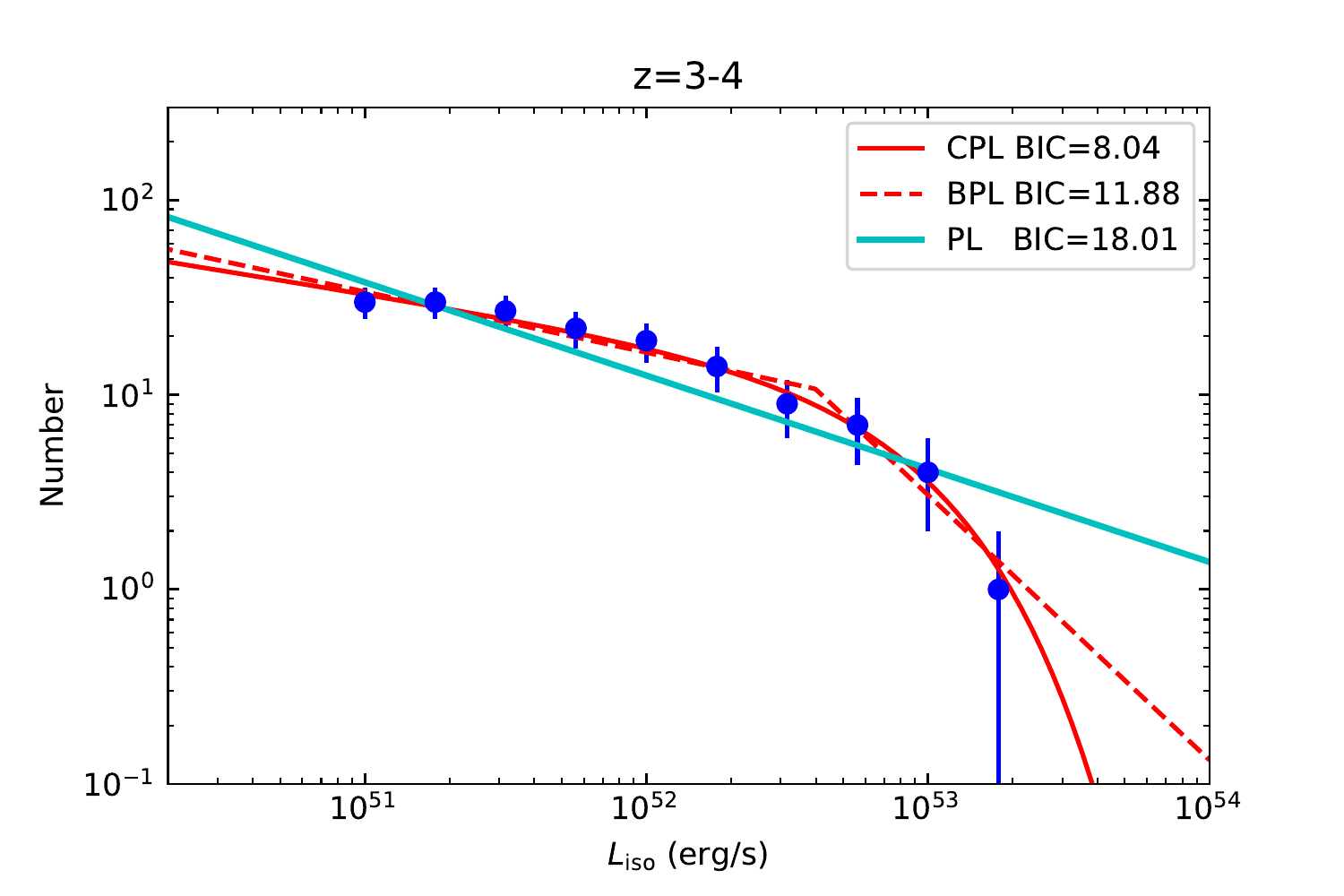}
\includegraphics    [angle=0,scale=0.55]     {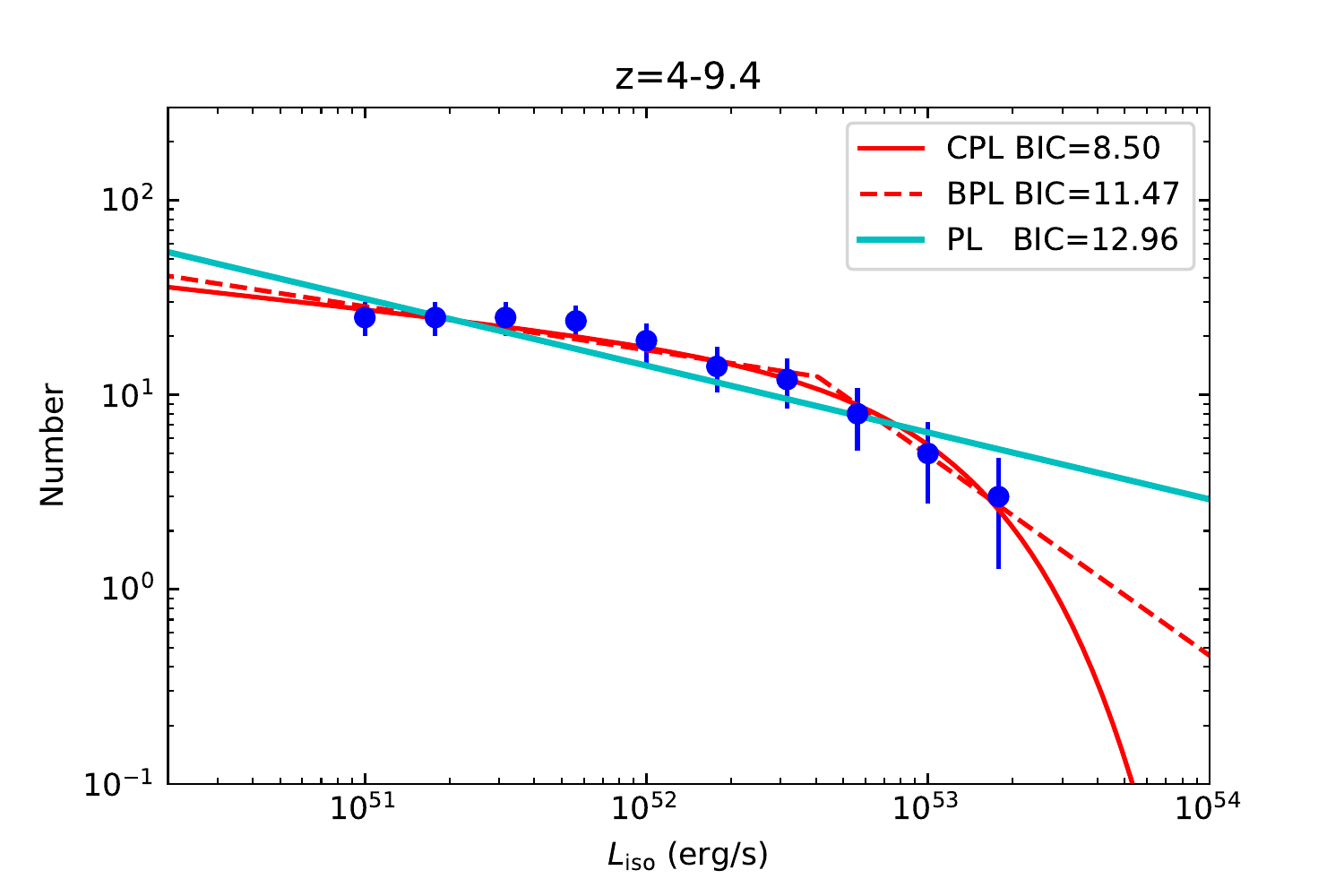}\\
\includegraphics    [angle=0,scale=0.55]     {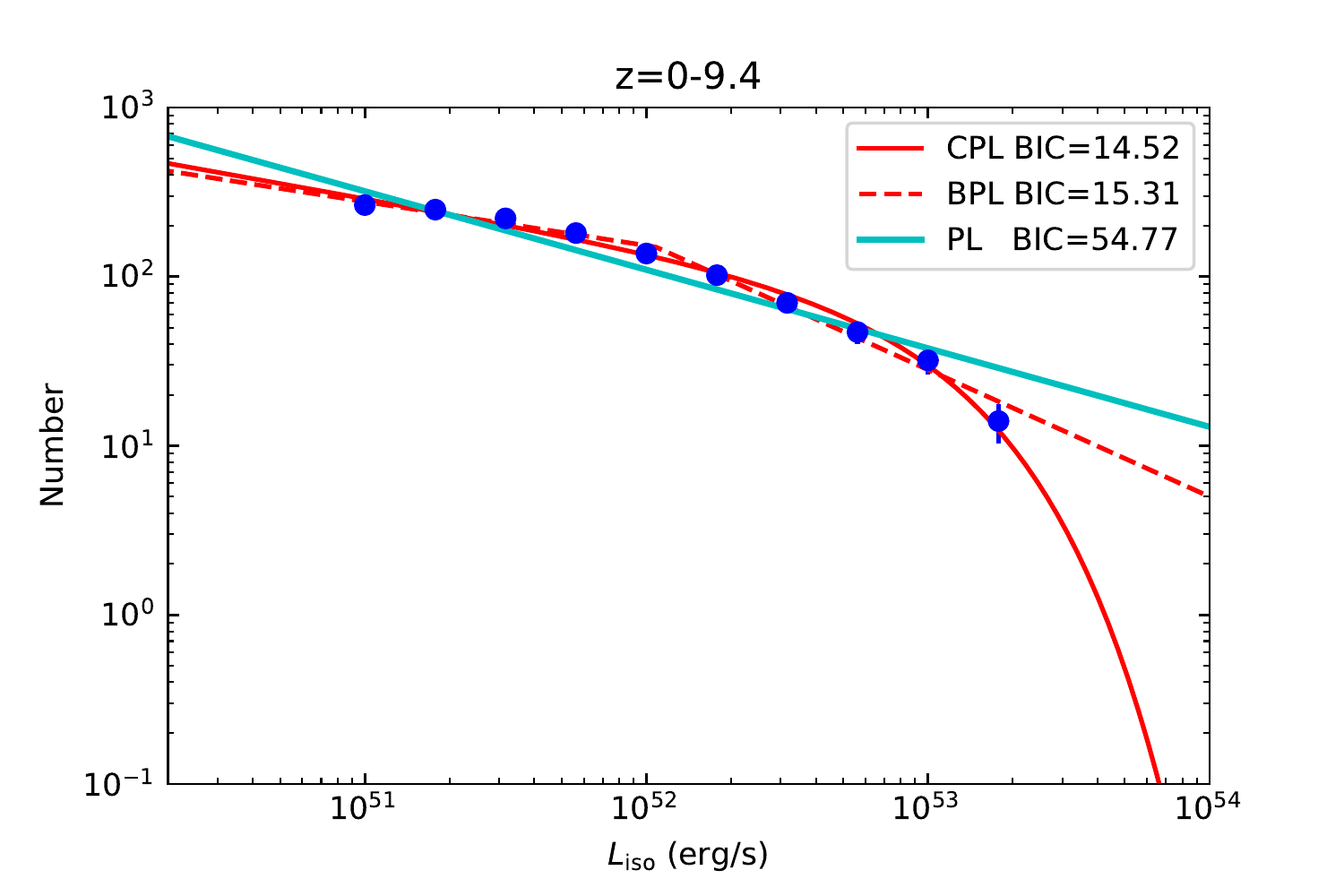}

\caption{Cumulative distribution of GRB isotropic luminosity for total samples in the different redshift bins. The blue circles show the observed distribution, and the red solid lines, red dashed lines, and blue solid lines represent the best-fitting CPL model, BPL model, and PL model, respectively.}
\label{fig:Liso}
\end{figure*}

The weak advantage of PL model performance in low and high redshift samples is likely due to the limited sample size. There is no clear evidence that more energetic GRBs are easily generated at low redshift. To further prove this, we normalize the best fitting CPL function of the total sample to compare with the distribution of $E_{\rm iso}$ at $0<z<0.5$ (see Figure \ref{fig:CPL-compare}). If the best fitting result of the total sample can represent the intrinsic distribution of the GRB energy function, we find that the occurrence of GRB 221009A is consistent with the expectation within 1.84 $\sigma$ Poisson fluctuation error.

\tablenum{2}
\begin{table*}[!htb] \scriptsize
\begin{center}
\caption{The best-fitting results of energy function for pure Konus-\emph{Wind} and total sample GRBs in different redshift bins.}
\label{table-2}
\begin{tabular}{lc|ccc|ccccccc}
\hline\hline

redshift interval &   \multicolumn{1}{c}{PL} & \multicolumn{3}{c}{BPL} & \multicolumn{2}{c}{CPL}\\\hline
Konus-\emph{Wind} GRBs & $\alpha$ & $\alpha$ & $\beta$ & $E_b$ ($10^{54}\rm erg$) & $\alpha$ & $E_c$ ($10^{54}\rm erg$)\\\hline

\object{[$0.0\sim0.5$]} & $0.38^{+0.08}_{-0.07}$ & $0.36^{+0.08}_{-0.07}$ & $1.06^{+1.21}_{-0.72}$ & $2.70^{+1.55}_{-1.58}$ & $0.31^{+0.10}_{-0.10}$  & $5.83^{+2.87}_{-3.15}$\\
\object{[$0.0\sim1.0$]} & $0.53^{+0.05}_{-0.04}$ & $0.41^{+0.09}_{-0.15}$ & $0.94^{+0.32}_{-0.24}$ & $0.38^{+0.90}_{-0.27}$ & $0.42^{+0.06}_{-0.08}$  & $2.77^{+1.98}_{-1.51}$\\
\object{[$1.0\sim2.0$]} & $0.39^{+0.03}_{-0.03}$ & $0.22^{+0.05}_{-0.07}$ & $1.22^{+0.60}_{-0.37}$ & $0.49^{+0.27}_{-0.26}$ & $0.17^{+0.06}_{-0.06}$  & $1.26^{+0.52}_{-0.34}$\\
\object{[$2.0\sim3.0$]} & $0.44^{+0.03}_{-0.03}$ & $0.17^{+0.06}_{-0.07}$ & $1.50^{+0.66}_{-0.41}$ & $0.45^{+0.25}_{-0.19}$ & $0.09^{+0.06}_{-0.07}$  & $0.87^{+0.27}_{-0.23}$\\
\object{[$3.0\sim4.0$]} & $0.32^{+0.05}_{-0.05}$ & $0.13^{+0.09}_{-0.11}$ & $1.08^{+0.56}_{-0.40}$ & $0.50^{+0.31}_{-0.25}$ & $0.11^{+0.09}_{-0.10}$  & $1.77^{+1.12}_{-0.74}$\\
\object{[$4.0\sim9.4$]} & $0.23^{+0.07}_{-0.06}$ & $0.17^{+0.09}_{-0.09}$ & $0.62^{+0.59}_{-0.33}$ & $0.60^{+0.29}_{-0.35}$ & $0.13^{+0.08}_{-0.09}$  & $3.47^{+1.65}_{-1.59}$\\
\object{[$0.0\sim9.4$]} & $0.52^{+0.02}_{-0.01}$ & $0.22^{+0.04}_{-0.05}$ & $1.23^{+0.24}_{-0.19}$ & $0.45^{+0.23}_{-0.18}$ & $0.18^{+0.03}_{-0.04}$  & $1.41^{+0.30}_{-0.26}$\\
\hline
Total sample GRBs\\
\hline
\object{[$0.0\sim0.5$]} & $0.40^{+0.08}_{-0.07}$ & $0.38^{+0.05}_{-0.06}$ & $0.86^{+0.72}_{-0.58}$ & $2.89^{+1.37}_{-1.32}$ & $0.33^{+0.08}_{-0.09}$  & $5.62^{+2.96}_{-3.16}$\\
\object{[$0.0\sim1.0$]} & $0.54^{+0.04}_{-0.03}$ & $0.49^{+0.05}_{-0.05}$ & $0.95^{+0.32}_{-0.27}$ & $0.97^{+0.64}_{-0.48}$ & $0.44^{+0.05}_{-0.05}$  & $3.19^{+1.22}_{-1.38}$\\
\object{[$1.0\sim2.0$]} & $0.47^{+0.02}_{-0.02}$ & $0.30^{+0.04}_{-0.06}$ & $1.21^{+0.53}_{-0.32}$ & $0.43^{+0.22}_{-0.22}$ & $0.26^{+0.05}_{-0.05}$  & $1.24^{+0.48}_{-0.34}$\\
\object{[$2.0\sim3.0$]} & $0.52^{+0.03}_{-0.03}$ & $0.25^{+0.05}_{-0.05}$ & $1.74^{+0.63}_{-0.43}$ & $0.45^{+0.16}_{-0.15}$ & $0.16^{+0.06}_{-0.06}$  & $0.73^{+0.22}_{-0.17}$\\
\object{[$3.0\sim4.0$]} & $0.45^{+0.04}_{-0.04}$ & $0.27^{+0.08}_{-0.09}$ & $1.14^{+0.51}_{-0.33}$ & $0.36^{+0.29}_{-0.17}$ & $0.25^{+0.07}_{-0.09}$  & $1.40^{+0.80}_{-0.56}$\\
\object{[$4.0\sim9.4$]} & $0.39^{+0.05}_{-0.04}$ & $0.30^{+0.07}_{-0.11}$ & $0.81^{+0.50}_{-0.27}$ & $0.46^{+0.37}_{-0.33}$ & $0.31^{+0.06}_{-0.07}$  & $3.38^{+2.23}_{-1.68}$\\
\object{[$0.0\sim9.4$]} & $0.57^{+0.01}_{-0.01}$ & $0.29^{+0.03}_{-0.04}$ & $1.19^{+0.15}_{-0.14}$ & $0.37^{+0.10}_{-0.13}$ & $0.27^{+0.03}_{-0.03}$  & $1.28^{+0.29}_{-0.24}$\\
\hline
\end{tabular}
\end{center}
\end{table*}

\tablenum{3}
\begin{table*}[!htb] \scriptsize
\begin{center}
\caption{The best-fitting results of luminosity function for total sample GRBs in different redshift bins.}
\label{table-3}
\begin{tabular}{lc|ccc|ccccccc}
\hline\hline

redshift interval &   \multicolumn{1}{c}{PL} & \multicolumn{3}{c}{BPL} & \multicolumn{2}{c}{CPL}\\\hline
 & $\alpha$ & $\alpha$ & $\beta$ & $L_b$ ($10^{54}\rm erg/s$) & $\alpha$ & $L_c$ ($10^{54}\rm erg/s$)\\\hline

\object{[$0.0\sim0.5$]} & $0.52^{+0.10}_{-0.09}$ & $0.51^{+0.08}_{-0.09}$ & $0.93^{+0.70}_{-0.72}$ & $0.10^{+0.07}_{-0.06}$ & $0.47^{+0.07}_{-0.09}$  & $0.45^{+0.35}_{-0.28}$\\
\object{[$0.0\sim1.0$]} & $0.60^{+0.05}_{-0.05}$ & $0.59^{+0.05}_{-0.05}$ & $1.11^{+0.59}_{-0.61}$ & $0.12^{+0.05}_{-0.05}$ & $0.56^{+0.06}_{-0.06}$  & $0.48^{+0.34}_{-0.28}$\\
\object{[$1.0\sim2.0$]} & $0.48^{+0.03}_{-0.03}$ & $0.40^{+0.07}_{-0.11}$ & $0.89^{+0.57}_{-0.22}$ & $0.03^{+0.07}_{-0.02}$ & $0.39^{+0.06}_{-0.06}$  & $0.21^{+0.29}_{-0.09}$\\
\object{[$2.0\sim3.0$]} & $0.45^{+0.03}_{-0.03}$ & $0.25^{+0.06}_{-0.07}$ & $1.21^{+0.40}_{-0.31}$ & $0.03^{+0.01}_{-0.01}$ & $0.19^{+0.07}_{-0.07}$  & $0.08^{+0.03}_{-0.02}$\\
\object{[$3.0\sim4.0$]} & $0.48^{+0.05}_{-0.04}$ & $0.31^{+0.08}_{-0.10}$ & $1.36^{+0.42}_{-0.41}$ & $0.04^{+0.03}_{-0.02}$ & $0.23^{+0.09}_{-0.10}$  & $0.09^{+0.05}_{-0.03}$\\
\object{[$4.0\sim9.4$]} & $0.34^{+0.05}_{-0.04}$ & $0.22^{+0.08}_{-0.09}$ & $1.03^{+0.55}_{-0.39}$ & $0.04^{+0.03}_{-0.02}$ & $0.16^{+0.07}_{-0.08}$  & $0.12^{+0.05}_{-0.04}$\\
\object{[$0.0\sim9.4$]} & $0.46^{+0.02}_{-0.02}$ & $0.26^{+0.05}_{-0.07}$ & $0.75^{+0.11}_{-0.08}$ & $0.01^{+0.01}_{-0.01}$ & $0.30^{+0.03}_{-0.03}$  & $0.11^{+0.03}_{-0.02}$\\
\hline
\end{tabular}
\end{center}
\end{table*}

\section{Statistic investigation}

With our collected sample, it is of great interest to investigate whether GRB 221009A and other energetic GRBs are systematically different from other GRBs in terms of statistics of various properties. Here we focus on the following aspects: $T_{90}$ distribution, minimum variability timescale distribution, Amati relation, Spectral lag distribution, X-ray and optical afterglow properties, the relation between the $E_{\rm \gamma,iso}$ and $E_{\rm X,iso}$, initial Lorentz factor distribution, and host galaxy properties.

\subsection{$T_{90}$ distribution} 
\begin{figure*}
\centering
\includegraphics    [angle=0,scale=0.8]     {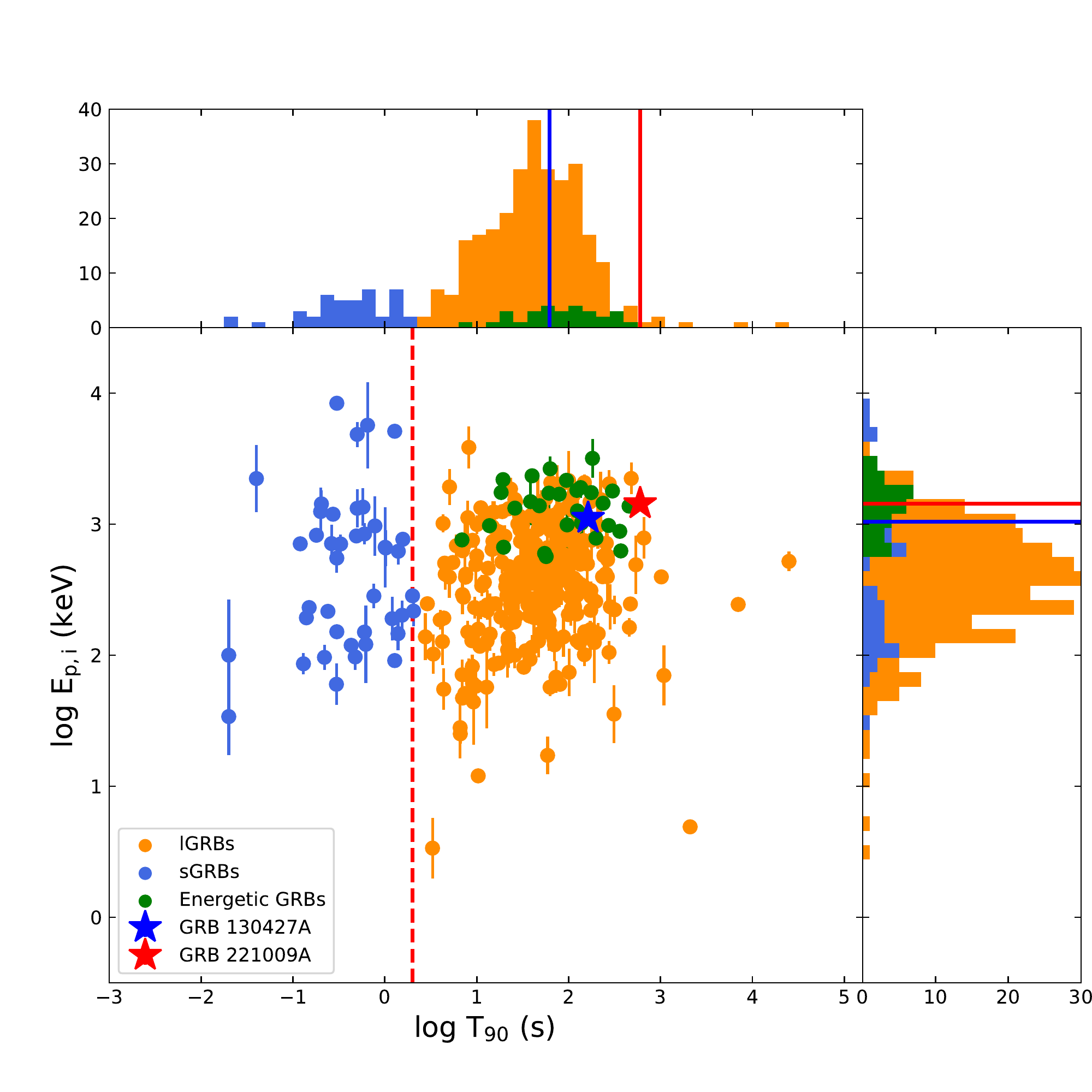}
\caption{lGRB/sGRB classification diagram in the $T_{90}-E_{\rm p}$ domain. The blue circles, orange circles, blue star, red star, and green circles represent the sGRBs, lGRBs, GRB 130427A, GRB 221009A,  and other energetic GRBs, respectively. The red vertical dotted line represent $T_{90}=2$ s. The histograms of top and right represent the $T_{90}$ and $E_{\rm p}$ distribution for all GRBs, and the blue and red vertical lines represent GRB 130427A and GRB 221009A location, respectively.}
\label{fig:T90}
\end{figure*}

Phenomenologically, GRBs fall into two classes: the long-duration, soft-spectrum class(duration $<2\; \rm s$; lGRBs) and the short-duration, hard-spectrum class(duration $>2\; \rm s$; sGRBs), based on the bimodal distribution of GRBs in the duration-hardness diagram \citep{Kouveliotou93}. In Figure \ref{fig:T90}, we show the GRBs in our sample in the duration $T_{90}$ versus intrinsic peak energy $E_{\rm p,i}$ diagram. We find that all the energetic GRBs fall into the distribution of lGRBs, and the distributions in $T_{90}$ and $E_{\rm p,i}$ are not significantly different with respect to the other lGRBs in our sample. Among the energetic bursts, GRB 221009A is distributed at the relatively large side of $T_{90}$ and the center of $E_{\rm p,i}$. 

\subsection{Minimum variability timescale distribution}

\begin{figure*}
\centering
\includegraphics    [angle=0,scale=1]     {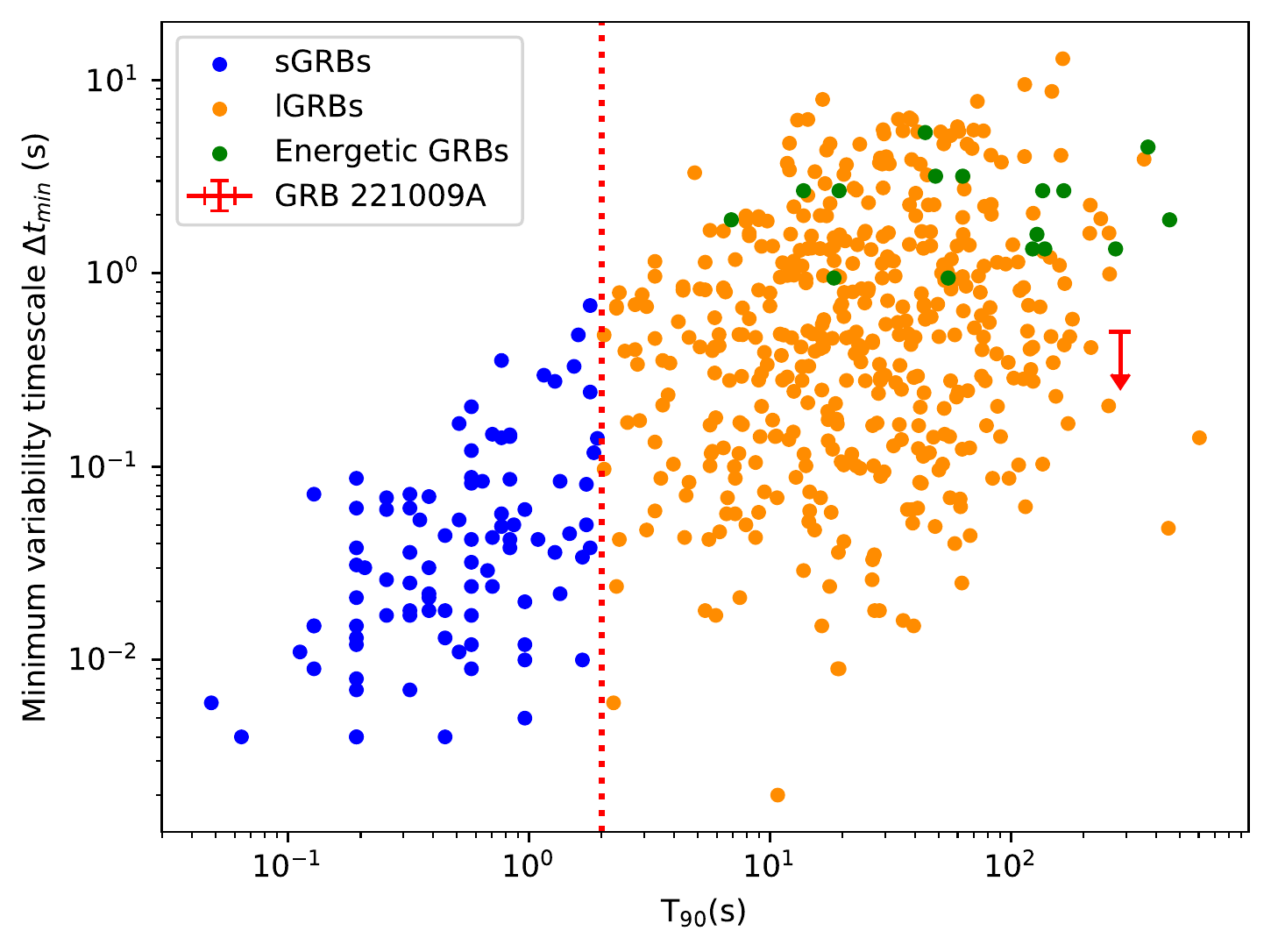}
\caption{The diagram in the $T_{90}-\Delta t_{\rm min}$ domain. The samples of sGRBs (blue circles) and lGRBs (orange circles) are from \cite{Golkhou15}. The red arrow and green circles represent GRB 221009A and other energetic GRBs, respectively. The red vertical dotted line represent $T_{90}=2$ s.}
\label{fig:Mvt}
\end{figure*}

The minimum timescale ($\Delta t_{\rm min}$) on which a GRB exhibits significant flux variations is believed to provide an upper limit as to the size of the radiation zone, the lower limit of the Lorentz factor and potentially shedding light on the nature of emission mechanism \citep{Schmidt78}. For typical lGRBs and sGRBs, the average minimum timescales in the rest frame (i.e. $\Delta t_{\rm min}$/(1+$z$)) are
45 and 10 ms, respectively \citep{Golkhou15}. In order to calculate the minimum timescale for the energetic GRB sample, we employ the Bayesian Blocks algorithm and define the 1/2 shortest significant structures of blocks as the duration of minimum time interval \citep{Vianello18,Xiao22c}. The median minimum timescale of the energetic GRB sample is about 2.32 s at 10-1000 keV, which is well consistent with typical lGRBs. The minimum timescale upper limit of GRB 221009A\footnote{GBM detectors have reached saturation for the extremely bright GRB 221009A in main emission phase, and we can only get the upper limit of the minimum variability timescale in main emission.} is 0.4 s at 10-1000 keV, which is no significant difference compared to typical lGRBs and other energetic GRBs. We also adopt the continuous wavelet transform (CWT) method \citep{Vianello18} and the results are consistent. The result is shown in Figure \ref{fig:Mvt}.

\subsection{Amati relation} 

\begin{figure*}
\centering
\includegraphics    [angle=0,scale=0.8]     {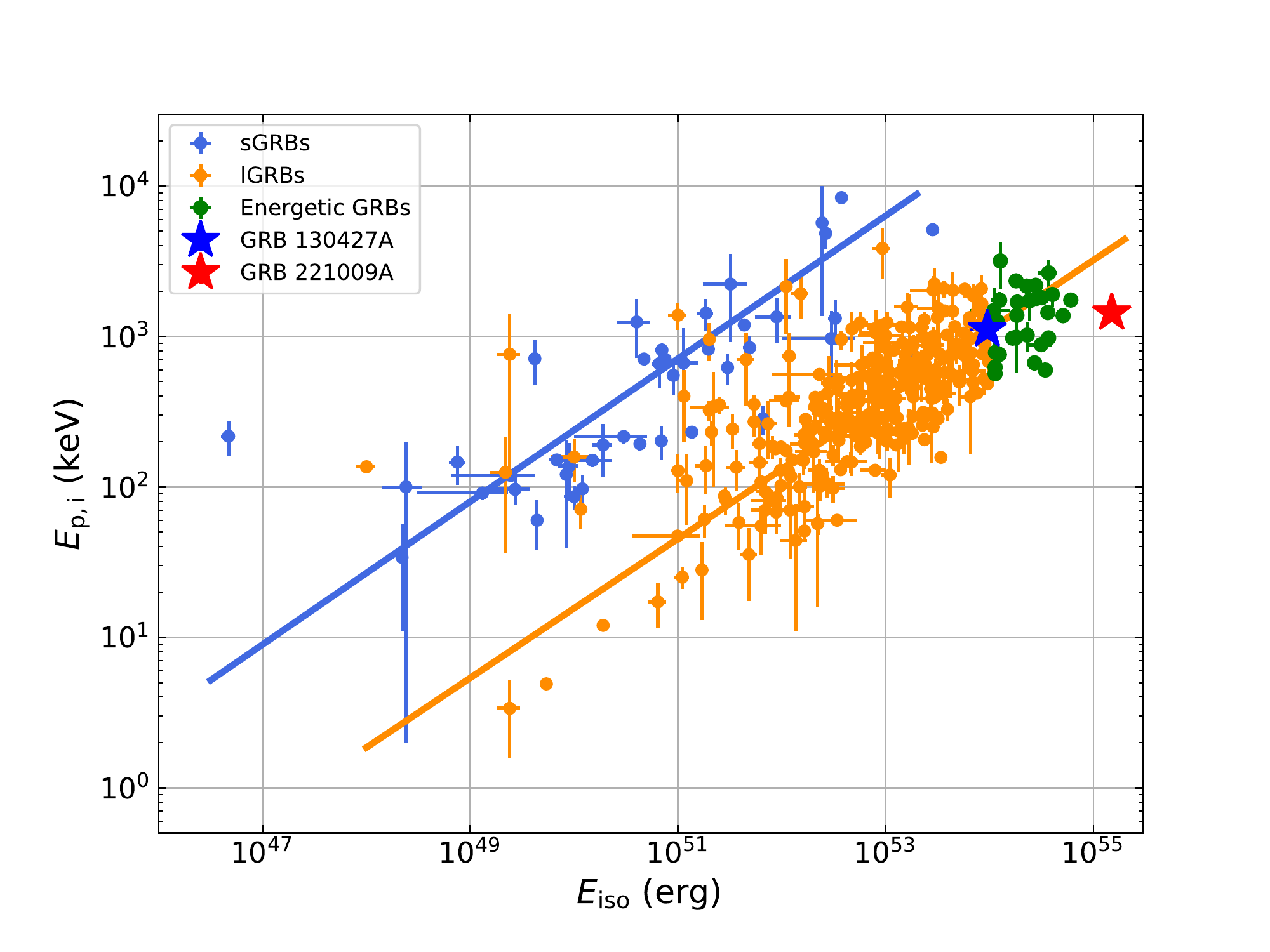}
\caption{$E_{\rm p,i}$ and $E_{\rm \gamma, iso}$ correlation diagram with known redshift data \citep{Zhang09,Qin13,Minaev20,Zou18,Jia22}. The orange and blue solid lines represent the best-fit correlations for lGRBs and sGRBs, respectively. The blue circles, orange circles, blue star, red star, and green circles represent the sGRBs, lGRBs, GRB 130427A, GRB 221009A, and other energetic lGRBs, respectively.}
\label{fig:Amati}
\end{figure*}

Some empirical correlations among GRB observational parameters have been discovered in the literature. The most famous one is the Amati relations \citep{Amati02},
which is a correlation between the GRB isotropic energy $E_{\rm\gamma,iso}$ and the rest-frame peak energy $E_{\rm p,i}=(1+z)E_{\rm p}$.
\cite{Amati02} discovered that higher energy lGRBs have a harder spectrum than that of lower energy lGRBs, and the sGRBs also follow the same trend between $E_{\rm p,i}$ and $E_{\rm \gamma,iso}$ but form distinct tracks \citep{Zhang09}. Here we plot the Amati diagram for both the lGRB and sGRB populations(see Figure \ref{fig:Amati}) with our collected sample. We find that most energetic GRBs are well located in the same trend as normal lGRBs, although their $E_{\rm \gamma,iso}$ are higher than those of most observed lGRBs. GRB 221009A slightly deviates from the Amati relation towards higher isotropic energy (see \cite{GECAM} for more details).

\subsection{Spectral lag distribution} 
The redshift distribution of energetic GRBs in our sample ranges from 0.151 to 6.318. For each GRB, the fixed rest-frame energy bands are selected to be 100–150 keV and 200–250 keV, corresponding to the observed energy of [200-250]/(1+z) keV and [100-150]/(1+z) keV, which is the same as in \citep{Ukwatta12}.
The purpose of the above selection of energy bands is to make full use of the data and ensure sufficient energy difference between these two bands.

Based on the high time-resolution initial light curves of \emph{Swift}/BAT or \emph{Fermi}/GBM, we utilize the Li-CCF method (see \cite{Li04} and \cite{Xiao21,Xiao22a,Xiao22b} for details) to calculate the spectral lags for energetic GRBs in our sample, which is defined as
\begin{equation}
\begin{split}\label{equ:MCCF}
{\rm MCCF}(k,\Delta t)=\frac{1}{M_{\Delta t}} \sum_{m=1}^{M_{\Delta t}} \sum_{i} u_m(i;\Delta t)\upsilon_{m+k}(i;\Delta t)/\sigma_ u \sigma_\upsilon,
\end{split}
\end{equation}
where the combination starts from the mth bin of the initial light curves, the phase factor $m$ = 1, 2, $\cdots$, $M_{\Delta t}$, and $u_m(\Delta t)$ and $\upsilon_m(\Delta t)$ are the background-subtracted series of $x_m(\Delta t)$ and $y_m(\Delta t)$ by re-binning the initial series, we obtain the lightcurves with an optimized time bin $\Delta t=M_{\Delta t}\delta t$ (from 1 to 100ms for sGRBs, from 1 ms to 1 s for lGRBs), respectively,
\begin{equation}
\begin{split}\label{equ:improved MCCF}
u_m(i;\Delta t)=x_m(i;\Delta t)-b_{x_m}(i;\Delta t),\\
\upsilon_m(i;\Delta t)=y_m(i;\Delta t)-b_{y_m}(i;\Delta t),\\
\sigma^{2}_u=\sum_{i} {u_m(i;\Delta t)}^{2},\\ \sigma^{2}_\upsilon=\sum_{i} {\upsilon_m(i;\Delta t)}^{2},
\end{split}
\end{equation}
where $b_{x_m}$ and $b_{y_m}$ are the estimated background counts of lightcurves $x_m(\Delta t)$ and $y_{m}(\Delta t)$, respectively. We can obtain a value $k_{\rm max}$ of $k$ that maximizes MCCF($k=k_{\rm max}$, $\Delta t$), then the relative time lag between two light curves $y_m(i;\Delta t)$ and $x_m(i;\Delta t)$ on $\Delta t$ is
\begin{equation}
\tau(\Delta t)=k_{\rm max}\delta t.
\end{equation}
We implement a Monte Carlo simulation of the observed light curves based on Poisson probability distribution to obtain the uncertainty of spectral lag.

\begin{figure*}
\centering
\includegraphics    [angle=0,scale=1]     {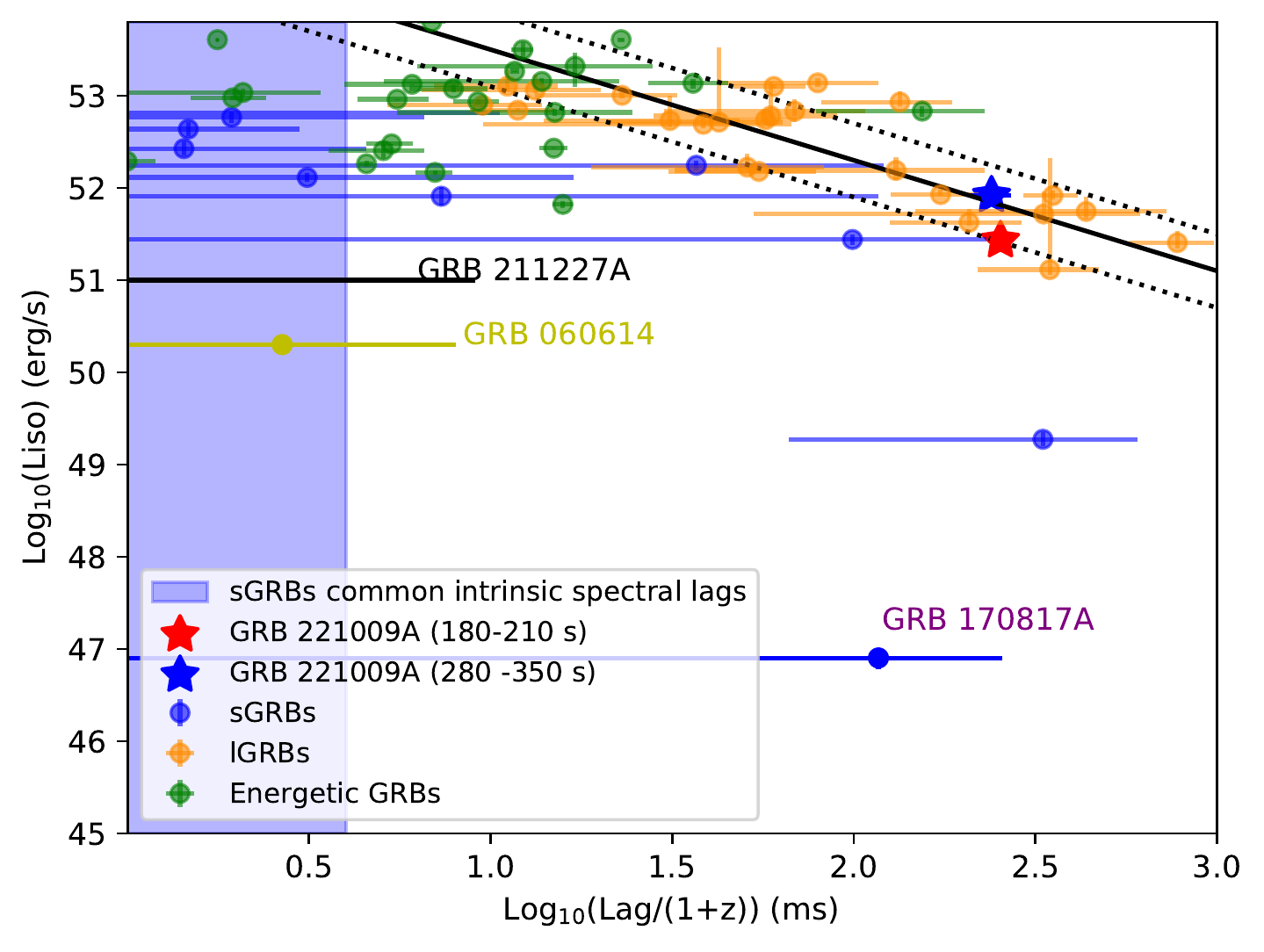}
\caption{The correlation between spectral lag and peak luminosity in the rest frame. The GRB sample is collected from the previous statistical investigations \citep{Li04,Gehrels06,Ukwatta12,Goldstein17,Lv22, Xiao22a}. The GRB 221009A for spectral lag of the two unsaturated main emission episodes ($T_0+180$ s $\sim$ $T_0+210$ s, $T_0+280$ s $\sim$ $T_0+350$ s) and other energetic GRBs are highlighted by red star, blue star and green circles, respectively. The blue shaded region represents the common intrinsic spectral lags (3$\sigma$) calculated from 46 sGRBs with redshift measurements \citep{Xiao22a}.}
\label{fig:Spectral-lag}
\end{figure*}

It has long been found that there is an anti-correlation between spectral lag and peak luminosity exists for lGRBs \citep{Norris00,Hakkila08}, namely short-lag and variable bursts having greater luminosities than long-lag and smooth bursts \citep{Norris02,Hakkila07,Ukwatta12,Shao17}. Here we found that the energetic sources systematically deviate from this relationship, and their average spectral lag is smaller than that of normal lGRBs (see Figure \ref{fig:Spectral-lag}). We adopt two unsaturated emission episodes ($T_0+180$ s $\sim$ $T_0+210$ s, $T_0+280$ s$\sim$ $T_0+350$ s) in the main burst to calculate the spectral lag for GRB 221009A. We find that the spectral lags of the two unsaturated emission episodes in 200-250 keV compared to that in 100-150 keV at rest frame are $292 \pm 15$ ms and $276\pm 36$ ms, respectively. In order to estimate the peak luminosity of these two unsaturated emission episodes, a time-resolved spectral fitting was performed by a Band spectrum, and the peak energy flux of these two episodes are $(4.31\pm0.03)\times10^{-5}{\rm ~erg~cm^{-2}~s^{-1}}$ and $(1.32\pm0.01)\times10^{-4}{\rm ~erg~cm^{-2}~s^{-1}}$ in the 10 keV–1000 keV energy range, respectively. As shown in Figure \ref{fig:Spectral-lag}, the two unsaturated main emission episodes of GRB 221009A are located in the long burst region for the anti-correlation between spectral lag and peak luminosity.

\subsection{X-ray and optical afterglow properties}

\begin{figure*}
\centering
\includegraphics    [angle=0,scale=0.45]     {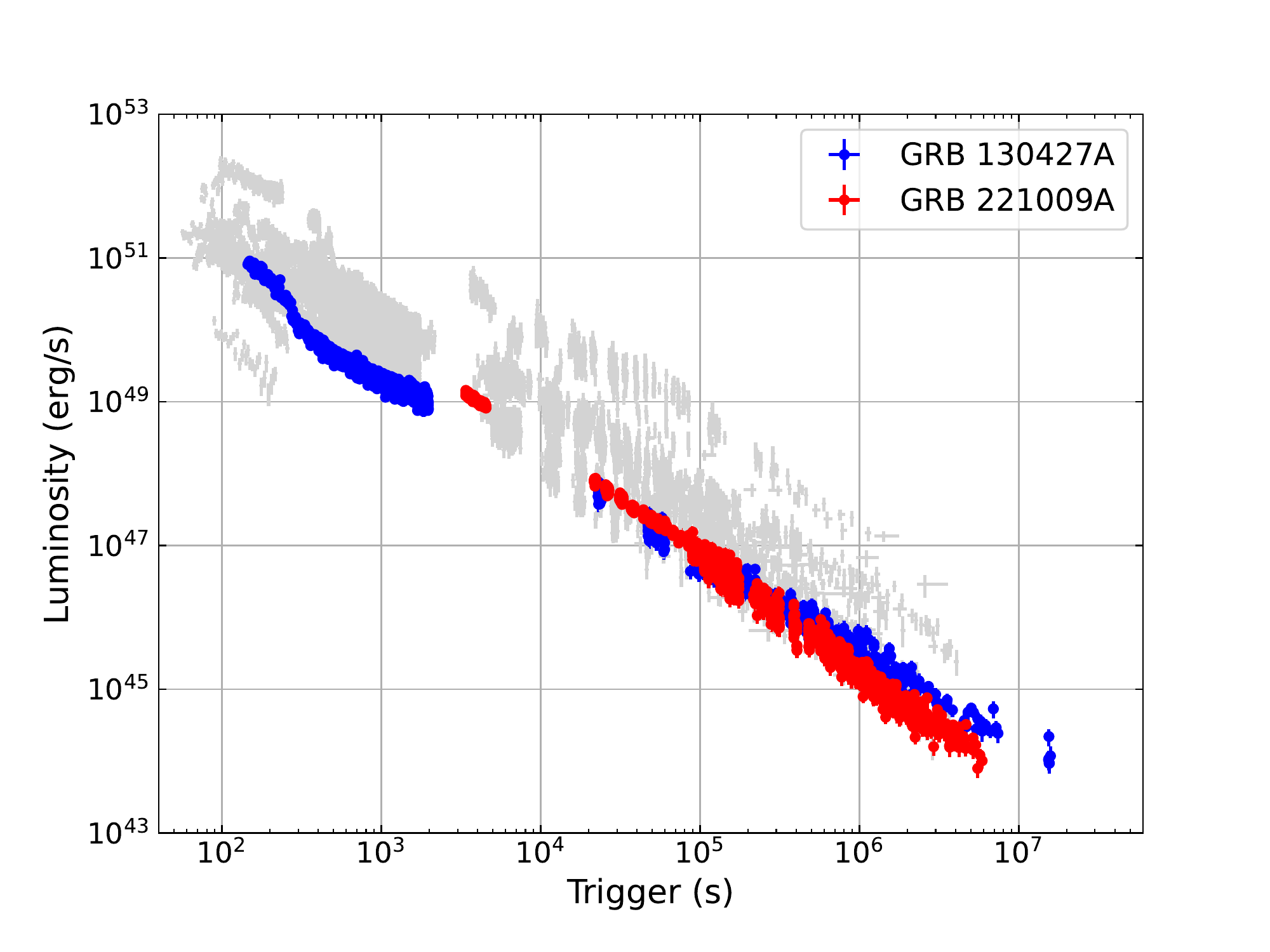}
\includegraphics    [angle=0,scale=0.52]     {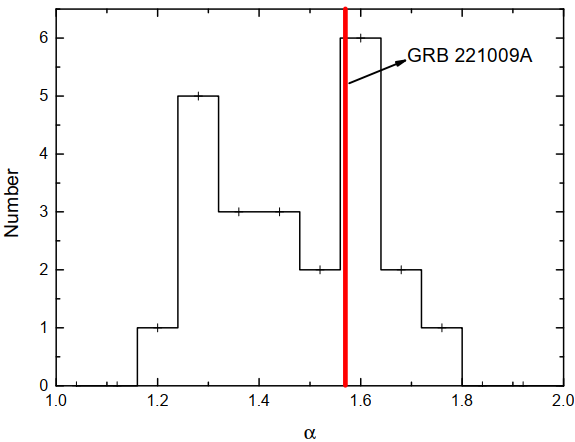}
\caption{Left panel: The X-ray luminosity lightcurves of energetic GRBs in our sample. The blue circles and red circles represent the GRB 130427A and GRB 221009A, respectively. The gray background circles show that other energetic GRBs. Right panel: The distribution of decay slope in X-ray late phase for energetic GRBs in our sample, and the red vertical line represent GRB 221009A location.}
\label{fig:XRT}
\end{figure*}

In order to compare whether GRB 221009A and other energetic GRBs are systematically different from other GRBs in X-ray and optical afterglow. Here, we
collected the X-ray and optical afterglow for GRB 221009A and other energetic GRBs. The XRT light curve is obtained by using the public data from the Swift archive\footnote{\url{https://www.swift.ac.uk/xrt_curves/}}. For each energetic GRB, we derived the X-ray luminosity, which is calculated by $L_{\rm X}=4k\pi D_{L}^2F_{\rm X}$, where $F_{\rm X}$ is observed X-ray flux, $D_L$ is the GRB luminosity distance, and the k-correction factor corrects the XRT-band (0.3-10 keV) flux to a wide band in the burst rest frame (0.1-1000 keV in this analysis), i.e.
\begin{equation}
k=\frac{\int^{10^3/{1+z}}_{0.1/{1+z}}EN(E)dE}{\int^{10}_{0.3}EN(E)dE}.
\end{equation}
Here $N(E)$ is the observed time-dependent X-ray photon spectrum, which could be best fitted by a power-law model (spectral parameters could be obtained from the Swift archive). To calculate $D_{L}(z)$, the concordance cosmology parameters $H_0 = 67.4$ km s$^{-1}$ Mpc $^{-1}$, $\Omega_M=0.315$, and $\Omega_{\Lambda}=0.685$ have been adopted according to the {\it Planck} results \citep{Planck20}. In the left panel of Figure \ref{fig:XRT}, we plot the X-ray luminosity curves for GRB 221009A and other energetic GRBs. We find that most energetic GRBs show simple power-law decay characteristics in the late phase with decay slopes systemically steeper compared to the so-called ``normal decay slope" (with a typical slope approximately -1.2, \cite{zhang2006}). The distribution of decay slope is shown in the right panel of Figure \ref{fig:XRT}, where the decay slope of GRB 221009A is located at the center. 

\begin{figure*}
\centering
\includegraphics    [angle=0,scale=0.7]     {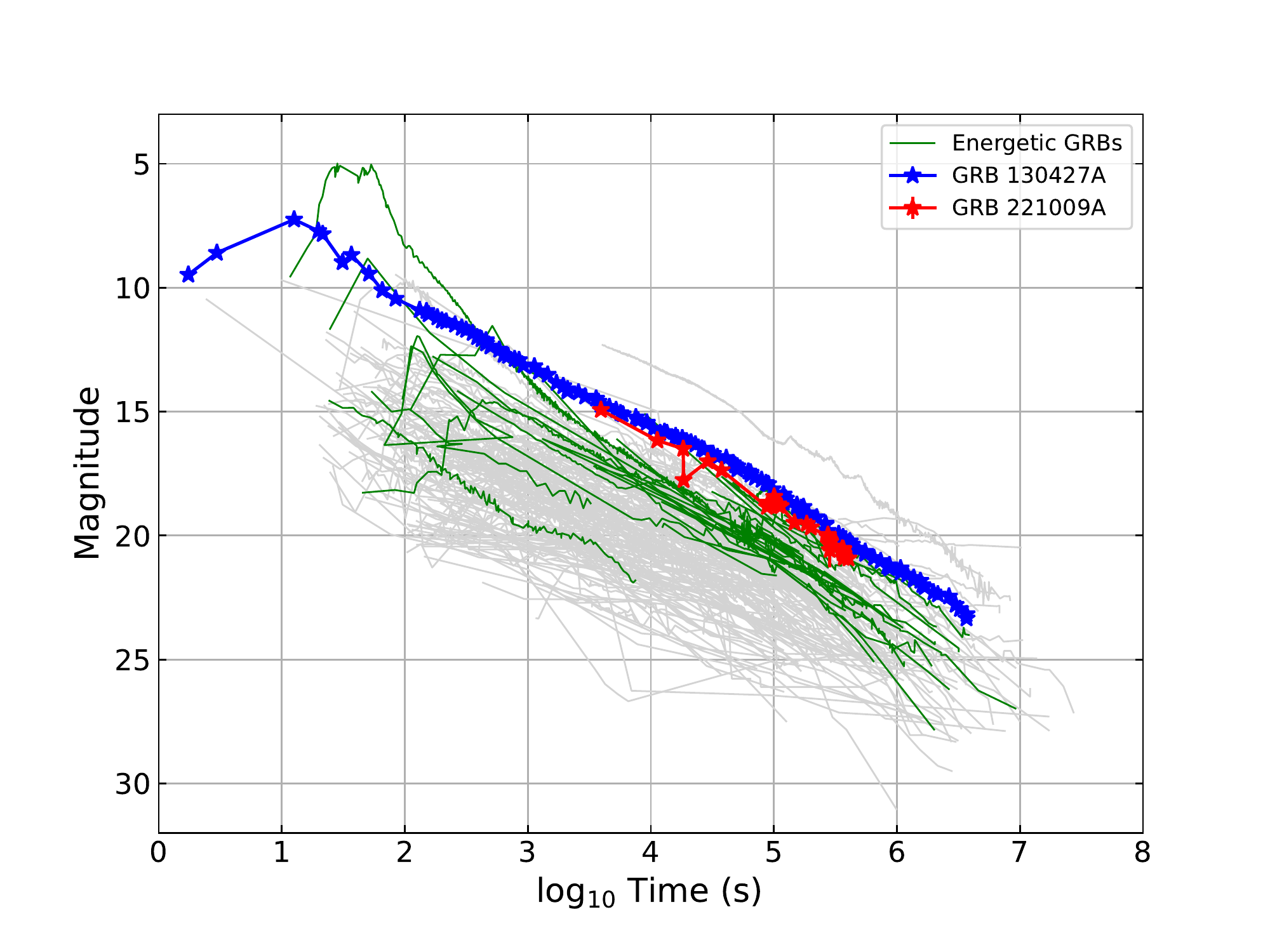}
\caption{Galactic extinction corrected R band afterglow lightcurves. The gray background lines show that we collected optical data, and the green lines show the optical afterglow of energetic GRBs in our sample. The blue star line and red star line represent the GRB 130427A and GRB 221009A, respectively.}
\label{fig:optic}
\end{figure*}

For optical afterglow, we extensively search for the optical data from published papers or from the Gamma-ray Coordinates Network (GCN) Circulars if no published paper is available. We found 358 GRBs in total with optical observations being reported from 1997 February to 2020 December, including 308 GRBs having well-sampled optical lightcurves, which contain at least three data points, excluding upper limits.
In figure \ref{fig:optic}, we show the optical lightcurves (in absolute magnitude) for energetic GRBs and other GRBs. We find that the optical afterglow is systematically brighter than other GRBs. Because of this, the optical afterglow observation of energetic GRBs generally starts earlier. Many of these bursts have been detected the reverse shock radiation and/or the onset bump of the forward shock radiation (unfortunately, the optical observation of GRB 221009A stars relatively late, because \emph{Swift}/BAT triggered is $\sim0.88$ h later than \emph{Fermi}/GBM). Considering that the host galaxy properties of energetic GRBs have no obvious distinctiveness (see section \ref{host galaxy} for details), the brighter optical afterglow should not be attributed to the impact of the circumburst environment but should be due to higher kinetic energy. This infers that the radiation efficiency of these energetic GRBs should not be special compared with other normal long bursts.

\subsection{Energy relation between the $E_{\rm \gamma,iso}$ and $E_{\rm X,iso}$}

\begin{figure*}
\centering
\includegraphics    [angle=0,scale=0.8]     {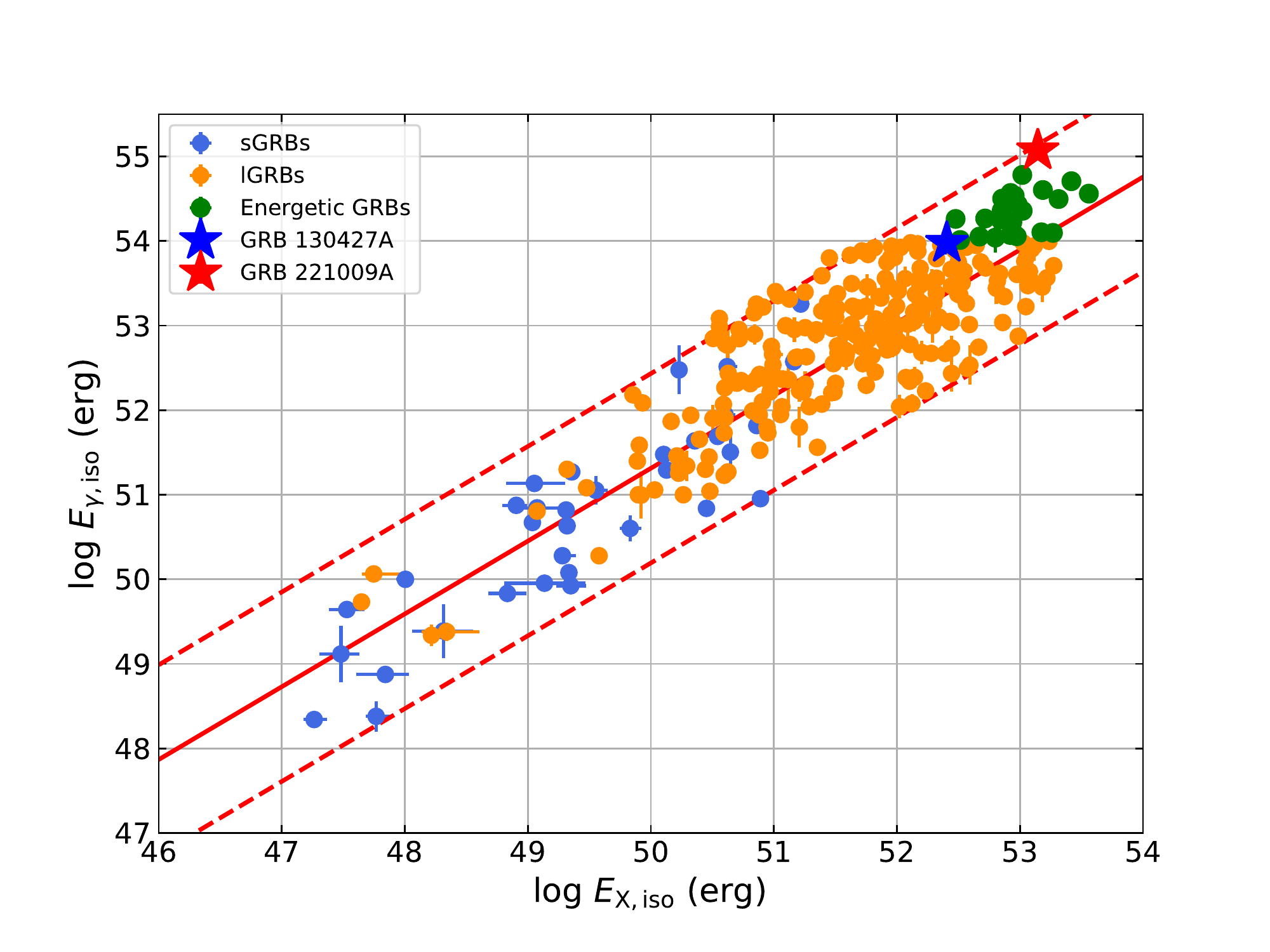}
\caption{The correlation between $E_{\rm X,iso}$ and $E_{\rm \gamma, iso}$.  The red solid line and dashed lines are the best-fit and the 95\% confidence level of the fits, respectively. The blue circles, orange circles, blue star, red star, and green circles represent the sGRBs, lGRBs, GRB 130427A, GRB 221009A, and other energetic lGRBs, respectively.}
\label{fig:Exiso}
\end{figure*}

It has been proposed that the energy partition between the prompt emission and afterglow may be quasi-universal, i.e. a GRB with more energetic prompt emission can power a more energetic afterglow emission \citep{Zou19}. Here it is of interest to investigate the relation between the energies released in the $\gamma$-ray band ($E_{\rm \gamma,iso}$) and in the X-ray band ($E_{\rm X,iso}$). $E_{\rm X,iso}$ can be calculated by $E_{\rm X,iso}=4k\pi D_{L}^2S_{\rm X}/(1+z)$, where $S_{\rm X}$ is observed
X-ray fluence and $k$ is the correction factor that corrects the XRT-band (0.3-10 keV) flux to a wide band in the burst rest frame (0.1-1000 keV). As shown in Figure \ref{fig:Exiso}, there is indeed a strong correlation between $E_{\rm \gamma, iso}$ and $E_{\rm X,iso}$.  Pearson’s correlation coefficient is $r=0.92$ and chance probability $p<10^{-4}$. Our linear fit with the least square regression algorithm gives
\begin{equation}
\log E_{\rm \gamma,iso}/{\rm erg}=(8.25\pm 1.12)+(0.86\pm 0.02)\times \log E_{\rm X,iso}/{\rm erg}.
\end{equation}
Our result strengthens that the energy partition between the prompt emission and afterglow is quasi-universal. We find that all energetic GRBs, including GRB 221009A, also satisfy this correlation very well.

\subsection{Initial Lorentz factor distribution}

It is well known that GRBs are powered by relativistic outflows. The initial Lorentz factor $\Gamma_0$ during the GRB prompt emission phase is very important to understand the physics of GRBs. Here, the collected optical afterglow data allows us to estimate the $\Gamma_0$ for our sample. Using the peak time $t_{\rm peak}$ of the onset of early afterglow as the deceleration time of the external forward shock, one can constrain the $\Gamma_0$ \citep{Sari99}. If the peak time is not detected due to without timely observations or pollution of other emission components (e.g. reverse shock emission), one can take the first optical observation time of the normal decay phase as the upper limit of the peak time and thus derive the lower limit of $\Gamma_0$. In the so-called `thin' shell case, the deceleration timescale $t_{\rm dec}\sim R_{\rm dec}/(2c\Gamma^2_{\rm dec})$ corresponds to the quantity $t_{\rm peak}/(1+z)$,
where $R_{\rm dec}$ is the deceleration radius, $c$ is the speed of light and $\Gamma_{\rm dec}$ is the fireball Lorentz factor at $t_{dec}$. We apply the standard afterglow model with a constant-density medium (i.e., the interstellar medium [ISM]) to derive the initial Lorentz factor, which is twice that of the Lorentz
factor at the deceleration timescale \citep{Sari99}
\begin{equation}\label{Gamma_0}
\Gamma_0=2\left[\frac{3E_{\rm \gamma,\ iso}(1+z)^3}{32\pi nm_pc^5 \eta t_{\rm peak}^3}
\right]^{1/8}\simeq 193(n\eta)^{-1/8}\times \left(\frac{E_{\rm
\gamma,iso,52}}{t_{\rm dec,2}^3} \right)^{1/8},
\end{equation}
and the deceleration radius
\begin{equation}
R_{\rm dec}=2c t_{\rm peak} \Gamma_{\rm dec}^2/(1+z)=2.25\times 10^{16}{\rm cm}\ \Gamma_{0,2}^2t_{\rm peak,2}/(1+z).
\end{equation}
Here we take $n=1$ cm$^{-3}$ and $\eta=0.2$.

Using the method above, we can constrain $\Gamma_0$ for the energetic 
GRBs with enough observational data. For GRB 221009A, the onset timescale is earlier than the very first optical detection at $\sim$ 3000 s. One can take the first optical observation time of the normal decay phase as the upper limit of the peak time and thus derive the lower limit $\Gamma_0>72$ for GRB 221009A.

\begin{figure*}
\centering
\includegraphics    [angle=0,scale=0.7]     {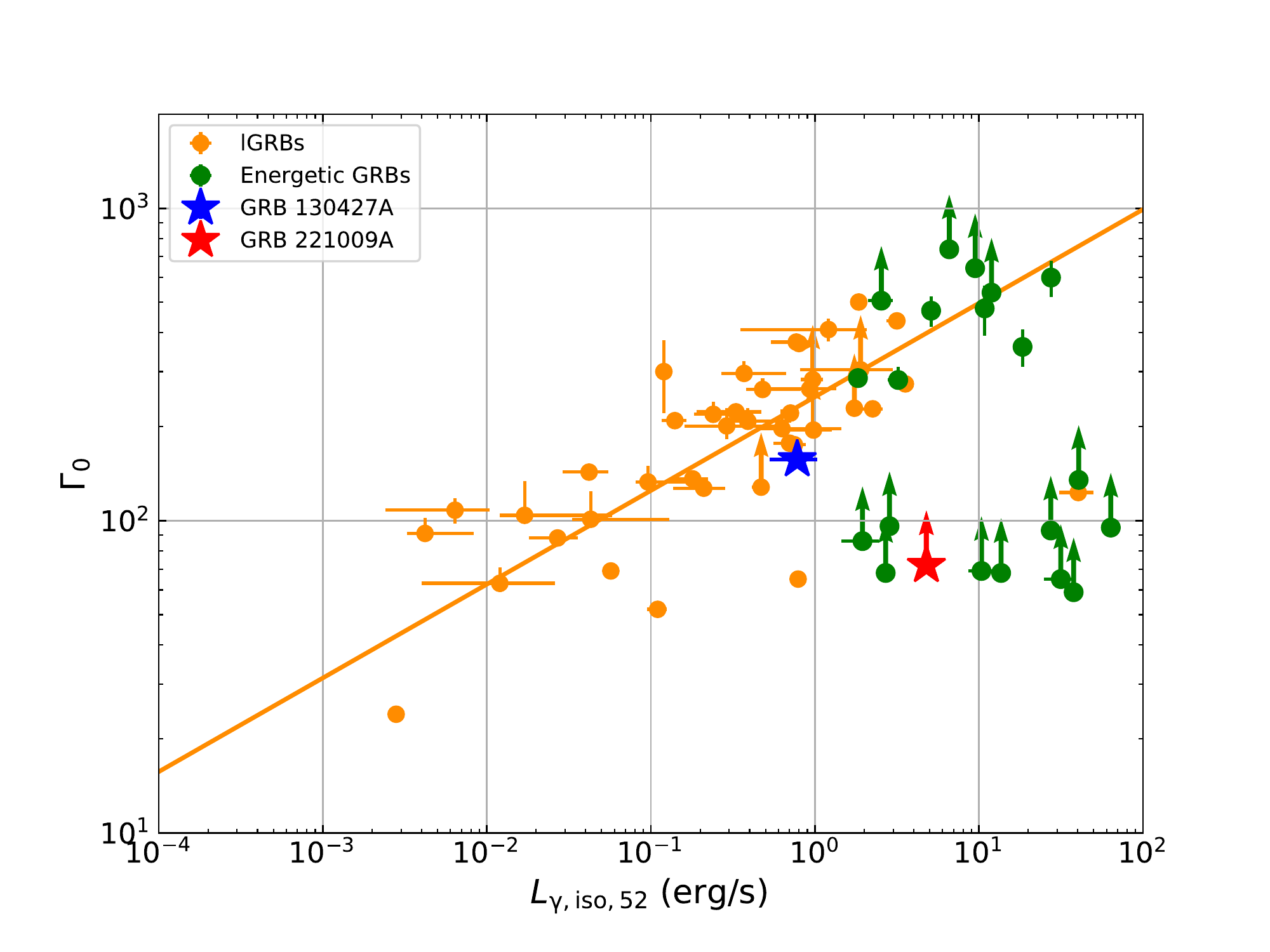}
\includegraphics    [angle=0,scale=0.7]     {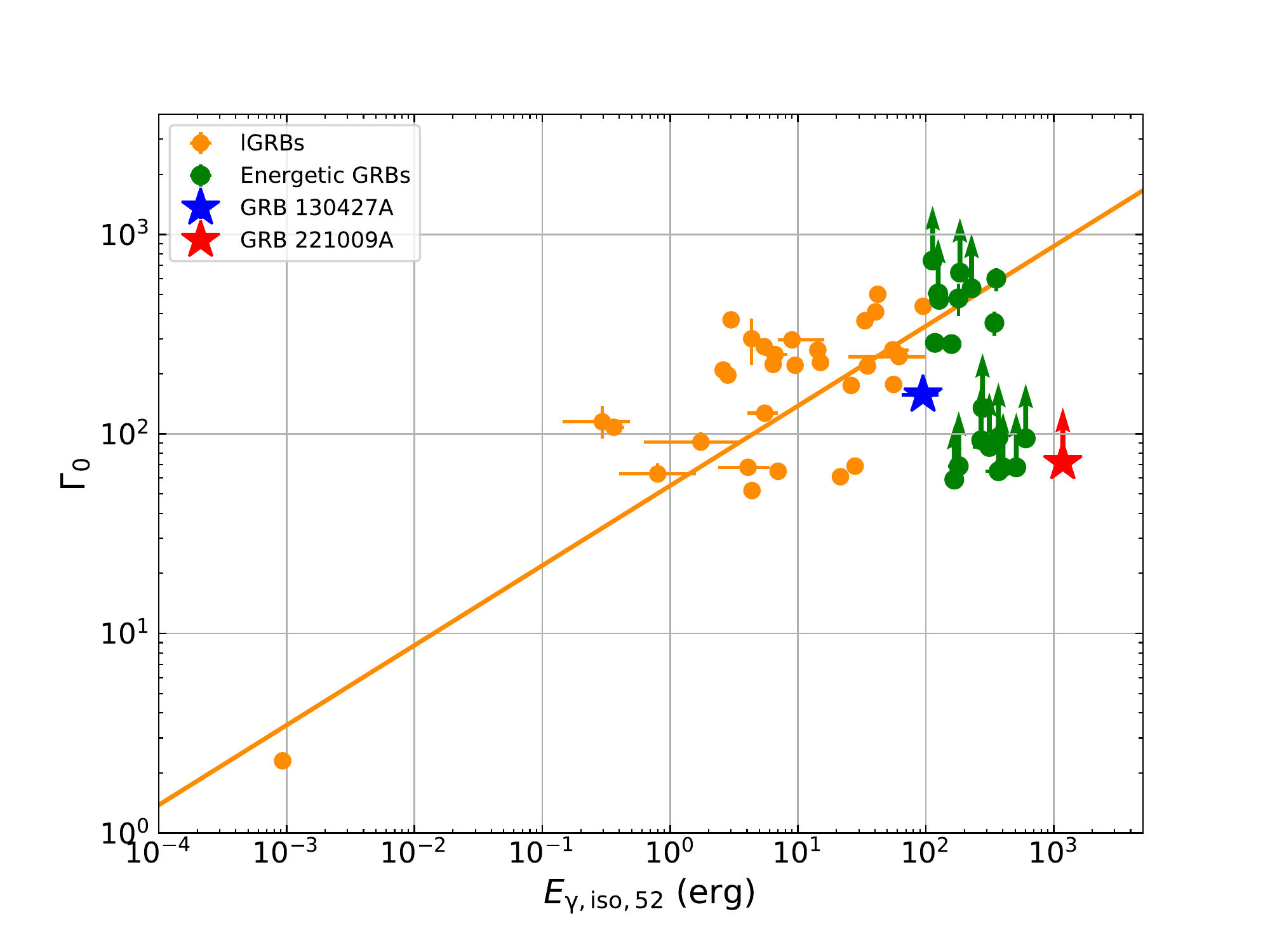}
\caption{The $L_{\gamma,\rm iso}$,  $E_{\gamma,\rm iso}$ and $\Gamma_0$ relation reported by \cite{Lv12,Liang10}. The orange circles data from \cite{Lv12}(top panel) and \cite{Liang15}(bottom panel), and the GRB 130427A, GRB 221009A and other energetic GRBs are marked with different colors and shapes in the plot. The orange solid line marks the relation.}
\label{fig:Liso-gamma0}
\end{figure*}

\cite{Liang10} proposed a correlation between the isotropic energy of prompt emission $E_{\gamma,\rm iso}$ and the initial Lorentz factor $\Gamma_0$. Later, \cite{Lv12} proposed a correlation between the average isotropic luminosity of prompt emission $L_{\gamma,\rm iso}$ and the initial Lorentz factor $\Gamma_0$. Here we plot the $L_{\gamma,\rm iso}-\Gamma_0$ and $E_{\gamma,\rm iso}-\Gamma_0$ diagram in Figure \ref{fig:Liso-gamma0}. We find that energetic GRBs with onset features well satisfy the correlation, and some energetic GRBs without onset features systematically deviate from this relationship (including GRB 221009A), which is mainly due to the lack of peak time observation.  

\begin{figure*}
\centering
\includegraphics    [angle=0,scale=0.8]     {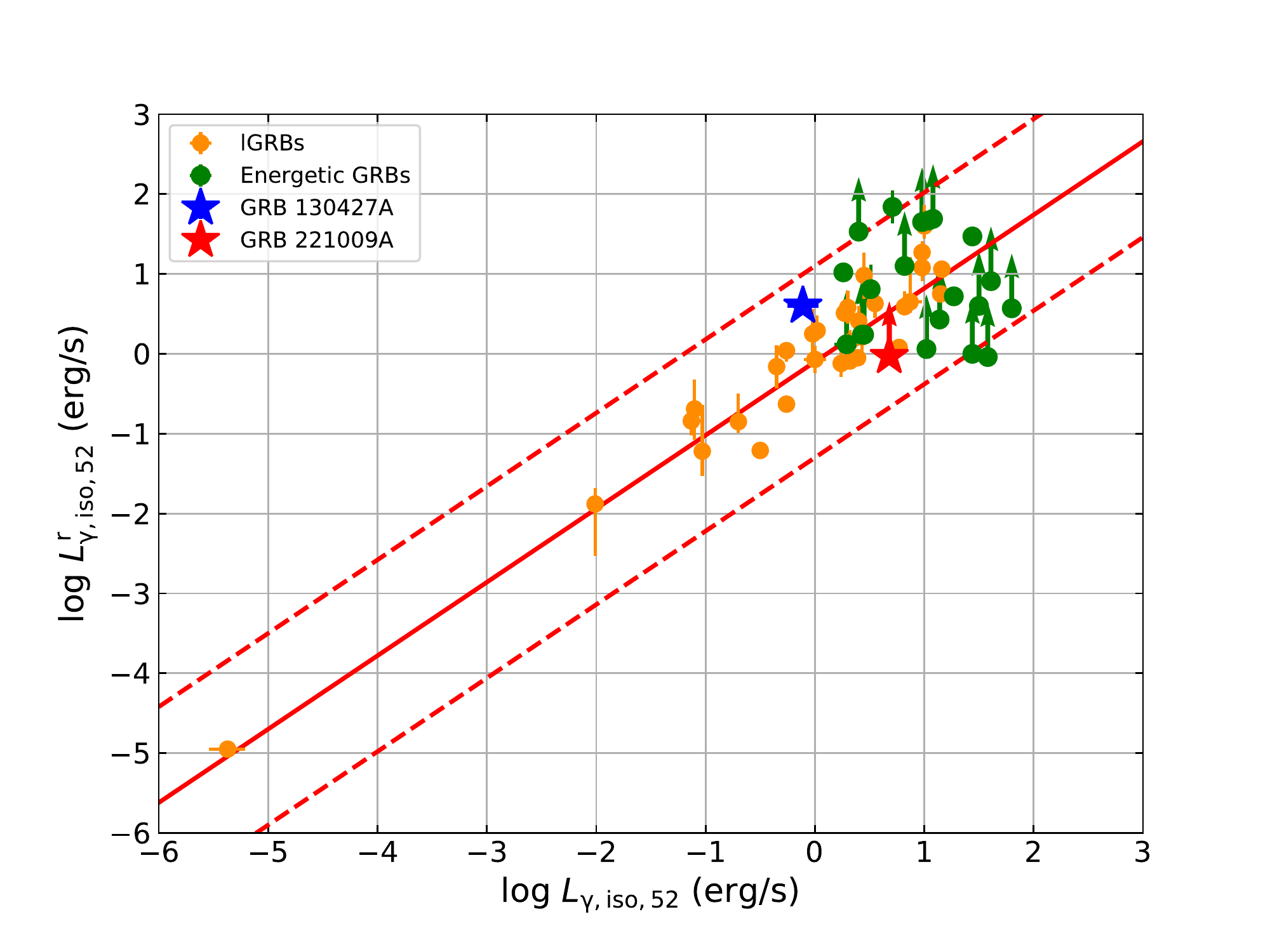}
\caption{Luminosity calculated with the $L_{\gamma, \rm iso}-E_{\rm p,z}-\Gamma_{\rm 0}$ relation reported by \cite{Liang15} as a function of the observed luminosity for GRB 130427A, GRB 221009A and other energetic GRBs as marked in the plot. The orange circles data from \cite{Liang15}, and the black solid and dashed lines mark the relation and its $3\sigma$ dispersion.}
\label{fig:Liso-Ep-gamma0}
\end{figure*}

\cite{Liang15} found another tight $L_{\gamma,\rm iso}-E_{\rm p,i}-\Gamma_0$ correlation, i.e., $L^{\rm r}_{\rm \gamma, iso}=10^{45.62\pm 0.35} ~{\rm erg~s^{-1}}~{(E_{p,i}/{\rm keV})}^{1.34\pm 0.14}\Gamma_{0}^{1.32\pm 0.19}$. This relation combines the GRB jet luminosity, the initial Lorentz factor, and the prompt emission spectrum. It significantly reduces the intrinsic scatters of the $L_{\gamma,\rm iso}-E_{\rm p,i}$ \citep{Liang04, Amati06} and $L_{\gamma,\rm iso}-\Gamma_0$ \citep{Liang10,Lv12} relations. Here we also plot the $L_{\gamma,\rm iso}-E_{\rm p,i}-\Gamma_0$ diagram in Figure \ref{fig:Liso-Ep-gamma0}. We find that the GRB 221009A and other energetic GRBs follow well this relation and tend to locate at the high luminosity end.

\subsection{Host galaxy properties}\label{host galaxy} 

The properties of the host galaxy can provide important information for the study of the properties of the progenitors of GRBs. For instance, the association of some lGRBs with type Ic SNe verifies that lGRBs likely originate from the collapse of massive stars. Therefore, the host galaxies of lGRBs are generally dwarf galaxies with actively star-forming, and lGRBs generally occur in regions with a high star formation rate (SFR) in galaxies \citep{Savaglio2009}. In order to keep enough mass and angular momentum when a massive star collapses, the metallicity should not be too large \citep{Li16}. On the other hand, sGRBs are believed to be formed from compact star mergers and have been confirmed by GW170817/GRB 170817A \citep{Abb2017a,Abb2017b}. Therefore, sGRB's host galaxies have more widely distributed parameters \cite[][and reference therein]{Li16}: sGRBs were detected in both early and late type galaxies; no metallicity limitation is required for sGRBs; some sGRBs are expected to be associated with the old stellar populations and no recent star formation is required; some sGRBs are expected to have a large offset from the original birth location in the host galaxy, since the explosion of SNe that formed the compact binary systems would have given the system two kicks. 

\begin{figure*}
\centering
\includegraphics    [angle=0,scale=0.6]     {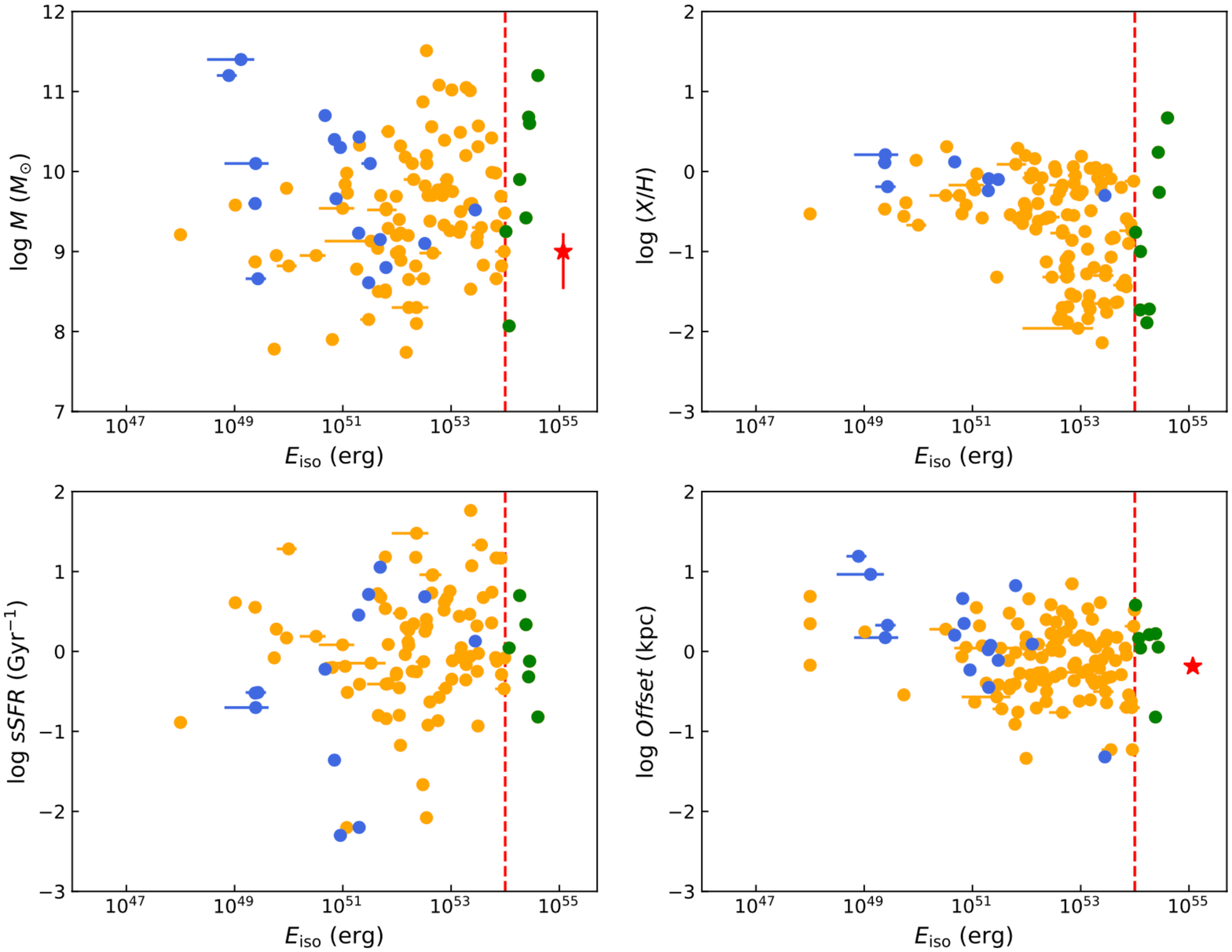}
\caption{Distribution of host galaxy properties of GRBs. The blue and orange circles represent sGRBs and lGRBs respectively. The green circles are energetic GRBs. The rad star mark GRB 221009A and the data is from \citet{Levan2023}. The red dashed line mark the position of $10^{54}$ erg. }
\label{fig:galaxy}
\end{figure*}

Here we compare the energetic GRBs with other GRBs in terms of four important host galaxy properties, including the stellar mass, the SFR, the metallicity, and the offset. The data are mainly taken from \cite{Li16}. Although there are few samples with good host galaxy measurements, it can be clearly seen from Figure \ref{fig:galaxy} that there is no systematic difference between the energetic GRBs and other normal lGRBs, indicating that the energetic GRBs are likely not from special progenitor systems.

\section{Conclusion and Discussion}

GRB 221009A is the closest and most energetic GRB (with $E_{\gamma,\rm iso}\sim10^{55} {\rm ergs}$) detected so far. Its emergence further strengthens our interest in the study of energetic GRBs. In this work, we extensively collect a good sample of GRBs with well measured redshifts and spectral parameters. The sample covers the redshift range from 0.0098 to 8.23, and the isotropic $\gamma$-ray energy range from $4.7\times10^{46}$ ergs to $\sim10^{55}$ ergs. 

With the collected sample, we have studied the GRB energy functions and luminosity functions at different redshifts in detail. We find that for the low redshift subsample with $0<z<0.5$, even though the best fitting model of the energy function is a PL (regardless of whether GRB 221009A is introduced or not), the results of the BIC analysis do not support that PL model is clearly better than CPL and BPL models. If we extend the redshift range from $z<0.5$ to $z<1$, the best fitting model of the energy function becomes a CPL (regardless of whether GRB 221009A is introduced or not), and again the results of the BIC analysis do not support that the CPL model is clearly better than the other two models. For the high redshift samples, we find that both the CPL and BPL models could well fit the observational data, but the PL model could be excluded with high significance as long as the sample size is large enough. Nevertheless, we find that the best fitting parameters for different redshift samples are in good agreement with each other. Based on our finding, we suggest that the energy function of GRBs does not evolve with redshift, and always follows the CPL or BPL model, namely, there is always a cutoff or break in the high energy end. Assuming that the best fitting result of the total sample can represent the intrinsic distribution of the GRB energy function, we find that the occurrence of GRB 221009A is consistent with the expectation within 1.84 $\sigma$ Poisson fluctuation error. 

On the other hand, with the collected sample, we have investigated whether GRB 221009A and other energetic GRBs are systematically different from other normal GRBs in terms of various statistical  properties, including the prompt emission, afterglow, and host galaxy properties. We find that the energetic GRBs, including GRB 221009A, do not show significant peculiarity compared with other normal lGRBs in the following aspects: $T_{90}$ distribution, minimum timescale distribution, Amati relation,  $E_{\rm \gamma,iso}$-$E_{\rm X,iso}$ relation, $L_{\gamma,\rm iso}-\Gamma_0$ relation, $E_{\gamma,\rm iso}-\Gamma_0$ relation, $L_{\gamma,\rm iso}-E_{\rm p,i}-\Gamma_0$ relation, and the distributions of host galaxy properties, including stellar mass,  SFR, metallicity and offset. 

There are some characteristics of energetic GRBs that differ somewhat from normal GRBs. However, they are all understandable. For example, the average spectral lag of energetic GRBs is smaller than that of normal lGRBs, but this is consistent with the luminosity - spectral lag correlation, \citep{Norris00,Gehrels06}. Their optical afterglows are sysmetically brighter than other GRBs, but this is expected if the GRB efficiency does not significantly depend on energy \citep{Lloyd-ronning04,Wang15}). Finally, most energetic GRBs show a simple power-law decay lightcurve with decay slopes systemically steeper compared to the so-called ``normal decay slope" (with a typical slope approximately -1.2, \cite{zhang2006}). This may be related to a structured jet viewed at the central core, which can explain their high isotropic energy \citep{Meszaros98,Dai01}.

The facts that GRB 221009A and other energetic GRBs follow the same energy function and luminosity function as normal lGRBs and that their statistical properties are consistent with normal lGRBs suggest that there is nothing special for these bursts except their apparent brightness ($E_{\rm \gamma,iso}$). This suggests that they likely share the similar progenitor systems and experience similar energy dissipation processes and radiation mechanisms as normal lGRBs.

The large apparent energies may be related to the properties of the central engine, such as the black hole mass and spin, or the accretion process of the central engine.
However, a more natural understanding would be that they are related to a special viewing angle of a quasi-universal structured jet, as has been proposed to account for the luminosity function of the entire lGRB population \citep{zhang2002,rossi2002}. Within this picture, the structured jet has a nearly uniform narrow core surrounded by a wing with a decreasing energy per unit solid angle with increasing viewing angle. Depending on the shape of the structured jet in the wing (e.g. power law or Gaussian, \citet{zhang2002}), the slope of the energy function / luminosity function could be different. When the line of sight enter the core, the luminosity would show a cutoff. The narrowness of the core ensures the rareness of energetic GRBs. GRB 221009A, with the record-breaking $E_{\rm \gamma,iso} \sim 10^{55}$ erg, suggests that central core can be very narrow. This is consistent with the LHAASO results \citep{cao23}. The structured jet wing can also help to interpret the relatively steep afterglow decay index in the X-ray band, see also \cite{sato2022}.

\section*{Acknowledgments}
This work is supported by the National Natural Science Foundation of China (Projects:12021003, U2038107, U1931203), and the National SKA Program of China (grant No. 2022SKA0130101).


\end{document}